\documentclass[longauth]{aa}
\usepackage{graphicx}
\usepackage[varg]{txfonts}
\usepackage[colorlinks=true, allcolors=blue]{hyperref}
\usepackage{multirow}
\usepackage{array}
\usepackage{CJK}
\usepackage[utf8]{inputenc}
\usepackage{silence}
\usepackage{ulem}
\usepackage{mathtools}
\usepackage{amsmath}
\usepackage{orcidlink}
\usepackage{booktabs}
\usepackage{longtable}
\usepackage{float}
\usepackage{tabularx}
\usepackage{graphicx}
\usepackage{threeparttable}
\usepackage{tabularx}
\usepackage{multirow}
\usepackage{amsmath}
\usepackage{amssymb}
\usepackage{flushend}
\usepackage{xcolor}

\newcolumntype{Y}{>{\centering\arraybackslash}X}

\newcommand{\Ha} {\mbox{H$\alpha$}\,}

\newcommand{\degree}{\mbox{$^\circ$}}

\begin{document} 
    \title{SN~2025aico: Early observations of a faint type IIb supernova with a low-mass envelope}

   \authorrunning{J.-W Zhao et al.} 
   \titlerunning{SN~2025aico}
   
\author{
        J.-W.~Zhao \inst{\ref{inst1},\ref{inst2}}\and
        A.~Pastorello\inst{\ref{inst3}}\and
        B.~Kumar\inst{\ref{inst1},\ref{inst2}}\corrauth{brajesh@ynu.edu.cn} \and
        Y.-Z.~Cai \inst{\ref{inst3},\ref{inst4},\ref{inst5}}\corrauth{yongzhi.cai@inaf.it (CYZ)} \and
        A.~Dutta\inst{\ref{inst6},\ref{inst7}} \and
        D.~K.~Sahu\inst{\ref{inst6}} \and
        A.~Reguitti\inst{\ref{inst3},\ref{inst8}} \and
        R.~S.~Teja\inst{\ref{inst9}} \and
        H.~Das\inst{\ref{inst6}} \and
        T.~J.~Moriya\inst{\ref{inst10}, \ref{inst11}, \ref{inst12}} \and
        N.~Pyykkinen\inst{\ref{inst13}, \ref{inst14}}\and
        K.~Valeckas\inst{\ref{inst15}, \ref{inst14}}\and
        G.~Valerin \inst{\ref{inst3}}\and
        X.-Z.~Zou\inst{\ref{inst1},\ref{inst2}} \and
        C.~Ashall\inst{\ref{inst16}}\and
        S.~Bijavara~Seshashayana\inst{\ref{inst17},\ref{inst14}}\and
        G.-W.~Du\inst{\ref{inst1},\ref{inst2}} \and
        G.~C.~Anupama\inst{\ref{inst6}} \and
        A.~L.~Bouquin\inst{\ref{inst14}, \ref{inst18}}\and
        S.~Campana\inst{\ref{inst8}} \and
        K.~Chatterjee\inst{\ref{inst1},\ref{inst2}} \and
        X.-L.~Chen\inst{\ref{inst1},\ref{inst2}} \and
        X.-L.~Du\inst{\ref{inst1},\ref{inst2}} \and
        N.~Elias-Rosa\inst{\ref{inst3},\ref{inst19}} \and
        Y.~Fang\inst{\ref{inst1},\ref{inst2}} \and
        M.~Fraser\inst{\ref{inst20}} \and
        W.~Hoogendam\inst{\ref{inst16}} \and
        E.~Hsiao\inst{\ref{inst21}} \and
        E.~Kankare\inst{\ref{inst13}} \and 
        E.~P.~Lagioia\inst{\ref{inst1},\ref{inst2}} \and
        W.-Y.~Li\inst{\ref{inst1},\ref{inst2}} \and
        X.-K.~Liu\inst{\ref{inst1},\ref{inst2}} \and
        P.~Lundqvist\inst{\ref{inst22}} \and
        K.~Matilainen\inst{\ref{inst13}, \ref{inst14}} \and
        J.~Martikainen\inst{\ref{inst23}, \ref{inst14}} \and
        K.~Medler\inst{\ref{inst16}} \and
        N.~Morrell\inst{\ref{inst24}}\and
        Y.~Pan\inst{\ref{inst1},\ref{inst2}} \and
        C.~Pfeffer\inst{\ref{inst16}}\and
        G.~Rameshan\inst{\ref{inst6}} \and
        T.~M.~Reynolds\inst{\ref{inst15}, \ref{inst25}, \ref{inst26}} \and
        M.~D.~Stritzinger\inst{\ref{inst27}} \and
        V.~Vuolteenaho\inst{\ref{inst28},\ref{inst14}} \and
        Z.-Y.~Wang\inst{\ref{inst29},\ref{inst30}} \and
        H.-F.~Xiao\inst{\ref{inst21}} \and
        J.-H.~Zhang\inst{\ref{inst1}, \ref{inst2}} \and
        X.-W.~Liu \inst{\ref{inst1}, \ref{inst2}}\corrauth{x.liu@ynu.edu.cn (LXW)} \and
        Y.-P.~Yang \inst{\ref{inst1}, \ref{inst2}}
    }
    
    \institute{
        \label{inst1}South-Western Institute for Astronomy Research, Yunnan University, Kunming 650500, P.R. China
        \and\label{inst2}Yunnan Key Laboratory of Survey Science, Yunnan University, Kunming, Yunnan 650500, P.R. China 
        \and\label{inst3}INAF - Osservatorio Astronomico di Padova, vicolo dell'Osservatorio 5, I-35122 Padova, Italy
        \and\label{inst4}Yunnan Observatories, Chinese Academy of Sciences, Kunming 650216, P.R. China
        \and\label{inst5}International Centre of Supernovae, Yunnan Key Laboratory, Kunming 650216, P.R. China
        \and\label{inst6}Indian Institute of Astrophysics, II Block, Koramangala, Bengaluru-560034, Karnataka, India
        \and\label{inst7}Pondicherry University, Chinna Kalapet, Kalapet, Puducherry 605014, India
        \and\label{inst8}INAF - Osservatorio Astronomico di Brera, Via E. Bianchi 46, 23807 Merate (LC), Italy
        \and\label{inst9}Tsung-Dao Lee Institute, Shanghai Jiao Tong University, No.1 Lisuo Road, Pudong New Area, Shanghai, P.R. China
        \and\label{inst10}National Astronomical Observatory of Japan, National Institutes of Natural Sciences, 2-21-1 Osawa, Mitaka, Tokyo, Japan
        \and\label{inst11}Graduate Institute for Advanced Studies, SOKENDAI, 2-21-1 Osawa, Mitaka, Tokyo 181-8588, Japan
        \and\label{inst12}School of Physics and Astronomy, Monash University, Clayton, VIC 3800, Australia
        \and\label{inst13}Department of Physics and Astronomy, University of Turku, FI-20014 Turku, Finland
        \and\label{inst14}Nordic Optical Telescope, Aarhus University, Rambla Jos\'e Ana Fern\'andez P\'erez 7, local 5, E-38711 San Antonio, Bre\~na Baja, Santa Cruz de Tenerife, Spain
        \and\label{inst15}Niels Bohr Institute, University of Copenhagen, Jagtvej 128, 2200 Copenhagen N, Denmark
        \and\label{inst16}Institute for Astronomy, University of Hawai’i, Honolulu, HI 96822, USA
        \and\label{inst17}Materials Science and Applied Mathematics, Malm\"o University, SE-205 06 Malm\"o, Sweden
        \and\label{inst18}Technical University of Denmark, DK-2800 Lyngby, Denmark
        \and\label{inst19}Institute of Space Sciences (ICE, CSIC), Campus UAB, Carrer de Can Magrans, s/n, E-08193 Barcelona, Spain 
        \and\label{inst20}School of Physics, O'Brien Centre for Science North, University College Dublin, Belfield, Dublin 4, Ireland
        \and\label{inst21}Department of Physics, Florida State University, 77 Chieftan Way, Tallahassee, FL32306, USA
        \and\label{inst22}The Oskar Klein Centre, Department of Astronomy, Stockholm University, AlbaNova, SE-10691 Stockholm, Sweden
        \and\label{inst23}Instituto de Astrof\'isica de Canarias (IAC), C/V\'ia L\'actea s/n, 38205 La Laguna, Tenerife, Spain
        \and\label{inst24}Carnegie Observatories, Las Campanas Observatory, Colina El Pino, Casilla 601, Chile
        \and\label{inst25}Tuorla Observatory, Department of Physics and Astronomy, University of Turku, FI-20014 Turku, Finland 
        \and\label{inst26}Cosmic Dawn Center (DAWN)
        \and\label{inst27}Department of Physics and Astronomy, Aarhus University, Ny Munkegade 120, DK-8000 Aarhus C, Denmark
        \and\label{inst28}Space Physics and Astronomy research unit, PO Box 3000, FI-90014 University of Oulu, Finland
        \and\label{inst29}School of Astronomy and Space Science, University of Chinese Academy of Sciences, Beijing 100049, P.R. China
        \and\label{inst30}National Astronomical Observatories, Chinese Academy of Sciences, Beijing 100101, P.R. China
    }

    \date{Received xx xx, 2026; accepted xx xx, 2026}
    
    \abstract
    {Originating from massive stars that have lost most of their hydrogen envelopes, type IIb supernovae are stellar explosions of partially stripped stars. Due to this extensive stripping, their spectra exhibit transitional signatures that bridge the gap between hydrogen-rich (type II) and helium-rich (type Ib) events, making them crucial astrophysical probes for investigating the final evolutionary stages of massive stars.}
    {We aimed to investigate the physical properties and the underlying explosion mechanisms of the type IIb SN~2025aico. Through a comprehensive analysis of early-phase optical light curves and spectroscopic data, we aimed to constrain the fundamental explosion parameters and evaluate the physical state of the event.}
    {We present early multi-band optical imaging and low-resolution optical spectroscopic follow-up observations of the type~IIb SN~2025aico, spanning $\sim70$~days from the explosion. We constrained the properties of SN~2025aico using a hybrid model that combines shock-cooling emission and radioactively powered diffusion, as well as through an analysis of its spectroscopic evolution. We used various approaches to constrain $^{56}$Ni mixing from early data and compared our spectra with models to constrain the properties of the progenitor.}
    {The explosion epoch of SN~2025aico is estimated to be Modified Julian Date (MJD)~$61032.69$, while the rise time in the $r_\mathrm{M}$ band is $22.30\pm0.70$~days. The peak pseudo-bolometric luminosity in the optical bands is $L_{\mathrm{opt}}=(4.07\pm0.10)\times10^{41}\,\mathrm{erg\,s^{-1}}$. The fitting yields a moderate to relatively low $^{56}$Ni mass of $M_{\mathrm{Ni}} = 0.033_{-0.004}^{+0.006}\,\mathrm{M_\odot}$ and an ejecta mass of $M_\mathrm{ej} = 2.79_{-0.18}^{+0.21}\,\mathrm{M_\odot}$. The photospheric velocity near the bolometric peak, measured from the \ion{Fe}{II}~$\lambda5169$ line, is $6450_{-160}^{+180}\,\mathrm{km\,s^{-1}}$. The derived envelope properties suggest a compact He-star progenitor possessing an H-rich envelope of $M_{\mathrm{env}} \approx0.02\,\mathrm{M_\odot}$ and a radius of $R_{\mathrm{env}} \approx 6\,$--$10\,\mathrm{R_\odot}$.}
    {The derived physical properties of SN~2025aico indicate an origin from a moderate-mass, stripped He-star in a compact binary system, characterized by a minimal residual hydrogen envelope. The explosion itself demonstrates weak to moderate $^{56}\mathrm{Ni}$ mixing throughout the ejecta.}

    \keywords{ supernovae: general -- supernovae: individual: SN~2025aico -- galaxies: individual: LEDA~35384} 
    
    \maketitle
    \nolinenumbers

\section{Introduction}

Massive stars ($M_\mathrm{ZAMS} \gtrsim 8\,\mathrm{M_\odot}$) end their evolution as core-collapse (CC) supernovae (SNe), leaving behind a compact remnant \citep{Woosley2002RevModPhys, Janka2012ARNPS}. Observationally, CC SNe are classified based on the presence or absence of hydrogen lines in their spectra. Events exhibiting prominent hydrogen lines are classified as type II SNe. Conversely, SNe lacking persistent hydrogen features are categorized as stripped-envelope (SE) SNe \citep{Clocchiatti1996ApJ}. This class encompasses type IIb events, which exhibit transient hydrogen features only in their early spectra; type Ib SNe, defined by the presence of helium; and type Ic SNe, which are distinguished by the absence of both hydrogen and helium signatures \citep{Filippenko1997ARAA, Modjaz2019NatAs}. These events originate from massive progenitors that have been partially or completely stripped of their hydrogen or helium envelopes, either by strong stellar winds \citep[e.g. Wolf-Rayet, WR, stars;][]{Smith2014ARAA, Crockett2008MNRAS} or via mass transfer processes in binary systems through Roche-lobe overflow (RLOF) \citep{Ritter1988AA, Maund2004Natur.427..129M}.

Type~IIb events belong to a subclass of SE SNe and are considered transitional objects between SNe~II and SNe~I. They exhibit hydrogen lines during their early evolution, whereas at later times, their spectra become dominated by helium lines, resembling those of SNe~Ib \citep{Filippenko1988AJ}. The photometric and spectroscopic diversity of SNe~IIb is fundamentally driven by the properties of their residual thin hydrogen envelopes. Based on progenitor characteristics, SNe~IIb can be classified into compact (cIIb) and extended (eIIb) subcategories \citep{Chevalier2010ApJL}. Supernova cIIb progenitors are characterized by relatively small radii ($\sim1$--$100\,\mathrm{R_\odot}$) and higher pre-explosion masses, resulting in single-peaked light curves or those with a negligible shock-cooling phase \citep{Barmentloo2024MNRAS}. Conversely, SNe eIIb progenitors possess more extended radii ($\gtrsim100$--$200\,\mathrm{R_\odot}$) and lower masses, resulting in double-peaked light curves. In the SN~eIIb scenario, the initial peak is governed by shock-cooling emission \citep{Nagy2016aap, Dessart2018AA}, whereas the broader secondary peak is sustained by radioactive decay \citep{Arnett1982apj, Arnett1989apj} and recombination processes \citep{Nagy2014AA, Nagy2016aap}. In contrast, in the SN~cIIb scenario, the shock-cooling phase is much shorter and fainter, or even entirely absent.

The degree of $^{56}$Ni mixing in SE~SNe significantly affects their early light curves and spectroscopic evolution, playing a crucial role in the overall development of these transients \citep{Ensman1988ApJ...333..754E, Woosley1997ASIC..486..821W, SN2011dh_Bersten2012ApJ...757...31B}. Such mixing can arise from large-scale asymmetric explosions and hydrodynamic instabilities \citep{Maund2009ApJ...705.1139M, Dessart2012MNRAS.424.2139D}. \cite{Dessart2015MNRAS.453.2189D} demonstrated that He features in SE~SN spectra result from the combined effects of $^{56}$Ni mixing and the presence of He-rich material. In cases of modest mixing, these features may be absent. Furthermore, \cite{Yoon2019ApJ...872..174Y} noted that $^{56}$Ni mixing strongly influences the early-phase colour evolution. Additionally, \cite{Moriya2020MNRAS.497.1619M, Ergon2024A&A} showed that varying degrees of $^{56}$Ni mixing result in different photospheric evolutions. This breaks the degeneracy among the parameters governing the early colour evolution, thereby providing more robust estimates of phenomena such as the shock-cooling phase. To accurately constrain the $^{56}$Ni mixing, early-phase data are crucial. Relying solely on peak and late-phase data makes such constraints highly difficult \citep{SN2017iro_Kumar2022ApJ...927...61K}. Moreover, because SE~SN explosions are often inherently asymmetric \citep{Baal2023MNRAS}, any mixing constraints derived from early-phase observations are strictly limited to the line of sight.

Constraining the progenitor properties of SE~SNe is important, as it provides critical insights into late-stage stellar evolution and mass transfer process in massive binary \citep{Dessart2015MNRAS.453.2189D, Dessart2016MNRAS, Yoon2017ApJ}. Furthermore, certain type cIIb events bridge the gap between typical SNe~IIb and SNe~Ib \citep{Medler2021MNRAS}, suggesting an evolutionary continuum characterized by the presence of only a residual amount of hydrogen. However, placing tight constraints on these progenitor properties requires early-phase observations with comprehensive coverage. Therefore, we present early multi-band photometric and low-resolution spectroscopic observations of the nearby cIIb SN~2025aico, obtained using various facilities. By analysing this event, we aim to constrain the properties of the progenitor and the degree of early $^{56}$Ni mixing. These findings yield a clearer picture of the early radiative transfer process of the explosion and the binary stellar evolutionary pathways leading to cIIb events.

In this work, we modelled the light curve by incorporating both the shock-cooling phase and radioactively powered diffusion \citep{P15_Piro2015ApJ...808L..51P, SW17_Sapir2017ApJ...838..130S, Arnett1989apj, Nagy2016aap}. Furthermore, we performed a semi-quantitative spectroscopic analysis of crucial features and investigated the early evolution to constrain $^{56}$Ni mixing, a critical parameter for early SE SNe evolution \citep{Yoon2019ApJ...872..174Y, Moriya2020MNRAS.497.1619M}. Drawing upon these modelling and analytical results, we discuss the nature of the progenitor and the explosion mechanism.
In Sect.~\ref{sec:info}, we report the fundamental properties of SN~2025aico and its host galaxy. In Sect.~\ref{sec:phot}, we present the photometric data and analyse the light curve features. In Sect.~\ref{sec:spec}, we describe the spectral sequence and perform a semi-quantitative spectroscopic analysis. In Sect.~\ref{sec:disc}, we examine the shock-cooling emission and non-thermal line evolution, alongside the proposed progenitor scenario. Finally, we summarize our findings and results in Sect.~\ref{sec:con}.

However, this article focuses primarily on standard early-time and photospheric phase spectroscopic analyses. For a detailed investigation into spectropolarimetry, we refer the reader to Pyykkinen et al. (in prep.).

\section{Discovery, distance, and extinction}
\label{sec:info}
SN~2025aico was first reported by the Gravitational-wave Optical Transient Observer \citep[GOTO;][]{GOTO_dev_Steeghs2022MNRAS.511.2405S} on December 24, 2025, corresponding to JD~2461033 \citep{ONeill2025TNSTR5147....1O}. The most recent non-detection was obtained on December 23 with a limiting magnitude of approximately $20.3$~mag. The discovery magnitude of SN~2025aico was $16.86\pm0.01$~mag in the GOTO-$L$ band. SN~2025aico was classified as a young SN~IIb on December 26, 2025, by \cite{Andrews2025TNSCR5179....1A}, based on a good match with SNe 1993J and 2013df. The redshift measured from the spectra is $z=0.00455$.

\begin{figure}[htbp]
\centering
\includegraphics[width=9cm]{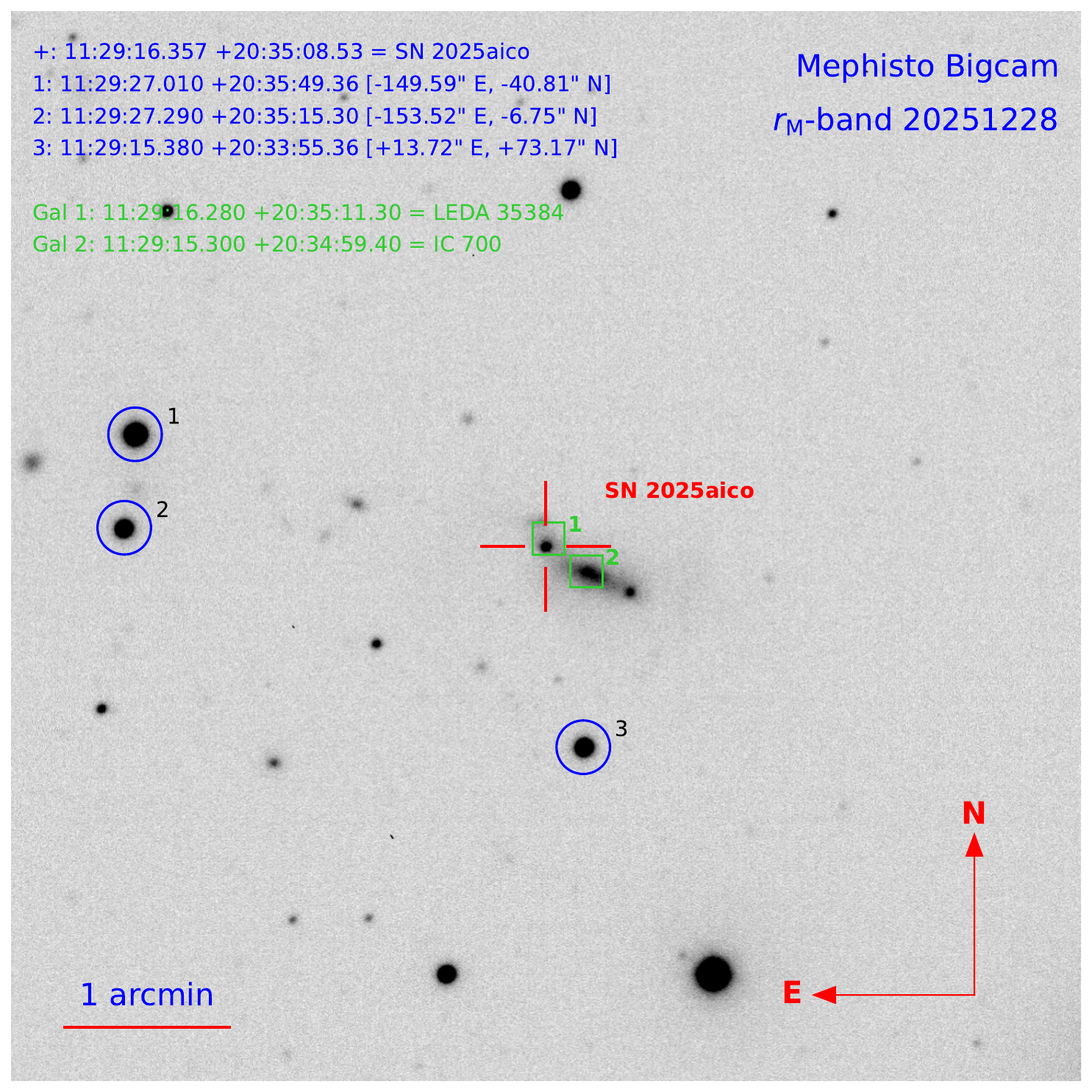}
\caption{$r_{\mathrm{M}}$~band optical image of SN~2025aico, the potential host dwarf galaxy LEDA 35384, along with IC 700, obtained by the Multi-channel Photometric Survey Telescope \citep[Mephisto;][]{Yuan2020SPIE11445E}. The red cross at the centre marks the position of the SN, and the blue circles indicate the stars for future blind-offset acquisition. The light-green square marks the possible host galaxy in the plot.}
\label{fig:phot:image}
\end{figure}

The position of SN~2025aico and its likely host are shown in Fig.~\ref{fig:phot:image}. SN~2025aico (RA=$11^{\rm h}29^{\rm m}16^{\rm s}.364$, DEC=$+20\degree35^{\rm m}08^{\rm s}.26$) is located in a field with several possible host galaxies, the largest of which is IC~700. We identify the closest galaxy, LEDA~35384 (RA = $11^{\rm h}29^{\rm m}16^{\rm s}.280$, DEC = $+20\degree35^{\rm m}11^{\rm s}.30$), as the likely host, which has a redshift of $z=0.004737\pm0.000117$. 
As the distance calculated from the redshift is small and susceptible to peculiar velocity effects, we adopted a distance $D_{\mathrm{L}}=21.5~\mathrm{Mpc}$ \citep{Bitsakis2011A&A...533A.142B} derived from the Tully-Fisher (T-F) relation \citep{Tully2016aj}. The measurement error was not reported, yielding $D_{\mathrm{L}} = 21.5\,\mathrm{Mpc}$. However, because the T-F relation introduces intrinsic uncertainties, we assumed a typical uncertainty of $\sim15\%$, yielding $D_{\mathrm{L}} = 21.5\pm3.2\,\mathrm{Mpc}$. The corresponding distance modulus is $\mu = 31.66\pm0.33$ mag.
We adopt the dust maps of \cite{Schlafly2011apj}, which imply a Galactic extinction along the line of sight of $E(B-V)_{\mathrm{MW}}=0.018\pm0.001~\mathrm{mag}$. Since we did not observe prominent \ion{Na}{I} D absorption in the early spectrum of SN~2025aico, the host extinction is neglected, and we assumed $E(B-V)_{\mathrm{Host}}=0\,\mathrm{mag}$\footnote{We find no prominent \ion{Na}{I}~D absorption features exceeding the $3\sigma$ noise level. Accounting for the local signal-to-noise ratio and instrumental resolution, we derived a conservative upper limit for the host-galaxy equivalent width of $EW(\text{\ion{Na}{I}~D}) \lesssim 0.38$~\AA, which corresponds to an upper limit on the host colour excess of $E(B-V)_{\rm host} \lesssim 0.06$~mag \citep{Cai2022A&A...667A...4C, Turatto2003}.}\footnote{However, for the limitations of relying solely on host \ion{Na}{I}~D absorption lines as a proxy for host extinction, see also \citet{Silva2026arXiv260723690S}. As a consistency check, we compared our data with the colour evolution template discussed in Sect~\ref{sec:phot:colorevo}.}. Thus, the total colour excess due to line-of-sight extinction is $E(B-V)_{\mathrm{Total}}=0.018\pm0.001~\mathrm{mag}$.

\section{Photometry}\label{sec:phot}

\subsection{Apparent light curve}
\label{sec:phot:applc}
The optical and ultraviolet (UV) apparent magnitude light curves are shown in Fig.~\ref{fig:phot:applc}. A detailed description of the observations is provided in Appendix~\ref{sec:app:dataobs}. The photometric data span from MJD~61033.2 to MJD~61100.9, providing a good coverage with a cadence of $\sim1$~day throughout the SN evolution. Only the magnitudes in the UV and the $U,B,V$ filters are reported in the Vega system, while those in all other bands are in the AB magnitude system. We estimated the explosion epoch by taking the average time between the last upper limit ($m_L>20.30\,\mathrm{mag}$) and the first detection ($m_L=16.86\pm0.01\,\mathrm{mag}$) in the GOTO-$L$ band, which yields MJD~$61032.69 \pm 0.51$. We notice that SN~2025aico exhibits a brief shock-cooling emission during the very early phase, similar to those observed in events such as SN~2011hs \citep{SN2011hs_Bufano2014MNRAS.439.1807B} and SN~2011dh \citep{SN2011dh_Arcavi2011ApJ...742L..18A}. The shock-cooling light-curve peak of SN~2025aico is prominent in the shorter-wavelengths bands (with the $U$~band declining by $2.3$~mags in just $3.7$~days). However, as the optical bands are poorly covered at early phases, we cannot estimate the decline rate in these bands.

To better constrain the epoch and the magnitude at maximum in all bands, which exhibit fairly good coverage, we applied polynomial fits to individual light curves; the uncertainties are estimated through a Monte-Carlo process. The apparent magnitude light curves of SN~2025aico reach the maximum brightness on MJD~$61052.90 \pm 0.28$ in the $g_{\mathrm{M}}$ band, and MJD~$61054.99\pm0.48$ in the $r_{\mathrm{M}}$ band. These provide rise times of $20.21\pm 0.58$~days and $22.30\pm0.70$~days in the $g_{\mathrm{M}}$ and $r_{\mathrm{M}}$ bands, respectively. These rise times are consistent with those observed in typical SNe IIb such as SN~2011dh \citep[$\sim21.9$~days in the $R$ band;][]{SN2011dh_Ergon2014A&A...562A..17E} and SN~2024abfo \citep[$\sim24.1$~days in the $r$ band;][]{Reguitti2025AA}. The  apparent magnitude at maximum are $15.72 \pm 0.02$~mag and $15.60 \pm 0.02$~mag in $g_{\mathrm{M}}$~band and $r_{\mathrm{M}}$~band, respectively.

\begin{figure*}[htbp]
\centering
\sidecaption
\includegraphics[width=12cm]{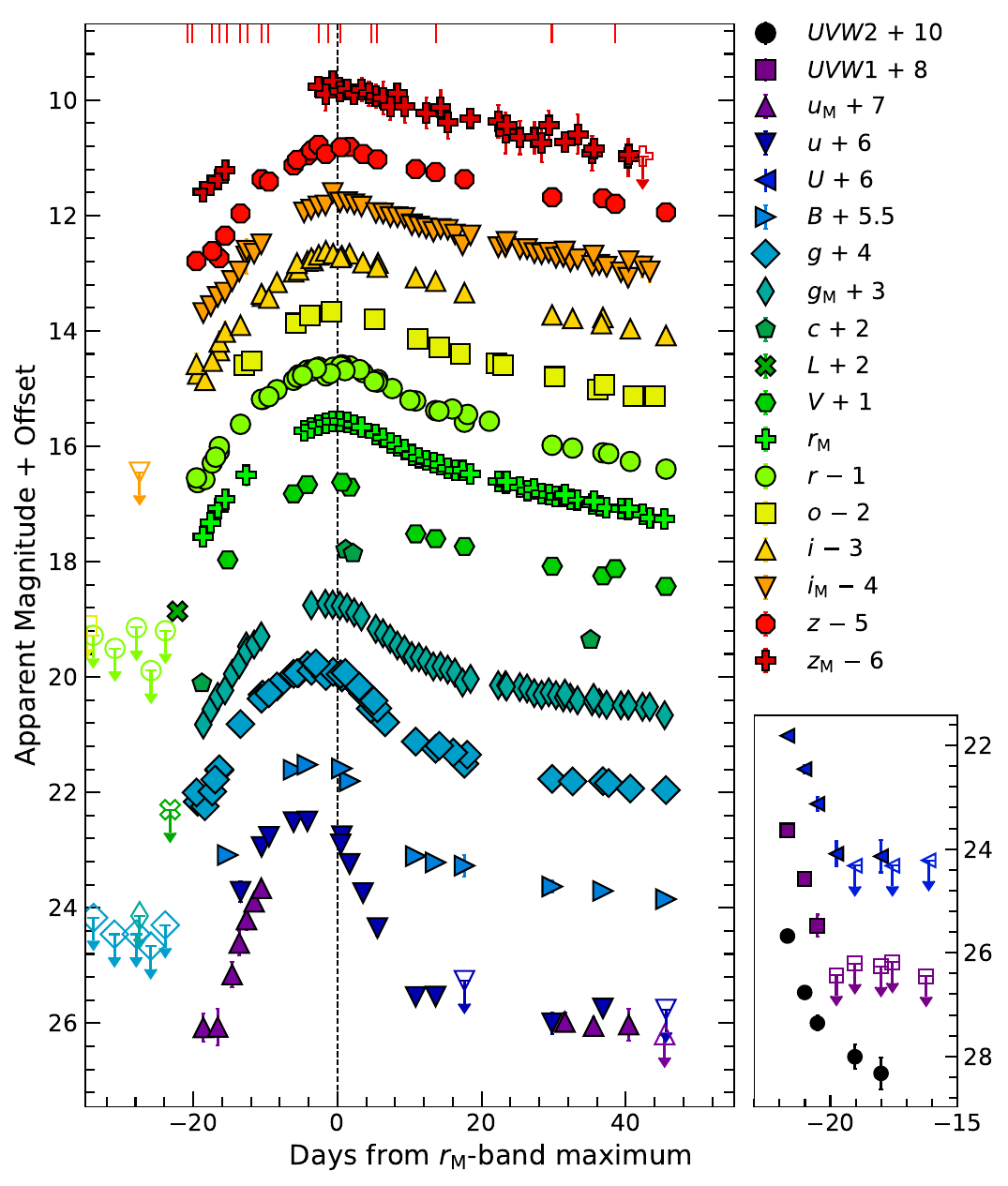}
\caption{Multi-band UV and optical apparent light curve of SN~2025aico. The vertical dashed line refers to the time of the $r_\mathrm{M}$-band maximum luminosity. The vertical red lines at the top mark the epochs of the spectra. Upper limits are plotted with empty symbols with arrows. The light curves for the different filters are shifted and marked with different colours. The early NUV and UV data obtained from \textit{Swift/UVOT} are shown in the bottom right panel with corresponding offsets.}
\label{fig:phot:applc}
\end{figure*}

We used a linear fitting procedure to the post-peak light curves to determine the decline rates in time different intervals ($0 \lesssim t \lesssim 15$~days from maximum and $t \gtrsim 20$~days after maximum). We note a rapid post-peak decline in the apparent light curves, which can be attributed to adiabatic expansion combined with radioactive decay. However, an alternative explanation involving the recession of the recombination front of helium in the co-moving frame, similar to the rapid decline following the plateau phase in SNe II-P \citep{Kumar2013mnras, Nagy2014AA}, cannot be entirely ruled out. A further discussion of this feature is presented in Sect.~\ref{sec:phot:lcmdl}.
The decline rates between 0 and 15 days is $\gamma_{0<t<15}=8.53\pm0.49\,\mathrm{mag\,(100\ d)^{-1}}$ in the $g_\mathrm{M}$~band, and $\gamma_{0<t<15}=6.33\pm0.39\,\mathrm{mag\,(100\ d)^{-1}}$ in the $r_\mathrm{M}$~band, suggesting a rapid decrease in temperature. 
After $\sim20$~days, the decline rates decrease to $\gamma_{t>20}=1.88 \pm 0.23\,\mathrm{mag\,(100\ d)^{-1}}$ in the $g_{\mathrm{M}}$~band and $\gamma_{t>20}=2.76\pm0.21\,\mathrm{mag\,(100\ d)^{-1}}$ in the $r_{\mathrm{M}}$~band. The higher decline rates in the redder bands suggest a receding of the photosphere in the co-moving frame, indicating that the inner regions of the ejecta with higher temperature are becoming gradually visible. This behaviour is consistent with most of SE SNe such as SN~2017iro \citep{SN2017iro_Kumar2022ApJ...927...61K} and SN~2022ngb \citep{SN2022ngb_Zhao2026A&A}. The main parameters inferred from the multi-band apparent light curves are reported in Table~\ref{apptab:phot:apparentlc}.

\subsection{Intrinsic colour evolution}
\label{sec:phot:colorevo}
\begin{figure}[htbp]
\centering
\includegraphics[width=\linewidth]{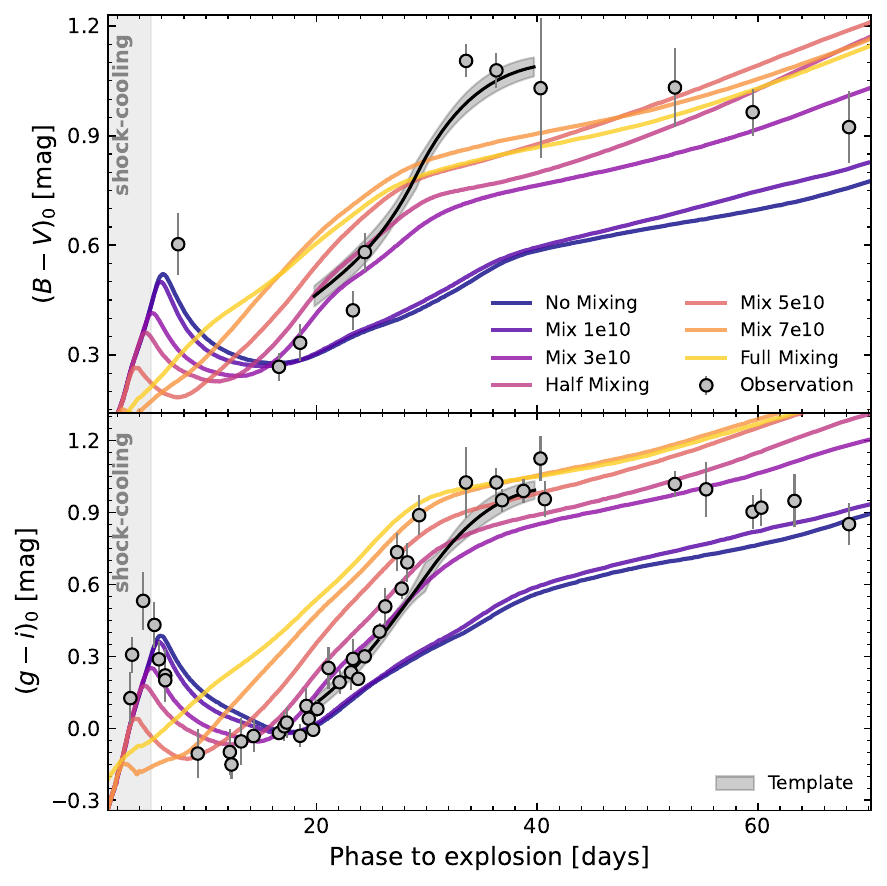}
\caption{Colour evolution of SN~2025aico compared with models with different degrees of $^{56}$Ni mixing (colour-coded). The grey region marks the shock-cooling phase of SN~2025aico. The solid black lines, with their corresponding light-grey regions, represent the colour evolution template of SNe~IIb from \cite{Stritzinger2018A&A}.}
\label{fig:phot:color}
\end{figure}

The intrinsic colour evolution of SN~2025aico is presented in Fig.~\ref{fig:phot:color}. The early-phase colour evolution provides a rough estimation of $^{56}$Ni mixing, a crucial parameter for SE SNe \citep{Dessart2015MNRAS.453.2189D, Dessart2016MNRAS, Yoon2019ApJ...872..174Y}. Therefore, we compared our observed colour evolution with models taken from \cite{Moriya2020MNRAS.497.1619M}, which include scenarios ranging from without mixing to full mixing.

During the very early phase ($t\lesssim5\,\mathrm{days}$), the colour rapidly reddens, with $g-i$ changing from $0.12\pm0.10\,\mathrm{mag}$ to $0.53\pm0.12\,\mathrm{mag}$. This rapid reddening marks efficient cooling driven by adiabatic expansion. Such a feature is common among SNe~IIb that exhibit a prominent shock-cooling phase, such as SN~1993J \citep{Richmond1994aj, Richmond1996aj} and SN~2011fu \citep{Kumar2013mnras, Morales-Garoffolo2015mnras}. From $t\sim7$~days to $t\sim17$~days after the explosion, the colour becomes bluer, revealing the contribution of radioactive heating. During this period, the colour reaches $0.26\pm0.04\,\mathrm{mag}$ in $B-V$ and $-0.15\pm0.06\,\mathrm{mag}$ in $g-i$. Between $t\sim20$ and $t\sim30$~days after the explosion, the colour becomes redder again as the cooling of the fast-expanding ejecta becomes more efficient than the radioactive energy input. The corresponding colour reaches $1.09\pm0.05\,\mathrm{mag}$ in $B-V$ and $1.02\pm0.15\,\mathrm{mag}$ in $g-i$. However, for epochs $t\gtrsim30$~days after the explosion, the colour index becomes stable and then gradually turns to blue. {In addition, as we assumed negligible host-galaxy extinction, we performed a consistency check by comparing the colour evolution of SN~2025aico to the empirical template of \cite{Stritzinger2018A&A} derived from a large sample of SNe~IIb. The observed colour evolution closely follows this template, indicating that our assumption of minimal host reddening is reasonable.}

To constrain $^{56}$Ni mixing in SN~2025aico, we compared our observational data against models featuring varying degrees of $^{56}$Ni mixing from \cite{Moriya2020MNRAS.497.1619M}. The models assumed an initial stripped He-star progenitor of $3.85\,\mathrm{M_\odot}$ that explodes with an ejecta mass of $M_\mathrm{ej} = 2.5\,\mathrm{M_\odot}$, a kinetic energy of $E_\mathrm{k} = 1\times10^{51}\,\mathrm{erg}$, and a nickel mass of $M_\mathrm{Ni} = 0.05\,\mathrm{M_\odot}$. We note that as the mixing becomes stronger, the early colour evolution becomes more monotonic. For SN~2025aico, we observe a prominent `U'-shaped colour evolution during the early phase ($t\lesssim25$~days after the explosion), which is also visible in SN~2024aecx \citep{SN2024aecx_Zou2026ApJ...997...77Z}. This shape serves as evidence of weak or negligible $^{56}$Ni mixing. From the early-phase comparison, we suggest that $^{56}$Ni mixing in SN~2025aico closely resembles models with no or minimal mixing, and{is unlikely to exceed half-mixing}. However, the early shock-cooling in the early phase could hamper us from accurately constraining the degree of $^{56}$Ni mixing. At later phases, we find that our observational data deviate significantly from the model predictions. This discrepancy likely arises from the lower synthesized $^{56}$Ni mass in SN~2025aico, which injects less energy into the ejecta and fails to maintain the temperature high. {Thus, considering the differences between the comparison models and the observational data, our constraints on the $^{56}$Ni mixing should be carefully considered as a qualitative estimation.} Furthermore, recombination during the post-peak evolution of SN~2025aico plays a crucial role, a topic we discuss in detail in Sect.~\ref{sec:phot:lcmdl}.

\subsection{Pseudo-bolometric light curve}
\label{sec:phot:pbollcs}
The pseudo-bolometric light curve of SN~2025aico, along with those of comparison events, is shown in Fig.~\ref{fig:phot:pbollcs}. We chose comparisons that are typically well studied or have similar luminosities. Each pseudo-bolometric light curve was constructed using photometry in the optical bands, ranging from the $B$- to the $z$ bands, utilizing the \textsc{superbol} code \citep{Nicholl2018RNAAS}. Specifically, due to the lack of early optical coverage, the first hollow data point was derived by scaling the $L$-band luminosity based on a blackbody temperature inferred from the \textit{Swift}/UVOT UV and NUV data. The pseudo-bolometric peak luminosity of SN~2025aico is $L_{\mathrm{opt}}=(4.07\pm0.10)\times10^{41}\,\mathrm{erg\,s^{-1}}$, and the corresponding rise time is $21.89\pm0.67\,\mathrm{days}$. The rising time of SN~2025aico is consistent with SN~1993J \citep{Richmond1994aj, Richmond1996aj} and SN~2008ax \citep{Pastorello2008mnras} but slightly shorter than SN~2022ngb \citep{SN2022ngb_Zhao2026A&A}. We note that SN~2025aico exhibits a lower luminosity compared to events such as SN~1993J and SN~2011fu \citep{Kumar2013mnras} but shares a similar peak luminosity with low-luminosity events such as SN~2011ei \citep{SN2011ei_Milisavljevic2013ApJ...767...71M}, SN~2011hs \citep{SN2011hs_Bufano2014MNRAS.439.1807B}, and SN~2022ngb. {In addition, if we consider the distance uncertainty introduced by the T-F relation, which is typically $\sim15$--17\%, the bolometric luminosity varies by $\sim0.15$~dex. {Furthermore, if we account for the upper limit of the host-galaxy extinction derived in Sect.~\ref{sec:info}, the bolometric luminosity only increases by $\sim0.1$--$0.2$~dex. Even when considering both of these factors, the bolometric luminosity of SN~2025aico remains firmly within the range of low-luminosity events, leaving our main conclusions unaffected.}
\begin{figure}[htbp]
\centering
\includegraphics[width=\linewidth]{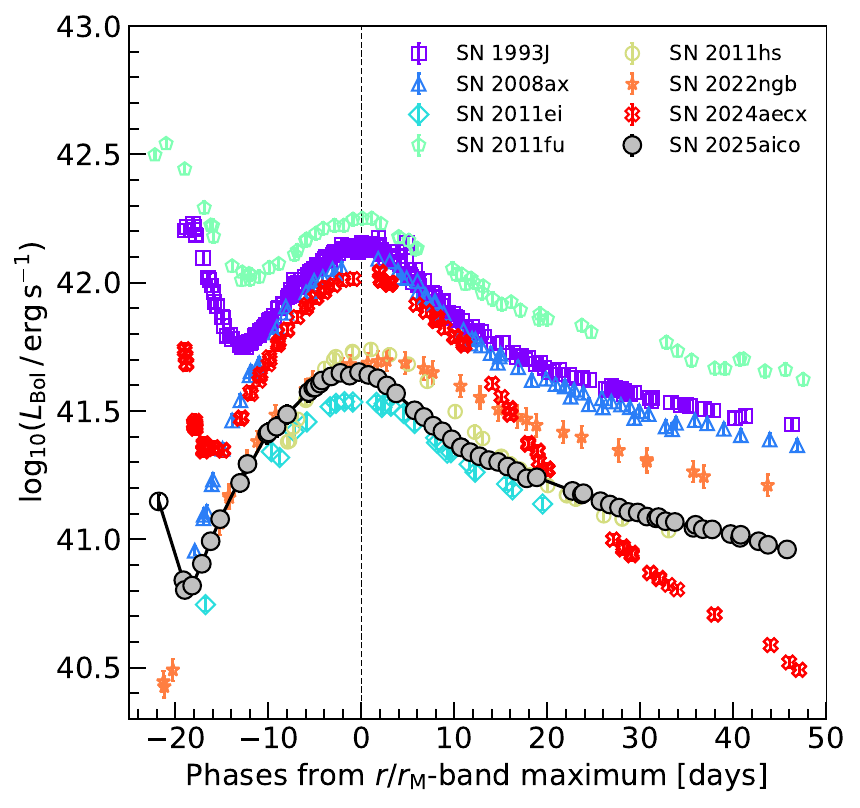}
\caption{Pseudo-bolometric light curve of SN~2025aico compared with those of other SNe IIb. All light curves are corrected for reddening. Phases are relative to either the $r$-band or the $r_\mathrm{M}$-band maximum.}
\label{fig:phot:pbollcs}
\end{figure}

SN~2025aico presents prominent shock-cooling emission during its very early phases (especially in bluer bands; see Sect.~\ref{sec:phot:applc}), a feature also visible in SN~1993J, SN~2011fu, and SN~2024aecx, indicating the remaining H-rich envelope. However, the luminosity of the shock-cooling emission in the optical bands is relatively faint when compared to other events and to the radioactively powered peak of SN~2025aico itself, reaching only $L_{\mathrm{opt}} \approx 1.5\times10^{41} \,\mathrm{erg\,s^{-1}}$ initially. This feature suggests that most of the emission is concentrated in the UV and bluest optical bands, which in turn suggests a compact envelope that produces a low-luminosity spectral energy distribution (SED) tail in the optical bands \citep{Rabinak2011apj, Nakar2010ApJ...725..904N}. Furthermore, the duration of the shock-cooling phase for SN~2025aico is shorter than that of comparison events, lasting only $\lesssim4$~days, which suggests a relatively compact and/or low-mass envelope \citep{Pastorello2008mnras, Roming2009ApJ}. Moreover, for the faint shock-cooling emission to remain visible, $^{56}$Ni mixing should be limited. This is because radioactive heating would otherwise obscure the very early shock-cooling features. This physical constraint is consistent with the modelling derived from the earlier colour evolution.

We also observe a rapid decline within $\sim20$~days after the $r_\mathrm{M}$-band maximum, a feature also visible in SN~1993J and SN~2011fu, which we also noticed in the apparent light curve in Sect.~\ref{sec:phot:applc}.
This feature suggests that the outer ejecta gradually become optically thin, resulting in a rapid recession of the photosphere in the co-moving frame. This rapid decline spans $\sim10$--$15$~days after the $r_\mathrm{M}$-band maximum, a duration slightly shorter than that observed in SN~1993J. Finally, the late-time decline rate of SN~2025aico is comparable to that of most SNe~IIb. However, it is slightly faster than the rates of SN~1993J and SN~2011fu, yet slower than that of fast-evolving events such as SN~2024aecx. This intermediate behaviour suggests a moderate degree of $\gamma$-ray leakage in SN~2025aico.

\subsection{Light curve modelling}
\label{sec:phot:lcmdl}
To constrain the explosion parameters of SN~2025aico, we constructed the bolometric light curve by fitting a blackbody function to the SED. We only kept the detection photometric data and ignored the upper-limit photometric data. Then, we implemented the model from \cite{Arnett1989apj}, which is a hybrid two-component light curve model, highly suitable for CC~SNe because it includes the contribution of recombination. 
This model accounts for the early shock-cooling emission and the radioactively powered light curve. We first fitted the portion of the light curve that is mainly powered by radioactive decay (for $t>5\,\mathrm{days}$). The heating rate follows the radioactive decay chain of $^{56}\mathrm{Ni} \rightarrow ^{56}\mathrm{Co} \rightarrow ^{56}\mathrm{Fe}$ \citep{Arnett1982apj, Valenti2008ApJ}. The $\gamma$-ray leakage during the post-peak decline phase is crucial for SE~SNe \citep{Clocchiatti1997ApJ, Chatzopoulos2012apj}. Taking this leakage into account, the total luminosity can be written as
\begin{equation}
L=L_\mathrm{rd}\times(1-e^{-\tau_{\gamma}}) + 4\pi r_{\mathrm{rec}}^2 Q\rho(r_{\mathrm{rec}},t)\frac{\mathrm{d}r_\mathrm{rec}}{\mathrm{d}t}\, .
\end{equation}
Here, $\tau_{\gamma}$ is the $\gamma$-ray optical depth, expressed as $\tau_{\mathrm{\gamma}} = A_\mathrm{\gamma}t^{-2}$. The term $A_\mathrm{\gamma}$ represents the characteristic $\gamma$-ray leakage timescale, which typically ranges from $10^{3}$ to $10^{5}\,\mathrm{day^{2}}$ for SE~SNe. Finally, $Q$ denotes the energy released per unit mass during recombination.

\begin{figure}[htbp]
\centering
\includegraphics[width=\linewidth]{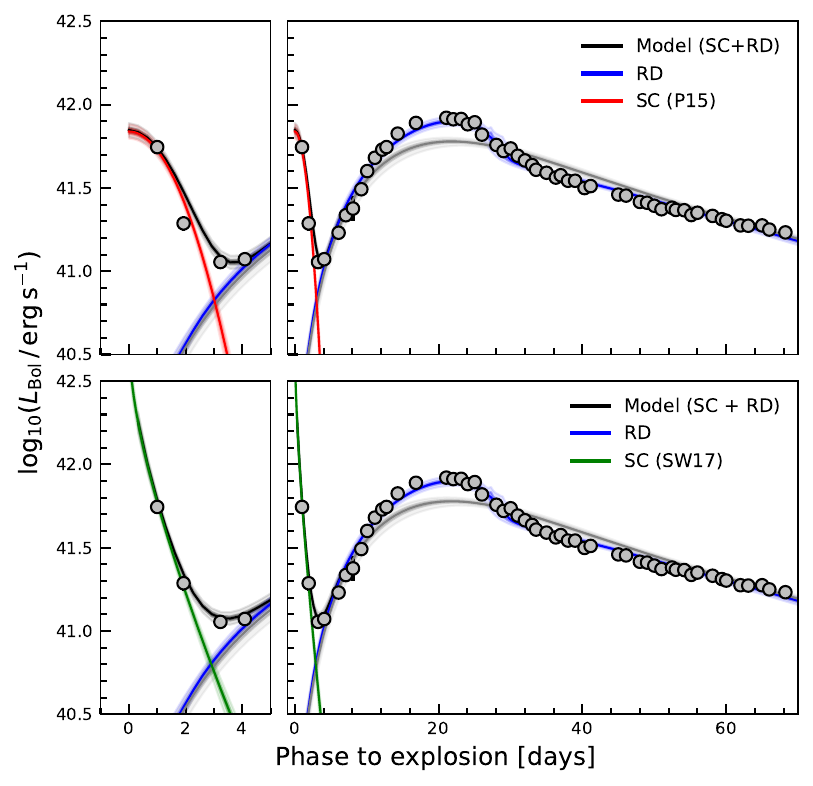}
\caption{Bolometric light curve model of SN~2025aico. The blue lines represent the radioactive powered diffusion model (RD), while the red and green lines represent different shock-cooling models (SC). The grey line represents the RD model but without the contribution of recombination. The region with the corresponding light colours represents the 1$\sigma$ and 3$\sigma$ confidence interval.}
\label{fig:phot:lcmdls}
\end{figure}
To avoid numerical instability, the minimum recombination radius in the co-moving frame was fixed at $x_\mathrm{min} = 0.1$, which is consistent with the core component described by \cite{Nagy2016aap}. The initial radius of the core was set to $R_0=10^{11} \, \mathrm{cm}$, a value typical for a compact helium-rich core, while the initial thermal energy was fixed at $E_{\mathrm{th}}=10^{50}\,\mathrm{erg}$, since the model of the core component is not sensitive to these parameters and these parameters are poor constrain in the fitting routine. Thus, we fixed them to ensure the convergence of the fitting. We note that the final results are not sensitive to these specific values. Furthermore, adopting a small initial radius and a modest initial thermal energy is consistent with the assumptions in \cite{Arnett1982apj} and \cite{Arnett1989apj}, suggesting that the choice of these parameters is reasonable. We adopted an optical opacity of $\kappa_{\mathrm{opt}} = 0.07 \, \mathrm{cm^{2} \, g^{-1}}$, which is typical for He-rich ejecta \citep{Taddia2018A&A}. In addition, we used the photospheric velocity measured from \ion{Fe}{II} spectral lines, applying $v_{\mathrm{ej}} = 6450_{-160}^{+190} \, \mathrm{km \, s^{-1}}$ as a Gaussian prior to fit the radioactively powered portion of the light curve. We did this in order to break the degeneracy between $M_\mathrm{ej}$ and $v_\mathrm{ej}$. Finally, we utilized \textsc{dynesty} \citep{dynesty_Speagle2020MNRAS.493.3132S} as the sampler to fit the bolometric light curve of SN~2025aico for epochs $\sim5$~days after the explosion. The posterior distributions are presented in Appendix~\ref{sec:app:ctables}.

To fit the data from the very early ($t<5\,\mathrm{days}$ and $T_{\mathrm{BB}} > 0.7\,\mathrm{eV}$) shock-cooling emission, we used the models of \cite{P15_Piro2015ApJ...808L..51P} (P15) and \cite{SW17_Sapir2017ApJ...838..130S} (SW17). Because the coverage of the early phases is very poor in the optical bands, we could not simultaneously fit the bolometric luminosity and the derived blackbody temperature. Therefore, we utilized multi-band light curve fitting. We note that the duration of the shock-cooling emission for SN~2025aico is shorter than in extended targets such as SN~1993J and SN~2011fu, appearing more similar to relatively compact scenarios such as SN~2016gkg \citep{Arcavi2017ApJ...837L...2A} and SN~2022ngb \citep{SN2022ngb_Zhao2026A&A}, which have a relatively smaller envelope radius. Consequently, we assumed a radiative envelope ($PV^{4/3}=\mathrm{const.}$) for the progenitor of SN~2025aico, a configuration common in blue supergiant (BSG) or low-mass Wolf-Rayet (WR) stars.

To account for the early shock-cooling emission peak in the light curve of SN~2025aico and isolate it from the rise to the broad light-curve second maximum, we modelled the total luminosity as the sum of shock-cooling ($L_{\mathrm{sc}}$) and radioactive decay ($L_{\mathrm{rd}}$) contributions: $L_{\mathrm{tot}}=L_{\mathrm{sc}} + L_\mathrm{rd}$. During the early phases, $L_{\mathrm{rd}}$ grows from an order of magnitude below $L_{\mathrm{sc}}$ to a value comparable to it. This combination can introduce significant differences in the SED\footnote{As a consistency check, we completely ignored the contribution from radioactive decay. The parameters derived using the model of \cite{P15_Piro2015ApJ...808L..51P} remain nearly identical, whereas the results based on \cite{SW17_Sapir2017ApJ...838..130S} remain consistent within one order of magnitude.}
Therefore, we utilized a double-blackbody function to characterize the total emission. For the radioactivity-powered component, we define the photospheric radius as $R_{\mathrm{ph}}=x_{\mathrm{rec}}\times(R_0+v_{\mathrm{ej}} t)$, where $x_\mathrm{rec}$ represents the position of the recombination front in the co-moving frame. The temperature of the radioactive powered component follows the Stefan-Boltzmann law. For the model of \cite{P15_Piro2015ApJ...808L..51P}, we assumed a homologous expansion of the ejecta, where $R(t)=R_\mathrm{env}+v_{\mathrm{env}}t$, and the temperature simply follows the Stefan-Boltzmann law based on the total luminosity $L_{\mathrm{sc}}$. For the model of \cite{SW17_Sapir2017ApJ...838..130S}, the temperature evolution follows the relation:
\begin{equation}
T_{\mathrm{sc}}\left(t\right) = 1.96 \times10^4 \times \left(\frac{v_{s,8.5}^2 t^2}{f_{\rho}M\kappa_{0.34}}\right)^{0.016} \frac{R_{13}^{0.25}}{\kappa_{0.34}^{0.25}}t^{-0.5}\, \mathrm{K}\,.
\end{equation}
The radius $R(t)$ in the \cite{SW17_Sapir2017ApJ...838..130S} model follows the Stefan-Boltzmann law derived from $T_{\mathrm{sc}}$ and $L_{\mathrm{sc}}$. Here, $f_\rho$ represents an empirical parameter for the density profile, and $M$ represents the total ejected mass; hence, $M=M_{\mathrm{ej}}+M_{\mathrm{env}}$. Because we adopted the assumption of a radiative envelope (where $n=3$, in the model of \citealt{SW17_Sapir2017ApJ...838..130S}), we have $f_\rho M=0.08(M_\mathrm{env}/M_{\mathrm{ej}}) \times (M_\mathrm{env}+M_{\mathrm{ej}})$. Since the envelope is H-rich, we adopted the opacity of $\kappa_{\mathrm{env}} = 0.34\,\mathrm{cm^2\,g^{-1}}$. The fitting parameters for both models are the same, consisting of $M_{\mathrm{env}}$, $R_\mathrm{env}$, and $v_{\mathrm{env}}$. 

We used synthetic photometry to convert the blackbody output from the models into AB magnitudes across all corresponding bands. For the fitting process, we utilized the Monte Carlo sampler \textsc{emcee} \citep{emcee_Foreman-Mackey2013PASP..125..306F}, employing a 3000-step burn-in phase to ensure convergence, followed by 1000 steps for the final sampling. The fitting routines were directly modified from the \textsc{lightcurvefitting} package \citep{Hosseinzadeh_2024_11405219}. Building upon the results from the previous fitting, we utilized a multi-band fitting routine to constrain the parameters of the shock-cooling phase. To avoid convergence difficulties, such as degeneracy in $f_\rho M$ -- parameters to which the results are not sensitive -- we applied the posterior distribution of $M_{\mathrm{ej}}$ from the earlier fitting as a Gaussian prior for this multi-band analysis.

The best-fit results with $\sim3\sigma$ confidence interval are shown in Fig.~\ref{fig:phot:lcmdls}. The fitting yields a moderate ejecta mass of $M_{\mathrm{ej}} = 2.79_{-0.18}^{+0.21} \, \mathrm{M_\odot}$, alongside a relatively low $^{56}$Ni mass, $M_{\mathrm{Ni}} = 0.033_{-0.004}^{+0.006} \, \mathrm{M_\odot}$. {Note that the reported uncertainties for the $^{56}$Ni mass here only reflect the statistical errors from the fit.} The explosion energy derived from the fitting result of SN~2025aico was $E_\mathrm{k} = 0.70_{-0.08}^{+0.09}\times10^{51}\,\mathrm{erg}$, which is in reasonable agreement with the models we used in Sect.~\ref{sec:phot:colorevo}, and is consistent with most of the type IIb events \citep{Taddia2018A&A}. The derived value of $M_\mathrm{ej}$ highly depends on the choice of opacity; therefore, varying the adopted optical opacity yields reasonable estimates ranging from $1.5\,\mathrm{M_\odot}$ to $2.8\,\mathrm{M_\odot}$. These results suggest a relatively high ejecta mass paired with a low nickel mass, a configuration similar to the parameters of SN~2024abfo \citep{Reguitti2025AA} and SN~2022ngb \citep{SN2022ngb_Zhao2026A&A}. Furthermore, accounting for the distance uncertainty and {the upper limit of the host reddening}, the $M_{\mathrm{Ni}}$ derived from SN~2025aico still falls within the range of $\sim0.02-0.04\,\mathrm{M_\odot}$, which is reasonable for SNe~IIb and indicating a low luminosity event, consistent with our previous conclusions. The best-fit recombination temperature is $T_{\mathrm{rec}} = 11.1_{-0.7}^{+0.8} \, \mathrm{kK}$, which is consistent with the typical temperature for He-recombination \citep[$\sim10$~kK;][]{Hatano1999ApJS..121..233H}. As a consistency check, we overplotted a model with the same parameters, but where we simply turned off the contribution of recombination along with the data. We notice that it fails to fit the peak luminosity and the rapid decline after the radioactive powered peak. Thus, we can confirm that recombination could possibly be an alternative explanation for the rapid post-peak in SN~2025aico. We observe that the $^{56}$Ni mass of SN~2025aico is located at the lower end of the distribution for SNe IIb \citep{Taddia2018A&A}. Furthermore, the $M_\mathrm{Ni}$ estimated from the peak luminosity can be affected by the potential recombination luminosity. To mitigate this effect and perform a consistency check, we fitted the tail of the bolometric light curve. Assuming that the tail luminosity is powered by radioactive decay with partial $\gamma$-ray leakage, the bolometric luminosity can be expressed as
\begin{equation}
\begin{split}
L_{\mathrm{Bol}}(t)&\approx L_{\mathrm{inp}}(t)\times\left(1-e^{-\tau_{\gamma}}\right)\\
&=M_{\mathrm{Ni}}\left[\left(\epsilon_{\mathrm{Ni}} - \epsilon_{\mathrm{Co}}\right)e^{-\frac{t}{\tau_{\mathrm{Ni}}}} + \epsilon_{\mathrm{Co}}e^{-\frac{t}{\tau_{\mathrm{Co}}}}\right] \times\left(1-e^{-\tau_{\gamma}}\right)\;.
\end{split}
\end{equation}
Fitting this model to the bolometric light curve data starting from $+$20~days after the $r_\mathrm{M}$-band maximum yields $M_\mathrm{Ni} = 0.032\pm0.001\,\mathrm{M_\odot}$. This result is consistent with our previous estimate, indicating that the modelling of the radioactive-powered phase is reliable.

The early-phase light curve evolution ($T_{\mathrm{BB}} > 0.7\,\mathrm{eV}$) is well described by shock-cooling models \citep[e.g.][]{P15_Piro2015ApJ...808L..51P, SW17_Sapir2017ApJ...838..130S}, and the fits convergence well. The fit based on the model of \cite{P15_Piro2015ApJ...808L..51P} yields an envelope mass of $M_{\mathrm{env}} = 0.017 \pm 0.001\,\mathrm{M_\odot}$, an envelope radius of $R_{\mathrm{env}} = 6.1_{-0.6}^{+0.7}\,\mathrm{R_\odot}$, and a bulk envelope velocity of $v_{\mathrm{env}} = 13930_{-800}^{+820}\,\mathrm{km\,s^{-1}}$. Additionally, we employed the model of \cite{SW17_Sapir2017ApJ...838..130S} as a consistency check. 
Because the derived configuration is relatively compact, we assumed a radiative envelope with a polytropic index of $n = 3$. The resulting best-fit parameters suggest $M_{\mathrm{env}} = 0.018 \pm 0.001\,\mathrm{M_\odot}$ and $R_{\mathrm{env}} = 9.1 \pm 1.1\,\mathrm{R_\odot}$. The shock velocity derived from this model is $v_{\mathrm{sh}} = 21850_{-1670}^{+1880}\,\mathrm{km\,s^{-1}}$, which is broadly consistent with the expansion velocity measured from the early P~Cygni profile of H$\alpha$  ($\sim 18000\,\mathrm{km\,s^{-1}}$; see below in Sect.~\ref{sec:spec:vt}). In brief, these results indicate a progenitor consistent with a compact He star or even a low-mass WR star, suggesting that SN~2025aico is a cIIb event similar to SN~2022ngb \citep{SN2022ngb_Zhao2026A&A} and SN~2008ax \citep{Pastorello2008mnras, Folatelli2015apj}. {Furthermore, when accounting for the distance uncertainty, the derived parameters of the envelope remain largely unaffected. Specifically, $M_{\rm env}$ remains almost constant because it depends on the evolution timescales, while $R_{\rm env}$ varies by only $\sim0.15$~dex, following the luminosity \citep[$L\propto R_{\rm env}$;][]{SN2022ngb_Zhao2026A&A, Arcavi2017ApJ...837L...2A, SN2024iss_Chen2026AA}.}

Prior and posterior distributions of each model utilized in the fitting process are presented in Table~\ref{apptab:lcfitting}. The corner plots and the stand-alone shock-cooling fit result for the initial phase are presented in Figs.~\ref{fig:phot:rdmodelcorner} and~\ref{fig:phot:earlylcmdlcorner}. 

\section{Spectroscopy}\label{sec:spec}
\subsection{Spectral sequence}\label{sec:spec:specevo}

\begin{figure*}[htbp]
\centering
\sidecaption
\includegraphics[width=0.85\linewidth]{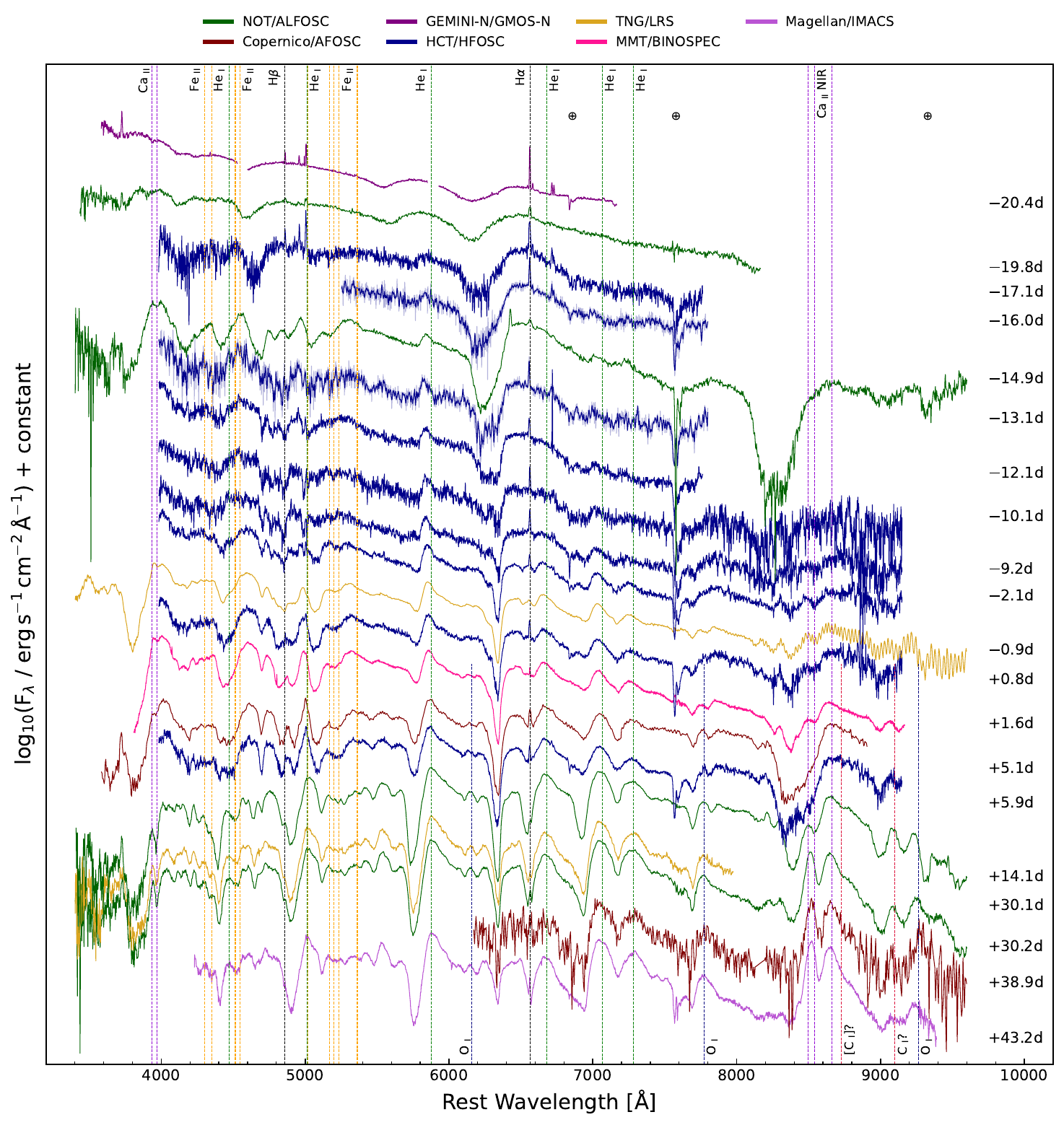}
\caption{Spectral sequence of SN~2025aico. Prominent spectral lines are marked with dashed coloured lines at each rest wavelength. The marked phases are calculated from the $r_\mathrm{M}$-band maximum. The telluric lines are indicated with encircled plus symbols. Data from different telescopes are marked using different colours. Some spectra with low S/N were smoothed with a Savitzky-Golay filter, and the original spectra are shown in the background with lighter colours.}
\label{fig:spec:evo}
\end{figure*}

The spectral evolution is presented in Fig.~\ref{fig:spec:evo}. A detailed description of the observations can be found in Appendix~\ref{sec:app:dataobs}. The spectroscopic data span a period from $-$20.4~days to $+$43.2~days relative to the $r_\mathrm{M}$-band maximum, with a fairly good coverage in the pre-peak phases. Narrow emission lines such as H$\alpha$, H$\beta$, [\ion{N}{II}], and [\ion{S}{II}] originate from \ion{H}{II} regions within the host galaxy, indicating active star formation in the local environment. All spectra were flux calibrated to match the photometric data and then corrected for reddening and reported to the rest frame. 

At very early epochs ($-$20.4 and $-$19.8~days relative to the maximum in the $r_{\mathrm{M}}$ band), the spectra exhibit a blue, relatively featureless continuum. During this period, the blackbody temperature decreases from $11.79\pm0.28$~kK to $8.20\pm0.27$~kK. Such temperatures are comparable to those observed in events with early shock-cooling emission, including SN~2024aecx \citep[$\sim15$~kK;][]{SN2024aecx_Zou2026ApJ...997...77Z}. Furthermore, broad P~Cygni profiles of the Balmer and \ion{He}{I}~$\lambda5876$ transitions display highly blueshifted absorption components, indicating a rapid expansion of the ejecta around $\sim18000\,\mathrm{km\,s^{-1}}$. Driven by adiabatic expansion, the temperature rapidly decreases to $7.66\pm0.16$~kK by $-$17.1~days. At this temperature, the continuum gradually reddens and low-ionization metal lines such as \ion{Fe}{II} begin to emerge. This spectral evolution reflects the cooling of the ejecta, coinciding with the end of the shock-cooling phase, in agreement with our photometric observations.

As the photosphere retreats from the outer envelope to the He core, the \ion{He}{I} $\lambda$5876 line with a P~Cygni profile emerges, showing a blueshifted emission peak. This blueshifted peak is primarily driven by the steep density profile and the dominance of electron scattering within the fast-expanding and optically thick ejecta, a combination that enables line formation inside the continuum photosphere \citep{Anderson2014MNRAS.441..671A}. Transitions of \ion{Ca}{II} and \ion{Fe}{II} become prominent in the blue region of the spectra. The broad P~Cygni profiles become gradually narrower in response to the decreasing photospheric velocity. Additionally, a new feature at the blue wing of the H$\alpha$ absorption appears at around $-$10~days. This feature could be explained by either a high-velocity hydrogen component \citep{Medler2021MNRAS} or the possible absorption of \ion{Si}{II} \citep{Gangopadhyay2018mnras, SN2022ngb_Zhao2026A&A}. However, since this feature appeared in the early phase, as most silicon is located in the inner region of the ejecta, {the possibility of the high-velocity hydrogen component cannot be entirely ruled out}, such as in SN~2020cpg \citep{Medler2021MNRAS}. In addition, we notice that in the early phase, the absorption of H$\beta$ line exhibits an asymmetric broad bottom. Around the epoch of $-13.1$~days, a possible high-velocity component emerged around the H$\beta$ line, resembling the H$\alpha$ line. Nevertheless, the low S/N of this spectral series and the line blending on the blue side of these spectra {such as \ion{N}{II}, \ion{Na}{I} and/or \ion{He}{I} lines} preclude a conclusive identification {\citep{Baron2000ApJ...545..444B}}. {Thus, we suggest that the identification of this high-velocity hydrogen component remains tentative.}

From $-$14.9~days, the blackbody temperature starts to rise again, increasing from $7.35_{-0.20}^{+0.16}$~kK to $9.13_{-0.21}^{+0.24}$~kK at $-$12.1~days, before finally dropping to $4.87\pm0.11$~kK at $+$30.2~days. This suggests that the energy input from the $^{56}$Ni radioactive decay acts as a primary heating source during the early phase. Following the adiabatic expansion of the ejecta, the photosphere recedes into the deeper layers, which are characterized by a lower velocity and a flatter density profile. In this phase, the emission peak of the \ion{He}{I} P~Cygni profile gradually shifts towards the rest wavelength \citep{Anderson2014MNRAS.441..671A}. The opacity in the blue region of the spectra is predominantly provided by metal lines such as \ion{Fe}{II}, \ion{Sc}{II}, and \ion{Ti}{II}. Because transitions of H$\alpha$ and \ion{He}{I} cannot accurately trace the velocity of the photosphere \citep{Dessart2005AA}, we estimated the photospheric velocity by fitting the P~Cygni absorption component of \ion{Fe}{II}~$\lambda5169$ in the spectrum obtained at $-0.9$~days with a Gaussian profile. This epoch is near the time of peak bolometric luminosity (MJD~$61053.27\pm0.38$). The fitting procedure yields $v_{\mathrm{ph}}=6450\pm170\,\mathrm{km\,s^{-1}}$, a value used as an indicative estimate of the ejecta velocity ($v_{\rm ej}$) in the light-curve modelling.

Following the peak luminosity, \ion{O}{I}~$\lambda7774$ and \ion{O}{I}~$\lambda6158$ emerge, features that originate from the deeper regions of the ejecta. The \ion{Ca}{II} near-infrared (NIR) emission features, along with a possible [\ion{C}{I}] line, gradually become prominent. In the spectra obtained at $+$14.0~days and $+$30.2~days, we observe the appearance of \ion{C}{I} and \ion{O}{I}~$\lambda9263$ transitions in the NIR region. The lines of the Balmer series gradually become weaker and narrower. At the same time, the spectra are dominated by the prominent P~Cygni profiles of the \ion{He}{I} transitions, consistent with the typical spectroscopic evolution of a SN IIb.

\subsection{Comparison with \textsc{synapps} models}
\label{sec:spec:id}
To identify the main spectral features, we utilized the \textsc{synapps} \citep{Thomas2013ascl} code to fit the spectra at three selected epochs ($-$19.8~days, $-$0.9~days, and $+$30.2~days relative to the $r_\mathrm{M}$-band maximum). The \textsc{synapps} code utilizes \textsc{syn++} as the parametrized generator of model spectra. It assumes spherical symmetry and homologously expanding ejecta. The photons in this code are emitted from a sharp photosphere. The results are presented in Fig.~\ref{fig:spec:id}. We note that in the early-phase spectrum, the broad Balmer lines appear to be significantly blueshifted, while the continuum is characterized by a blackbody temperature of approximately $15.1$~kK, which is briefly consistent with our result measured from the blackbody fit. The blue wing of H$\alpha$ may {possibly} result from a blend with \ion{Si}{II} {and/}or represent a high-velocity absorption component of H$\alpha$ itself{, as mentioned in Sect~\ref{sec:spec:specevo}} \citep{SN2022ngb_Zhao2026A&A, Gangopadhyay2018mnras, Medler2021MNRAS}. Furthermore, a broad \ion{He}{I} $\lambda$5876 is also visible in the early spectrum, likely excited by the shock. This feature suggests that some helium is mixed into the envelope. However, because mixing in the outer region remains limited \citep{Jerkstrand2015aap}, these \ion{He}{I} lines rapidly become narrow and no longer visible at later phases.

\begin{figure}[htbp]
\centering
\includegraphics[width=\linewidth]{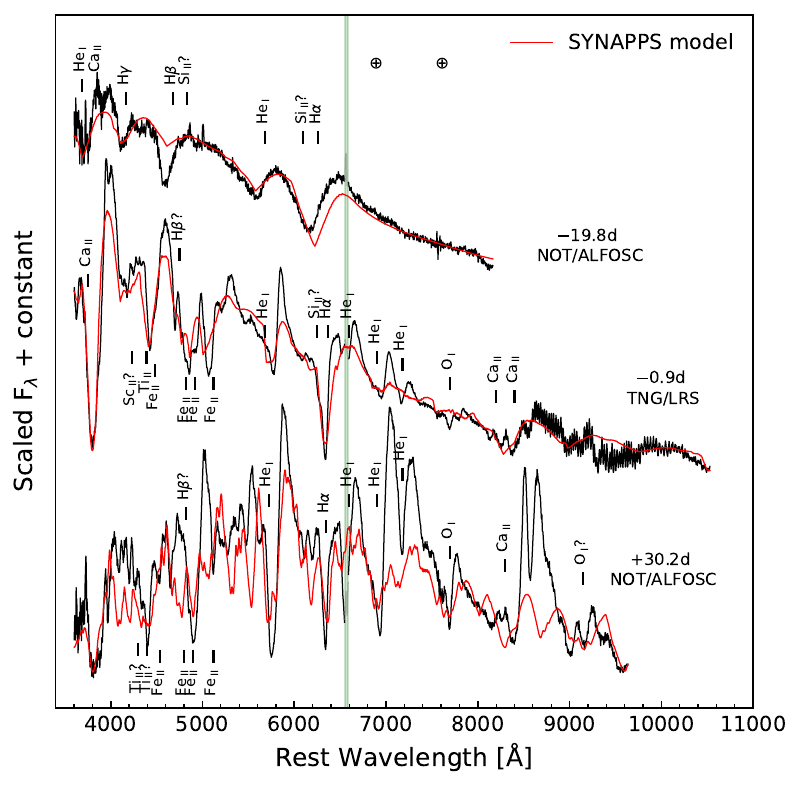}
\caption{\textsc{synapps} models compared with SN~2025aico spectra. The models are over-plotted with red lines. The green region marks the narrow \Ha emission line from the host galaxy, which is the only marked narrow emission line from the host. Phases were calculated using the $r_\mathrm{M}$-band maximum.}
\label{fig:spec:id}
\end{figure}

In the spectrum at peak ($-$0.9~days), the best-fit blackbody temperature drops to $6580$~K. Metal lines on the blue side of the spectra begin to emerge, including \ion{Ti}{II}, \ion{Sc}{II}, and \ion{Fe}{II}. The Balmer lines become narrower and less prominent, suggesting that the line-forming region gradually becomes H-poor. Furthermore, the high velocity component of H$\alpha$ gradually becomes invisible. We note that we cannot achieve a satisfactory fit for the \ion{He}{I} P~Cygni profiles because the emission peaks are highly intense. In the red and NIR regions of the spectra, \ion{O}{I} lines are identified, revealing the inner part of the ejecta. Furthermore, the absorption component of the \ion{Ca}{II}~NIR triplet flattens and subsequently evolves into a prominent emission feature. Finally, the red wing of the \ion{Ca}{II}~NIR profile may also be blended with \ion{C}{I} \citep{Jerkstrand2015aap}.

\subsection{Line velocities and temperature evolution}
\label{sec:spec:vt}
We measured the line velocities and the blackbody temperatures of the spectra, and the results are presented in Fig.~\ref{fig:spec:vtline}. The velocity of each line exhibiting a P~Cygni profile was estimated by applying a Gaussian fit to the absorption trough, while the temperature was derived by fitting a blackbody function to the continuum of each spectrum. Furthermore, we compare the velocity of \ion{Fe}{II}, which represents the photospheric velocity \citep{Dessart2005AA}, and the blackbody temperature against the $^{56}$Ni mixing models from \cite{Moriya2020MNRAS.497.1619M}. To facilitate a more accurate comparison and ensure clarity, the phases were estimated relative to the time of bolometric luminosity maximum.

\begin{figure}[htbp]
\centering
\includegraphics[width=\linewidth]{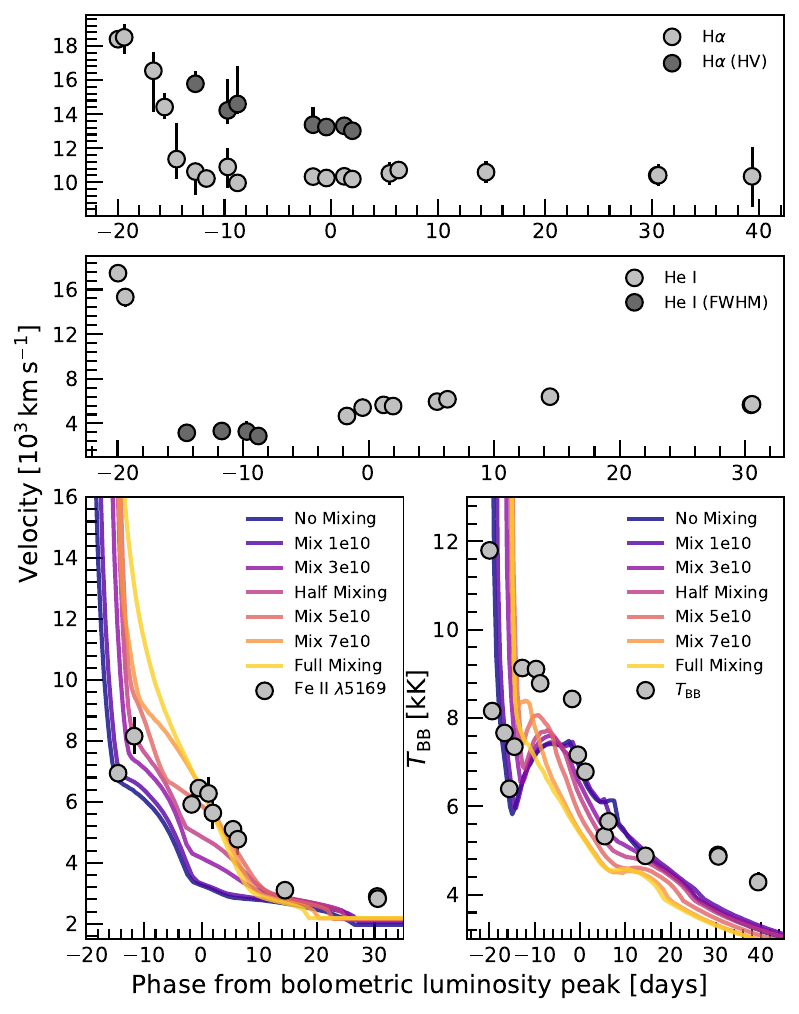}
\caption{Evolution of the line velocities and blackbody temperature from SN~2025aico spectra. The models from \cite{Moriya2020MNRAS.497.1619M} with different intensities of $^{56}$Ni mixing are over-plotted along with the measured data points. Phases were calculated using the bolometric luminosity peak.}
\label{fig:spec:vtline}
\end{figure}

The velocity of  H$\alpha$ at very early phases ($t\approx-20$~days) is $\sim18500\,\mathrm{km\,s^{-1}}$, which is indicative of the expansion velocity of the outer envelope. To constrain the velocity of the outermost region, we measured the blue side wing of H$\alpha$ P~Cygni absorption and cutoff at 3~$\sigma$. The velocity estimated from the bluer wing is $\sim35500\,\mathrm{km\,s^{-1}}$, which can be an upper-limit of the shock velocity. At approximately $t\approx-15$~days, this velocity decreases to $\sim10600\,\mathrm{km\,s^{-1}}$ as the photosphere recedes into the inner ejecta following the end of the shock-cooling phase. The velocity of the lower velocity H$\alpha$ component remains nearly constant throughout the subsequent evolution, which is consistent with typical SNe~IIb (e.g. see the velocities evolution in \citealt{Medler2022MNRAS, SN2022ngb_Zhao2026A&A}). 
Furthermore, {an absorption feature on the blue edge of the main H$\alpha$ profile appears at $t\approx-12$~days. If interpreted as a high-velocity component of H$\alpha$}\footnote{{While we tentatively refer to this feature as high-velocity H$\alpha$, such absorption can also be reliably attributed to \ion{Si}{II}. A detailed discussion of this spectral blending is presented in Sect.~\ref{sec:spec:specevo}.}}, its velocity decreases from $\sim15800\,\mathrm{km\,s^{-1}}$ to $\sim13000\,\mathrm{km\,s^{-1}}$. This component evolves in velocity more slowly than the main H$\alpha$ line, which, under this assumption, would suggest that it forms within the fast-expanding envelope.

The velocity of the \ion{He}{I}~$\lambda5876$ line during the early phase ($t\approx-20\,\mathrm{days}$) is $\sim17460\,\mathrm{km\,s^{-1}}$, comparable to that of H$\alpha$. The broad \ion{He}{I}~$\lambda5876$ feature disappears after $t\approx-17$~days, and the absence of an evident P~Cygni absorption component is noticeable between $t\approx-15$~days and $t\approx-8$~days. Instead, only relatively narrow \ion{He}{I} emission lines are present with a full width at half maximum (FWHM) of $\sim3200\,\mathrm{km\,s^{-1}}$. Because SN~2025aico does not exhibit any features indicative of circumstellar medium interaction, unlike, for example, the type Ibn/IIb SN~2018gjx \citep{SN2018gjx_Prentice2020MNRAS.499.1450P}, these low-velocity \ion{He}{I} lines can only form within the ejecta. 
At $t\approx-2$~days, the P~Cygni profile of \ion{He}{I}~$\lambda5876$ reappears with a velocity of $\sim4670\,\mathrm{km\,s^{-1}}$. This velocity then increases to $\sim6100\,\mathrm{km\,s^{-1}}$ before declining to $\sim5770\,\mathrm{km\,s^{-1}}$ at $t\approx30$~days. This evolution suggests that the line-forming region of \ion{He}{I}~$\lambda5876$ initially moves outwards in the co-moving frame, likely driven by the expanding illuminated region by $\gamma$ rays from the weakly mixed $^{56}$Ni as the ejecta density decreases, rather than simply tracking the recession of the continuum photosphere. A detailed discussion can be found in Sect.~\ref{sec:disc:lineevo}.

The velocity of the \ion{Fe}{II}~$\lambda5169$ line and the blackbody temperature evolution are presented in the lower panel of Fig.~\ref{fig:spec:vtline}. The P~Cygni profile of \ion{Fe}{II}~$\lambda5169$ appears at $t\approx-15$~days with a velocity of $\sim6940\,\mathrm{km\,s^{-1}}$. It then increases to $\sim8150\,\mathrm{km\,s^{-1}}$ before decreasing to $\sim5630\,\mathrm{km\,s^{-1}}$ prior to $t\approx8$~days. Subsequently, the velocity rapidly drops to $\sim3000\,\mathrm{km\,s^{-1}}$. This non-monotonic evolution, characterized by an early-phase rise and a subsequent plateau, matches models featuring weak or at most half $^{56}$Ni mixing, a configuration also observed in the type Ib SN~2007Y \citep{SN2007Y_Stritzinger2009ApJ...696..713S}. Furthermore, we note that the temperature evolution from $t\approx-15$~days (following the shock-cooling phase) increases before declining again. This non-monotonic behaviour suggests that weakly mixed radioactive material cannot fully heat the entire ejecta within a short timescale. This trend aligns with models of weak mixing scenarios, maintaining consistency with our previous analysis. In this case, the recombination front is pushed forward and the photosphere expands in the co-moving frame \citep{Moriya2020MNRAS.497.1619M}.

\subsection{Spectral comparisons}
\label{sec:spec:cmp}
The spectral comparison of SN~2025aico with other SNe~IIb/Ib is presented in Fig.~\ref{fig:spec:cmp}. We chose SN~1993J, SN~2008ax, SN~2020acat, and the type~Ib SN~2007Y as comparisons, representing an SN~IIb with an extended envelope, an SN~IIb with a compact envelope, an SN~IIb with intense mixing \citep{Ergon2024A&A}, and an H-poor, moderately mixed SN~Ib \citep{SN2007Y_Stritzinger2009ApJ...696..713S, Moriya2020MNRAS.497.1619M}, respectively. At a very early phase, both SN~2025aico and SN~1993J exhibit a very blue continuum, which is nearly featureless or contains only a few broad lines with P~Cygni profiles. However, the continuum of SN~2025aico is redder than that of SN~1993J and displays more features, such as prominent H$\alpha$ and \ion{He}{I} lines. Furthermore, compared to SNe lacking a prominent shock-cooling phase, the continuum of SN~2025aico is much bluer and exhibits fewer features, such as \ion{Fe}{II} lines. This suggests that SN~2025aico represents a transitional case between compact and extended SNe~IIb. Specifically, the progenitor is much more compact than that of SN~1993J, which is presumed to be an RSG \citep{Maund2004Natur.427..129M}, yet slightly more extended than typical compact stripped He-stars. 

\begin{figure}[htbp]
\centering
\includegraphics[width=\linewidth]{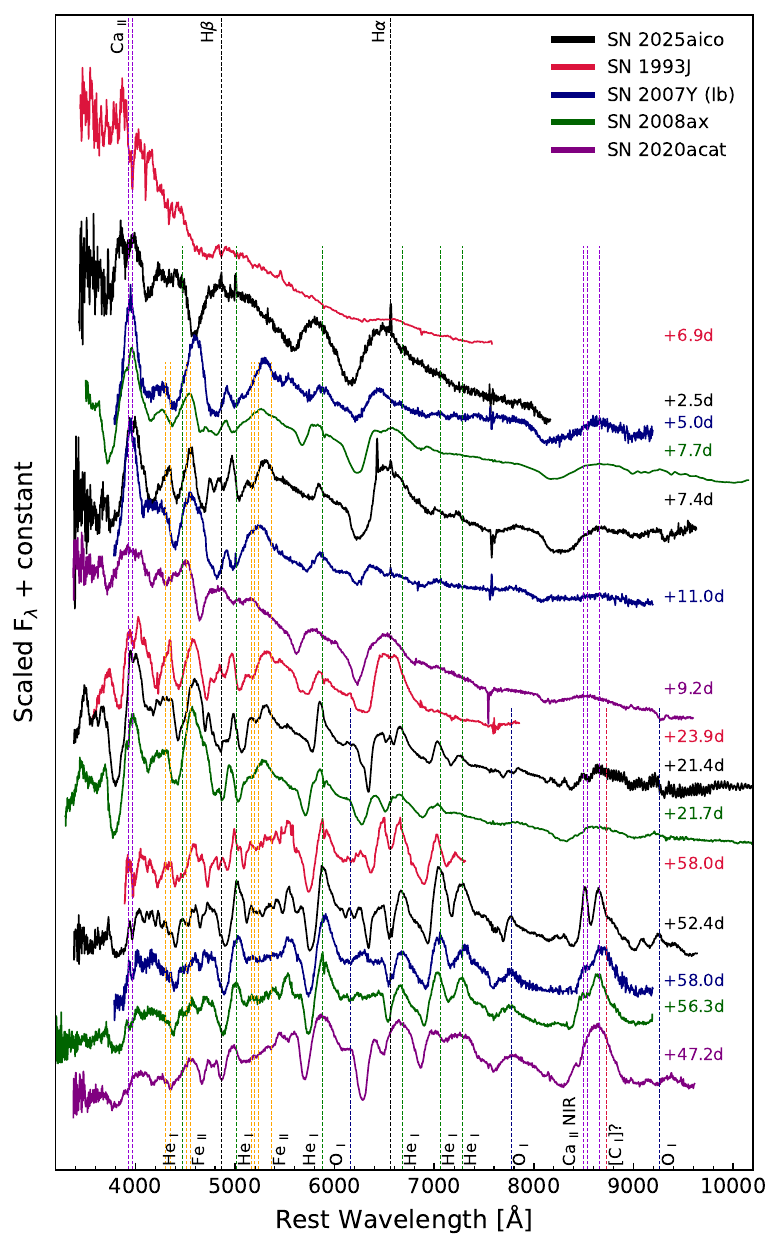}
\caption{Comparisons between spectra of SN~2025aico and those of other SNe IIb/Ib obtained at similar phases. Prominent spectral features are marked with dashed lines in various colours. Phases were calculated using the {explosion epoch.}}
\label{fig:spec:cmp}
\end{figure}

During the pre-peak rising phase ($-14.9\lesssim t \lesssim 0$~days), when the contribution from shock-cooling is weak and the light curve is dominated by the radioactive heating, we note that lines such as \ion{Fe}{II}, \ion{Ca}{II}, and \ion{Ti}{II} are present in the spectra. In general, the spectrum of SN~2025aico is similar to that of SN~2008ax, with the expected properties of the material ejected in the CC of a He core with a residual outer hydrogen envelope. The \ion{He}{I}~$\lambda5876$ line in SN~2025aico spectra does not exhibit a prominent P~Cygni profile. In particular, the absorption component is weak and only a relatively narrow emission peak is visible. 
This morphology is the direct result of weak $^{56}$Ni mixing (see Sects.~\ref{sec:phot:colorevo} and \ref{sec:spec:vt}). Furthermore, SNe with weak to moderate mixing scenarios (e.g. SN~2007Y) present similar features. In contrast, the strong-mixing event SN~2020acat exhibits a prominent \ion{He}{I}~$\lambda5876$ P~Cygni profile. These observational results are consistent with our earlier findings and the conclusions of \cite{Dessart2015MNRAS.453.2189D}, suggesting that a weak mixing may fail to excite prominent He-rich features, even when the ejecta environment is He-rich.

The peak spectrum ($\sim-0.9$~days) of SN~2025aico is similar to that of typical SNe~IIb. The blue side of the spectrum is dominated by metal lines. As the ejecta expand, certain features, such as the \ion{Ca}{II}~NIR triplet, evolve from absorption features into emission lines. Since the $\gamma$-ray photons were able to diffuse to the outer part of the ejecta from the weakly to moderately mixed core, a non-thermally excited \ion{He}{I}~$\lambda5876$ line exhibiting a P~Cygni profile is visible in SN~2025aico. Compared to SN~1993J, we note that the H$\alpha$ absorption trough of SN~2025aico is sharper and features a double-absorption component in the blue wing. This morphology can be explained either by blending with lines such as \ion{C}{I} and \ion{Si}{II} or by the presence of a high-velocity component of H$\alpha$. Additionally, a broad wing appears on the red side of the \ion{He}{I}~$\lambda5876$ emission peak, which could originate from higher-velocity helium in the outer ejecta. Nevertheless, since the photosphere moves deeper into the ejecta, the potential for blending caused by the gradually strengthening \ion{Na}{I}~D lines cannot be entirely ruled out.

In the post-peak declining phase ($\sim30.1$~days), the spectra of the various events are quite similar. The P~Cygni profile of \ion{He}{I}~$\lambda5876$ becomes broad, resembling those usually exhibited by other He-rich events. Additionally, the P~Cygni profiles of other \ion{He}{I} lines become more prominent. The primary difference during this epoch lies in the presence of the H$\alpha$ line. In SN~1993J, SN~2020acat, and SN~2025aico, the H$\alpha$ feature remains visible, although it is much narrower than at earlier phases. Conversely, for SN~2008ax and the H-poor event SN~2007Y, the H$\alpha$ feature is absent at later phases. This suggests a lower amount of residual hydrogen in the progenitors of SN~2008ax and SN~2007Y than in the progenitor of SN~2025aico. This provides further support to the conclusion that SN~2025aico represents a transitional event between compact and extended SNe~IIb.

\section{Discussion}
\label{sec:disc}
\subsection{Host galaxy of SN~2025aico}
\label{sec:disc:host}
To obtain a better constraint on the progenitor scenario, it is crucial to investigate the host environment of SN~2025aico. The metallicity of the progenitor significantly impacts the mass-loss rate of the star \citep{Yoon2017ApJ}, and we assume that the metallicity of the progenitor is similar to the local metallicity of the host galaxy. Consequently, we can estimate the metallicity using the host emission lines present in the early spectra. Due to the non-detection of the [\ion{O}{III}]~$\lambda4363$ line, which is used to determine the oxygen abundance in \ion{H}{II} regions \citep{Pilyugin2003A&A...399.1003P, Pilyugin2005ApJ}, we must rely on empirical strong-line metallicity estimations. Therefore, we utilized the empirical relations of the \texttt{O3N2} and \texttt{N2} indices, where
\begin{equation}
\begin{split}
O3N2&=\mathrm{log}\left(\frac{\mathrm{[O_{III}]\lambda5007}}{\mathrm{H\beta}}\times\frac{\mathrm{H\alpha}}{\mathrm{[N_{II}]\lambda6583}}\right)\\
12+\mathrm{log(O/H)}&=8.533[\pm0.012]-0.214[\pm0.012]\times O3N2
\end{split}
\end{equation}
for the \texttt{O3N2} relation, and
\begin{equation}
\begin{split}
N2&=\mathrm{log}\left(\frac{\mathrm{[N_{II}]\lambda6583}}{\mathrm{H\alpha}}\right)\\
12+\mathrm{log(O/H)}&= 8.743[\pm0.027]+0.462[\pm0.024]\times N2
\end{split}
\end{equation}
for the \texttt{N2} relation, following the methods established by \cite{Marino2013A&A...559A.114M, Pettini2004MNRAS.348L..59P}. To improve reliability, we use both of these relations as a consistency check.

We iteratively fitted the continuum of the first spectrum and clipped the prominent emission and absorption lines using polynomial functions. Subsequently, we subtracted the continuum from the spectrum to isolate the fluxes of the emission lines from the host. We applied Gaussian fits to the H$\alpha$, H$\beta$, [\ion{O}{III}]~$\lambda5007$, and [\ion{N}{II}]~$\lambda6583$ lines, adopting the integrated flux of the best-fit profile as the total flux of each line. This yields the indices of $O3N2=1.344\pm0.061$ and $N2=-0.994\pm0.054$. The corresponding metallicities, expressed as the oxygen abundance, are estimated to be $12+\mathrm{log(O/H)}=8.245\pm0.024$ and $12+\mathrm{log(O/H)}=8.284\pm0.044$, respectively. We note that these results are similar and that they point to a sub-solar metallicity environment (which $12+\mathrm{log(O/H)}=8.69\pm0.04$, \citealt{Asplund2021A&A...653A.141A}) and comparable to the LMC metallicity (which $12+\mathrm{log(O/H)}\sim8.30$, \citealt{Andrievsky2001A&A...367..605A}), suggesting that the host environment of the progenitor of SN~2025aico is likely to be relatively metal poor. This result suggests that the stellar wind of the progenitor of SN~2025aico should not be so strong, supporting the evolution path of the binary interaction.

To verify this result, we calculated the metallicity using the same approach on the spectrum around the core region of IC~700, obtained from the SDSS survey \citep{Ahumada2020ApJS..249....3A}. Utilizing the $N2$ index, the estimated metallicity in the core region is $12+\mathrm{log(O/H)}=8.366\pm0.036$, which is comparable to the metallicity at the location of SN~2025aico. This result suggests that the host galaxy region of SN~2025aico is generally metal-poor, which is consistent with our previous estimation. 

\subsection{Early shock-cooling emission}
\label{sec:disc:sc}
To perform a consistency check on the shock-cooling phase, we compared our results with those of several other SNe~IIb and the theoretical models provided by \cite{Ouchi2017apj}. The comparisons are presented in Fig.~\ref{fig:phot:env}. The green line represents the analytical solution of the model, while the purple dots represent the numerical solution of the model. We note that for SN~2025aico, the derived radius is much larger than the model predictions. This discrepancy might be caused by mass-transfer instabilities, which could result in a more extended H-rich envelope. The results derived from the models of \cite{P15_Piro2015ApJ...808L..51P} and \cite{SW17_Sapir2017ApJ...838..130S} show broad consistency. We observe that SN~2025aico is similar to cIIb events such as SN~2022ngb \citep[$M_\mathrm{env} = 0.035\,\mathrm{M_\odot}$, $R_\mathrm{env} = 4.5\,\mathrm{R_\odot}$;][]{SN2022ngb_Zhao2026A&A}, and SN~2016gkg \citep[$M_\mathrm{env} = 0.0272\,\mathrm{M_\odot}$, $R_\mathrm{env} = 41.8\,\mathrm{R_\odot}$;][]{Arcavi2017ApJ...837L...2A}. However, it differs significantly from events possessing highly extended envelopes during the shock-cooling phase, such as SN~1993J, SN~2013df \citep{VanDyk2014aj}, SN~2024iss \citep{SN2024iss_Chen2026AA}, and SN~2017ckj \citep{Li2025A&A...704A.233L}. Ultimately, these comparisons suggest that the progenitor of SN~2025aico is likely a compact stripped He star or a low-mass WR progenitor, similar to that of SN~2022ngb, and even more compact than the progenitor of SN~2016gkg.

\begin{figure}[htbp]
\centering
\includegraphics[width=\linewidth]{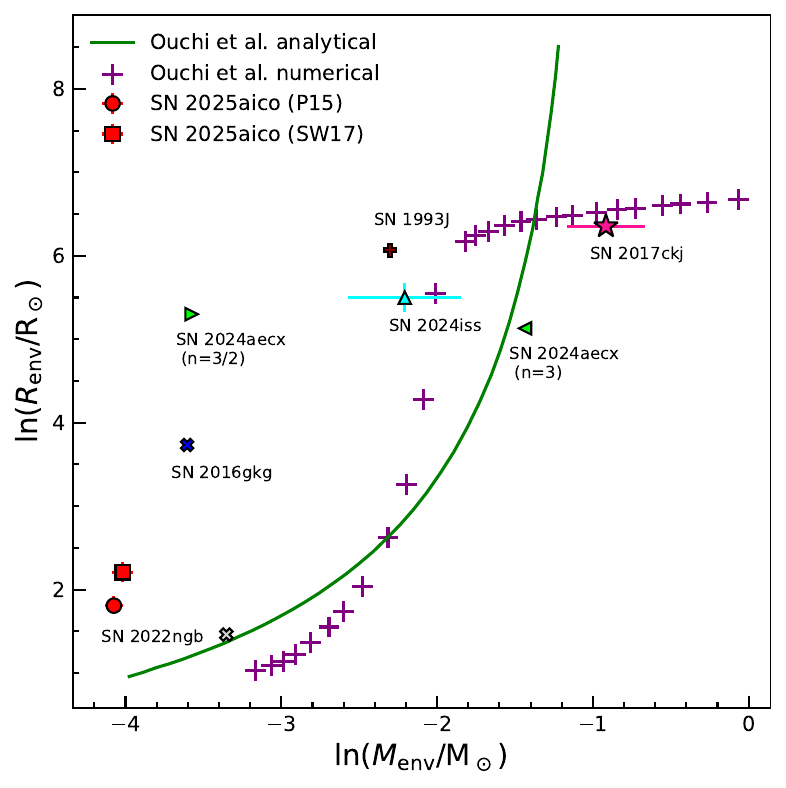}
\caption{Comparisons of the envelope properties of SN~2025aico and other SNe IIb, overplotted using the models of \cite{Ouchi2017apj}. The events are plotted by colour. The same events, when plotted with different models, are represented using the same colours but with different markers.}
\label{fig:phot:env}
\end{figure}

We also checked the {\it Swift}/XRT data for evidence of early (and/or) late X-ray emission. {\it Swift}/XRT collected 14 ks during 21 observations. The SN is not detected with an overall $3\,\sigma$ upper limit of $2.5\times 10^{-3}$ cts s$^{-1}$. Assuming a power-law spectrum with a photon index of $\Gamma=2$ and the Galactic column density of $1.3\times 10^{20}$ cm$^{-2}$, we derive an upper limit on the average unabsorbed 0.3-10 keV luminosity of $\sim 5\times 10^{39}$ erg s$^{-1}$. Concentrating on the 1.8 ks data taken on the same day of the GOTO discovery, we can derive an upper limit of $\sim 4\times 10^{40}$ erg s$^{-1}$.

\subsection{Evolution of \ion{He}{I}}\label{sec:disc:lineevo}

We present the evolution of the profiles for the \ion{He}{I}~$\lambda5876$ and \ion{He}{I}~$\lambda7065$ lines in Fig.~\ref{fig:spec:lineevo}.
During the early phase, the \ion{He}{I}~$\lambda5876$ feature exhibits a broad P~Cygni profile, common in SNe~IIb spectra during the shock-cooling phase \citep{SN2024aecx_Zou2026ApJ...997...77Z}. This morphology suggests an origin in the outer region of the ejecta but within the envelope, at a velocity of $\sim18000\,\mathrm{km\,s^{-1}}$. Following the shock-cooling phase, the P~Cygni profile of the \ion{He}{I}~$\lambda5876$ line becomes no longer detectable. Between $-$19.8~days and $-$17.1~days from the $r_\mathrm{M}$-band maximum, only the P~Cygni profiles of Balmer lines are identified in the spectra. After $-$16.0~days, both \ion{He}{I} lines show a weak, blueshifted emission peak with an offset of $-2000\,\mathrm{km\,s^{-1}}$ from the rest wavelength. To confirm that this feature does not simply result from a He-poor environment, we compare our spectra with those of the cIIb SN~2008ax in Sect.~\ref{sec:spec:cmp}, which also exhibits a prominent P~Cygni profile featuring an absorption component. The P~Cygni profile of the \ion{He}{I}~$\lambda5876$ line emerges at $-0.9$~days from the $r_\mathrm{M}$-band maximum, as the absorption component becomes prominent relative to the continuum. At later phases, the emission peaks of the two \ion{He}{I} lines shift towards the rest wavelength gradually, while the overall profile broadens. The absorption minimum of the \ion{He}{I}~$\lambda5876$ line also exhibits an increasing shift towards shorter wavelengths. These features reflect the typical characteristics of a He-rich environment and follow the standard evolutionary sequence of SNe~IIb. 

\begin{figure}[htbp]
\centering
\includegraphics[width=\linewidth]{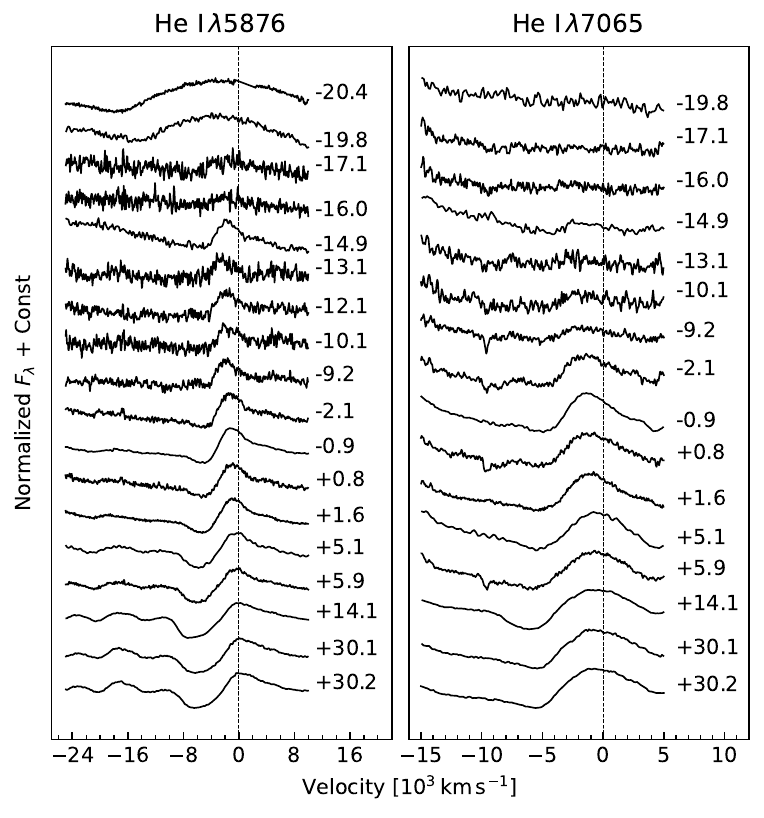}
\caption{Evolution of \ion{He}{I}~$\lambda5876$ and \ion{He}{I}~$\lambda7065$ lines. The zero velocity, coincident with the rest wavelength of the transition, is marked with dashed lines. The phases from the $r_\mathrm{M}$-band maximum are reported on the right side of each panel, near the corresponding spectra.}
\label{fig:spec:lineevo}
\end{figure}

The weak mixing scenario can explain at the same time the absence of the \ion{He}{I} P~Cygni absorption and the peculiar emission profile following the shock-cooling phase. Because \ion{He}{I} lines are primarily excited non-thermally, weak $^{56}$Ni mixing results in the rapidly expanding outer He-rich layer receiving inadequate illumination from $\gamma$-ray photons. Furthermore, as the temperature of the outer region drops below the He recombination temperature \citep[8000--10000~K;][]{Hatano1999ApJS..121..233H}, the outer ejecta becomes more optically thin to the \ion{He}{I}~$\lambda5876$ transition. This lack of non-thermal excitation and low opacity in the outer layers naturally explains the absence of the P~Cygni absorption trough \citep{Dessart2015MNRAS.453.2189D}. Meanwhile, deep within the ejecta near the $^{56}$Ni core, the material is adequately excited by $\gamma$ rays, allowing the \ion{He}{I} emission to form. As noted by \cite{Anderson2014MNRAS.441..671A}, because electron scattering dominates the continuum opacity, line photons have a low probability of destruction, enabling this emission to emerge from layers even deeper than the continuum photosphere. Also, the emission line from the inside of the ejecta exhibit a blueshifted peak, which is consistent with our observation. During the later phase, as $\gamma$ rays excited the outer \ion{He}{I} ejecta of SN~2025aico, the P~Cygni profiles of both the \ion{He}{I} lines formed via non-thermal excitation. Furthermore, the free-electron density at $-0.9$~days is higher, as the material is ionized by radioactively powered $\gamma$-ray photons. Because the opacity is dominated by electron scattering, this enhanced electron density makes the absorption components of the \ion{He}{I} lines visible, presenting the typical features of He-rich ejecta. In addition, the rising opacity along the line of sight causes the photosphere to shift outward in the co-moving frame. This outward movement results in the increased velocity and width of the \ion{He}{I}~$\lambda5876$ line, and hence an increased photospheric velocity \citep{Moriya2020MNRAS.497.1619M}. The late phase decrease in the \ion{He}{I}~$\lambda5876$ line velocity can be explained as an effect of the receding photosphere.

\subsection{Progenitor and explosion scenario}\label{sec:disc:prog}

Stripped-envelope SNe originate from either single stars that lose their envelope via strong stellar winds or binary systems through mass transfer processes. Given that the estimated ejecta mass of SN~2025aico is between $1.5$ and $2.8\,\mathrm{M_\odot}$, this yields an upper limit for the pre-SN progenitor mass of $M_{\mathrm{preSN}} \lesssim 4.5\,\mathrm{M_\odot}$. As a consequence of the relatively low mass, it is unlikely that the progenitor was stripped solely by stellar winds \citep{Folatelli2015apj, Yoon2017ApJ}. Therefore, binary mass transfer is the strongly favoured stripping mechanism for the progenitor of SN~2025aico. According to the models of \cite{Yoon2017ApJ}, forming such a compact progenitor requires either a short-period binary with Large Magellanic Cloud (LMC) like metallicity or a longer-period binary with a more massive progenitor and solar metallicity, the latter implying stronger stellar winds. However, based on our light curve fitting results, the massive progenitor scenario can be confidently ruled out, as it overestimates the pre-SN mass ($M_\mathrm{preSN} \gtrsim 6\,\mathrm{M_\odot}$). 
It is important to emphasize that, within this specific evolutionary framework, we cannot provide robust constraints for a high-metallicity progenitor that also satisfies our low-mass criteria. Thoroughly evaluating the role of metal-rich environments would require alternative theoretical models. Consequently, we can only place constraints based on the models utilizing LMC metallicity. Under this assumption and considering our previous fitting results, we propose that the progenitor of SN~2025aico was a stripped He star reasonably consistent with models having $M_\mathrm{preSN}$ ranging from $3.53$ to $4.56\,\mathrm{M_\odot}$ and a radius ranging from $10\,\mathrm{R_\odot}$ to $30\,\mathrm{R_\odot}$, which correspond to the models ranging from \texttt{Lm11p10} to \texttt{Lm13p400} in \cite{Yoon2017ApJ}. These models reside in binary systems with orbital periods between $10$ and $400$~days.

\begin{figure}[htbp]
\centering
\includegraphics[width=\linewidth]{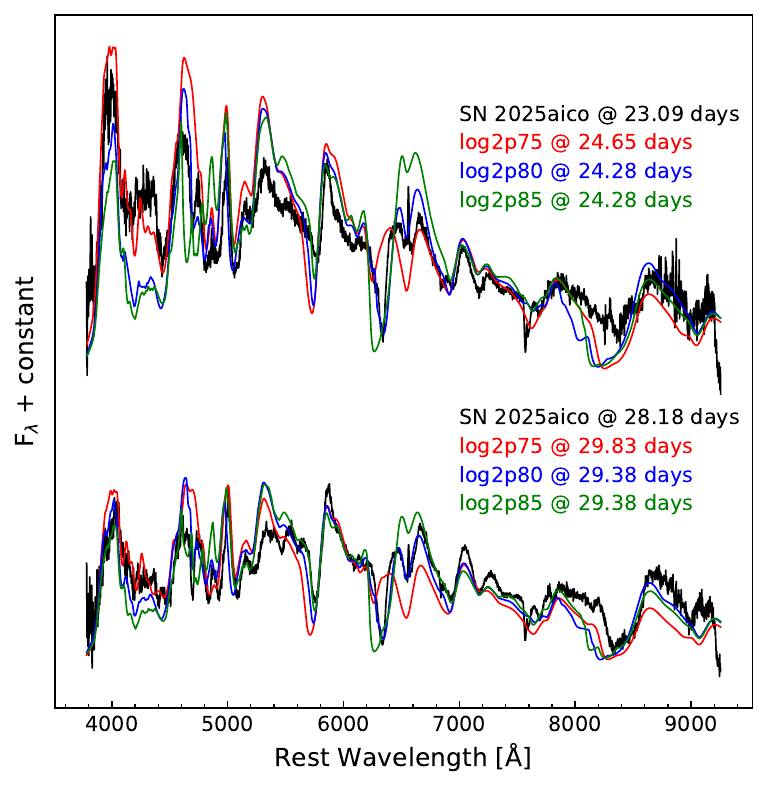}
\caption{Comparisons of the peak spectra of SN~2025aico and synthetic spectra models from \cite{Dessart2024A&A...685A.169D} for different orbital periods. Different models are overplotted in different colours, as shown in the legend. The epoch is calculated from the explosion epoch.}
\label{fig:spec:binarymt}
\end{figure}

To perform a robust consistency check of our results and provide a limit on the orbital period of the progenitor, we compared our high-S/N spectra with the synthetic models from \cite{Dessart2024A&A...685A.169D}, which have a similar companion initial mass compared to models of \cite{Yoon2017ApJ}. To minimize the influence of $^{56}$Ni mixing, we analysed the spectra obtained at the time of the maximum light. Furthermore, we scaled the synthetic spectra to match the luminosity of our spectra. We utilized models featuring relatively short periods ranging from 560 to 708~days. Because the models of \cite{Dessart2024A&A...685A.169D} assume solar metallicity, and because metallicity is highly degenerate with the orbital period, it is reasonable that the periods within these models are slightly longer. Therefore, we adopted the \texttt{log2p75}, \texttt{log2p80}, and \texttt{log2p85} models. Because the estimated $M_\mathrm{ej}$ falls within the reasonable range of $1.5$ to $2.8\,\mathrm{M_\odot}$, these models remain entirely valid under our assumptions. The comparisons are presented in Fig.~\ref{fig:spec:binarymt}.

The best-fit model is \texttt{log2p80}, which exhibits an H$\alpha$ profile with an intensity comparable to that of our observations. In the \texttt{log2p75} model, the hydrogen features are undetectable. This absence contradicts our observations and instead indicates a type~Ib event, making this model more suitable for objects such as SN~2007Y (see Sect.~\ref{sec:spec:cmp}). Conversely, in the \texttt{log2p85} model, the H$\alpha$ profile is excessively broad and intense compared to our spectra. Furthermore, the corresponding theoretical light curve of the \texttt{log2p85} model exhibits a much more prominent shock-cooling phase, inconsistent with our photometric data. Therefore, if we assumed a solar metallicity, the binary orbital period would be consistent with approximately $600$~days. However, we notice that the residual H-rich envelope in the \texttt{log2p80} model is much more massive than the outcome of the light curve fits ($\sim0.1\,\mathrm{M_\odot}$ in the model). Thus, the estimated orbital period in a regime of solar metallicity can only be an upper limit. Ultimately, these findings indicate that the progenitor of SN~2025aico was stripped within a close, interacting binary system via early-time Case~B mass transfer. In summary, the progenitor system had an orbital period between $10$ and $400$~days in the LMC metallicity regime, which is the scenario we favour and is consistent with our analysis of the host environment (see Sect.~\ref{sec:disc:host}). In addition, we can establish an upper limit of approximately $600$~days for the orbital period under the assumption of a solar metallicity environment.

\section{Conclusions}
\label{sec:con}
In this work, we presented multi-band photometric and spectroscopic observations of SN~2025aico. The photometric observations range from $\sim-25$~days to $\sim+45$~days relative to the $r_\mathrm{M}$-band maximum, while the spectroscopic observations range from $\sim-20$~days to $\sim+43$~days. {During the very early phase, the shock-cooling emission feature is clearly visible is the bluer bands.
Furthermore, the well-sampled} photometric data show a rise time in the $r_\mathrm{M}$ band of $\sim22.3$~days, similar to those observed in typical SNe~IIb, featuring a relatively low-luminosity peak of $L_{\mathrm{opt}}=(4.07\pm0.10)\times10^{41}\,\mathrm{erg\,s^{-1}}$. 
The spectroscopic observations generally follow the evolution of typical SNe~IIb, with initial broad P Cygni profiles in the Balmer and \ion{He}{I}~$\lambda5876$ lines. However, starting from about two weeks before maximum light, the profiles of these lines change. In particular, \ion{He}{I}~$\lambda5876$ no longer shows a P~Cygni profile, and H$\alpha$ exhibits a double-absorption trough. Furthermore, we note that the velocity derived from the \ion{He}{I}~$\lambda5876$ P~Cygni profile increases in the later spectra. Combined with the intrinsic colour evolution, these features are consistent with a weak to moderate $^{56}$Ni mixing.

By fitting theoretical models to the light curve, we find that the progenitor of SN~2025aico is consistent with a compact He star that has a residual H-rich envelope of $M_{\mathrm{env}} \approx 0.02\,\mathrm{M_\odot}$ and $R_{\mathrm{env}} \approx 6$--$10\,\mathrm{R_\odot}$. The ejecta mass is $M_{\mathrm{ej}} = 2.79_{-0.18}^{+0.21}\,\mathrm{M_\odot}$, accompanied by a relatively low nickel mass of $M_{\mathrm{Ni}} = 0.033_{-0.004}^{+0.006}\,\mathrm{M_\odot}$. Accounting for the dependence on the adopted optical opacity, the estimated $M_\mathrm{ej}$ reasonably falls within the range of $1.5\,\mathrm{M_\odot}$ to $2.9\,\mathrm{M_\odot}$. These values indicate a pre-SN progenitor mass of $M_{\mathrm{preSN}} \approx 3$--$4.5\,\mathrm{M_\odot}$ that retains only a trace of hydrogen, likely stripped via binary interaction. From the model comparison, we assumed that, for an LMC-like metallicity, the initial orbital period should reasonably lie in the range of 10--400~days due to early Case B mass transfer, which we favour in this article. An upper limit can be given at $\sim600$~days, assuming solar metallicity.

Early-phase data (e.g. the shock breakout, the shock-cooling phase, and the early rise time) of SNe~IIb significantly enhance our understanding of the explosion mechanism and provide crucial constraints on the progenitor. The Mephisto system, comprising a 1.6-metre telescope and two 50-centimetre telescopes with a large field of view, will enable simultaneous observations in both the \textit{ugi} and \textit{vrz} bands. This capability is crucial for obtaining real-time colour information of SNe during their earliest epochs, enabling improved modelling of the shock-cooling phase and offering tighter constraints on the $^{56}$Ni mixing, a parameter of remarkable importance in SE SNe. Furthermore, combining optical observations with X-ray data from the Einstein Probe (EP; \citealt{Yuan_EP2025SCPMA..6839501Y}) will allow us to capture the first light from the shock breakout. Ultimately, the synergistic use of these facilities will yield a much deeper understanding of such transient events.

\section{Data availability}
UV and optical photometric measurements of SN~2025aico are available at the Strasbourg astronomical Data Centre (CDS) via \url{https://cdsarc.cds.unistra.fr/viz-bin/cat/J/A+A/715/A126}. Our observations of the spectra are available through the Weizmann Interactive SN Data Repository (WISeREP; \citealt{Yaron2012pasp}) via \url{https://www.wiserep.org/object/29614}.

\bibliographystyle{aa}
\bibliography{ref.bib}

@article{Tully2016aj,
   author = {Tully, R. Brent and Courtois, H{\'e}l{\`e}ne M. and Sorce, Jenny G.},
    title = "{Cosmicflows-3}",
  journal = {\aj},
     year = 2016,
    month = aug,
   volume = {152},
   number = {2},
    pages = {50},
      doi = {10.3847/0004-6256/152/2/50}
}

@article{Woosley2002RevModPhys,
   title = "{The evolution and explosion of massive stars}",
  author = {Woosley, S. E. and Heger, A. and Weaver, T. A.},
 journal = {Rev. Mod. Phys.},
  volume = {74},
  number = {4},
   pages = {1015--1071},
    year = {2002},
   month = {Nov},
     doi = {10.1103/RevModPhys.74.1015}
}

@article{Janka2012ARNPS,
   author = {Janka, Hans-Thomas},
    title = "{Explosion Mechanisms of Core-Collapse Supernovae}",
  journal = {Annual Review of Nuclear and Particle Science},
     year = 2012,
    month = nov,
   volume = {62},
   number = {1},
    pages = {407-451},
      doi = {10.1146/annurev-nucl-102711-094901}
}

@article{Schlafly2011apj,
   author = {Schlafly, Edward F. and Finkbeiner, Douglas P.},
    title = "{Measuring Reddening with Sloan Digital Sky Survey Stellar Spectra and Recalibrating SFD}",
  journal = {\apj},
     year = 2011,
    month = aug,
   volume = {737},
   number = {2},
    pages = {103},
      doi = {10.1088/0004-637X/737/2/103}
}

@article{Richmond1994aj,
   author = {Richmond, Michael W. and Treffers, Richard R. and Filippenko, Alexei V. and Paik, Young and Leibundgut, Bruno and Schulman, Eric and Cox, Caroline V.},
    title = "{UBVRI Photometry of SN 1993J in M81: The First 120 Days}",
  journal = {\aj},
     year = 1994,
    month = mar,
   volume = {107},
    pages = {1022},
      doi = {10.1086/116915}
}

@article{Richmond1996aj,
   author = {Richmond, Michael W. and Treffers, Richard R. and Filippenko, Alexei V. and Paik, Young},
    title = "{UBVRI Photometry of SN 1993J in M81: Days 3 to 365}",
  journal = {\aj},
     year = 1996,
    month = aug,
   volume = {112},
    pages = {732},
      doi = {10.1086/118048}
}

@article{Pastorello2008mnras,
   author = {Pastorello, A. and Kasliwal, M.~M. and Crockett, R.~M. and Valenti, S. and Arbour, R. and Itagaki, K. and Kaspi, S. and Gal-Yam, A. and Smartt, S.~J. and Griffith, R. and Maguire, K. and Ofek, E.~O. and Seymour, N. and Stern, D. and Wiethoff, W.},
    title = "{The Type IIb SN 2008ax: spectral and light curve evolution}",
  journal = {\mnras},
     year = 2008,
    month = sep,
   volume = {389},
   number = {2},
    pages = {955},
      doi = {10.1111/j.1365-2966.2008.13618.x}
}

@article{Morales-Garoffolo2015mnras,
   author = {Morales-Garoffolo, A. and Elias-Rosa, N. and Bersten, M. and Jerkstrand, A. and Taubenberger, S. and Benetti, S. and Cappellaro, E. and Kotak, R. and Pastorello, A. and Bufano, F. and Domínguez, R. M. and Ergon, M. and Fraser, M. and Gao, X. and García, E. and Howell, D. A. and Isern, J. and Smartt, S. J. and Tomasella, L. and Valenti, S.},
    title = "{SN 2011fu: a type IIb supernova with a luminous double-peaked light curve}",
  journal = {\mnras},
   volume = {454},
   number = {1},
    pages = {95-114},
     year = {2015},
    month = {09},
      doi = {10.1093/mnras/stv1972}
}

@article{VanDyk2014aj,
    doi = {10.1088/0004-6256/147/2/37},
   year = 2014,
  month = {jan},
 volume = {147},
 number = {2},
  pages = {37},
 author = {Schuyler D. Van Dyk and WeiKang Zheng and Ori D. Fox and S. Bradley Cenko and Kelsey I. Clubb and Alexei V. Filippenko and Ryan J. Foley and Adam A. Miller and Nathan Smith and Patrick L. Kelly and William H. Lee and Sagi Ben-Ami and Avishay Gal-Yam},
  title = "{The Type IIb Supernova 2013df and Its Cool Supergiant Progenitor}",
journal = {\aj}
}

@article{Gangopadhyay2018mnras,
    DOI = {10.1093/mnras/sty478},
  title = "{SN 2015as: a low-luminosity Type IIb supernova without an early light-curve peak}",
 volume = {476},
 number = {3},
journal = {\mnras},
 author = {Gangopadhyay, Anjasha and Misra, Kuntal and Pastorello, A and Sahu, D K and Tomasella, L and Tartaglia, L and Singh, Mridweeka and Dastidar, Raya and Srivastav, S and Ochner, P and Brown, Peter J and Anupama, G C and Benetti, S and Cappellaro, E and Kumar, Brajesh and Kumar, Brijesh and Pandey, S B},
   year = {2018},
  month = feb,
  pages = {3611-3630}
}

@ARTICLE{Medler2022MNRAS,
       author = {{Medler}, K. and {Mazzali}, P.~A. and {Teffs}, J. and {Ashall}, C. and {Anderson}, J.~P. and {Arcavi}, I. and {Benetti}, S. and {Bostroem}, K.~A. and {Burke}, J. and {Cai}, Y. -Z. and {Charalampopoulos}, P. and {Elias-Rosa}, N. and {Ergon}, M. and {Galbany}, L. and {Gromadzki}, M. and {Hiramatsu}, D. and {Howell}, D.~A. and {Inserra}, C. and {Lundqvist}, P. and {McCully}, C. and {M{\"u}ller-Bravo}, T. and {Newsome}, M. and {Nicholl}, M. and {Padilla Gonzalez}, E. and {Paraskeva}, E. and {Pastorello}, A. and {Pellegrino}, C. and {Pessi}, P.~J. and {Reguitti}, A. and {Reynolds}, T.~M. and {Roy}, R. and {Terreran}, G. and {Tomasella}, L. and {Young}, D.~R.},
        title = "{SN 2020acat: an energetic fast rising Type IIb supernova}",
      journal = {\mnras},
         year = 2022,
        month = jul,
       volume = {513},
       number = {4},
        pages = {5540-5558},
          doi = {10.1093/mnras/stac1192},
archivePrefix = {arXiv},
       eprint = {2201.06991},
 primaryClass = {astro-ph.HE},
       adsurl = {https://ui.adsabs.harvard.edu/abs/2022MNRAS.513.5540M}
}

@ARTICLE{Ergon2024A&A,
       author = {{Ergon}, Mattias and {Lundqvist}, Peter and {Fransson}, Claes and {Kuncarayakti}, Hanindyo and {Das}, Kaustav K. and {De}, Kishalay and {Ferrari}, Lucia and {Fremling}, Christoffer and {Medler}, Kyle and {Maeda}, Keiichi and {Pastorello}, Andrea and {Sollerman}, Jesper and {Stritzinger}, Maximilian D.},
        title = "{Light curve and spectral modelling of the type IIb SN 2020acat. Evidence for a strong Ni bubble effect on the diffusion time}",
      journal = {\aap},
         year = 2024,
        month = mar,
       volume = {683},
          eid = {A241},
        pages = {A241},
          doi = {10.1051/0004-6361/202346718},
archivePrefix = {arXiv},
       eprint = {2308.07158},
 primaryClass = {astro-ph.HE},
       adsurl = {https://ui.adsabs.harvard.edu/abs/2024A&A...683A.241E}
}

@article{Kumar2013mnras,
   author = {Kumar, Brajesh and Pandey, S. B. and Sahu, D. K. and Vinko, J. and Moskvitin, A. S. and Anupama, G. C. and Bhatt, V. K. and Ordasi, A. and Nagy, A. and Sokolov, V. V. and Sokolova, T. N. and Komarova, V. N. and Kumar, Brijesh and Bose, Subhash and Roy, Rupak and Sagar, Ram},
    title = "{Light curve and spectral evolution of the Type IIb supernova 2011fu}",
  journal = {\mnras},
   volume = {431},
   number = {1},
    pages = {308-321},
     year = {2013},
    month = {02},
      doi = {10.1093/mnras/stt162}
}

@article{Nagy2016aap,
   author = {Nagy, A.~P. and Vink{\'o}, J.},
    title = "{A two-component model for fitting light curves of core-collapse supernovae}",
  journal = {\aap},
     year = 2016,
    month = may,
   volume = {589},
    pages = {A53},
      doi = {10.1051/0004-6361/201527931}
}

@article{Arnett1982apj,
   author = {Arnett, W.~D.},
    title = "{Type I supernovae. I - Analytic solutions for the early part of the light curve}",
  journal = {\apj},
     year = 1982,
    month = feb,
   volume = {253},
    pages = {785-797},
      doi = {10.1086/159681}
}

@article{Nicholl2018RNAAS,
   author = {Nicholl, Matt},
    title = "{SuperBol: A User-friendly Python Routine for Bolometric Light Curves}",
  journal = {Research Notes of the American Astronomical Society},
     year = 2018,
    month = dec,
   volume = {2},
   number = {4},
    pages = {230},
      doi = {10.3847/2515-5172/aaf799}
}

@article{Chen2024apjl,
    doi = {10.3847/2041-8213/ad62f7},
   year = {2024},
  month = {aug},
 volume = {971},
 number = {1},
  pages = {L2},
 author = {Chen, Xinlei and Kumar, Brajesh and Er, Xinzhong and Guo, Helong and Yang, Yuan-Pei and Lin, Weikang and Fang, Yuan and Du, Guowang and Liu, Chenxu and Zhao, Jiewei and Zhang, Tianyu and Bao, Yuxi and Zou, Xingzhu and Pan, Yu and Wang, Yu and Zhu, Xufeng and Chatterjee, Kaushik and Liu, Xiangkun and Liu, Dezi and Lagioia, Edoardo P. and Rangwal, Geeta and Zhong, Shiyan and Zhang, Jinghua and Lian, Jianhui and Cai, Yongzhi and Zhang, Yangwei and Liu, Xiaowei},
  title = "{Early-phase Simultaneous Multiband Observations of the Type II Supernova SN 2024ggi with Mephisto}",
journal = {\apjl}
}

@article{Chatzopoulos2012apj,
   author = {Chatzopoulos, E. and Wheeler, J. Craig and Vinko, J.},
    title = "{Generalized Semi-analytical Models of Supernova Light Curves}",
  journal = {\apj},
     year = 2012,
    month = feb,
   volume = {746},
   number = {2},
    pages = {121},
      doi = {10.1088/0004-637X/746/2/121}
}

@article{Arnett1989apj,
   author = {Arnett, W. David and Fu, Albert},
    title = "{The Late Behavior of Supernova 1987A. I. The Light Curve}",
  journal = {\apj},
     year = 1989,
    month = may,
   volume = {340},
    pages = {396},
      doi = {10.1086/167402}
}

@article{Rabinak2011apj,
   author = {Rabinak, Itay and Waxman, Eli},
    title = "{The Early UV/Optical Emission from Core-collapse Supernovae}",
  journal = {\apj},
     year = 2011,
    month = feb,
   volume = {728},
   number = {1},
    pages = {63},
      doi = {10.1088/0004-637X/728/1/63}
}

@article{Folatelli2015apj,
    doi = {10.1088/0004-637X/811/2/147},
   year = {2015},
  month = {sep},
 volume = {811},
 number = {2},
  pages = {147},
 author = {Folatelli, Gastón and Bersten, Melina C. and Kuncarayakti, Hanindyo and Benvenuto, Omar G. and Maeda, Keiichi and Nomoto, Ken’ichi},
  title = "{The Progenitor of the Type IIb SN 2008ax Revisited}",
journal = {\apj}
}

@article{Ouchi2017apj,
   author = {Ouchi, Ryoma and Maeda, Keiichi},
    title = "{Radii and Mass-loss Rates of Type IIb Supernova Progenitors}",
  journal = {\apj},
     year = 2017,
    month = may,
   volume = {840},
   number = {2},
    pages = {90},
      doi = {10.3847/1538-4357/aa6ea9}
}

@article{Jerkstrand2015aap,
   author = {Jerkstrand, A. and Ergon, M. and Smartt, S. J. and Fransson, C. and Sollerman, J. and Taubenberger, S. and Bersten, M. and Spyromilio, J.},
    title = "{Late-time spectral line formation in Type IIb supernovae, with application to SN 1993J, SN 2008ax, and SN 2011dh}",
      DOI = {10.1051/0004-6361/201423983},
  journal = {\aap},
     year = 2015,
   volume = 573,
    pages = {A12}
}

@ARTICLE{Yaron2012pasp,
       author = {{Yaron}, Ofer and {Gal-Yam}, Avishay},
        title = "{WISeREP{\textemdash}An Interactive Supernova Data Repository}",
      journal = {\pasp},
         year = 2012,
        month = jul,
       volume = {124},
       number = {917},
        pages = {668},
          doi = {10.1086/666656},
archivePrefix = {arXiv},
       eprint = {1204.1891},
 primaryClass = {astro-ph.IM},
       adsurl = {https://ui.adsabs.harvard.edu/abs/2012PASP..124..668Y}
}

@ARTICLE{Dessart2016MNRAS,
       author = {{Dessart}, Luc and {Hillier}, D. John and {Woosley}, Stan and {Livne}, Eli and {Waldman}, Roni and {Yoon}, Sung-Chul and {Langer}, Norbert},
        title = "{Inferring supernova IIb/Ib/Ic ejecta properties from light curves and spectra: correlations from radiative-transfer models}",
      journal = {\mnras},
         year = 2016,
        month = may,
       volume = {458},
       number = {2},
        pages = {1618-1635},
          doi = {10.1093/mnras/stw418},
archivePrefix = {arXiv},
       eprint = {1602.06280},
 primaryClass = {astro-ph.SR},
       adsurl = {https://ui.adsabs.harvard.edu/abs/2016MNRAS.458.1618D}
}

@ARTICLE{Barmentloo2024MNRAS,
       author = {{Barmentloo}, Stan and {Jerkstrand}, Anders and {Iwamoto}, Koichi and {Hachisu}, Izumi and {Nomoto}, Ken'ichi and {Sollerman}, Jesper and {Woosley}, Stan},
        title = "{Nebular nitrogen line emission in stripped-envelope supernovae - a new progenitor mass diagnostic}",
      journal = {\mnras},
         year = 2024,
        month = sep,
       volume = {533},
       number = {2},
        pages = {1251-1280},
          doi = {10.1093/mnras/stae1811},
archivePrefix = {arXiv},
       eprint = {2403.08911},
 primaryClass = {astro-ph.HE},
       adsurl = {https://ui.adsabs.harvard.edu/abs/2024MNRAS.533.1251B}
}

@ARTICLE{Dessart2005AA,
       author = {{Dessart}, L. and {Hillier}, D.~J.},
        title = "{Quantitative spectroscopy of photospheric-phase type II supernovae}",
      journal = {\aap},
         year = 2005,
        month = jul,
       volume = {437},
       number = {2},
        pages = {667-685},
          doi = {10.1051/0004-6361:20042525},
archivePrefix = {arXiv},
       eprint = {astro-ph/0504028},
 primaryClass = {astro-ph},
       adsurl = {https://ui.adsabs.harvard.edu/abs/2005A&A...437..667D}
}

@ARTICLE{Valenti2008ApJ,
       author = {{Valenti}, S. and {Elias-Rosa}, N. and {Taubenberger}, S. and {Stanishev}, V. and {Agnoletto}, I. and {Sauer}, D. and {Cappellaro}, E. and {Pastorello}, A. and {Benetti}, S. and {Riffeser}, A. and {Hopp}, U. and {Navasardyan}, H. and {Tsvetkov}, D. and {Lorenzi}, V. and {Patat}, F. and {Turatto}, M. and {Barbon}, R. and {Ciroi}, S. and {Di Mille}, F. and {Frandsen}, S. and {Fynbo}, J.~P.~U. and {Laursen}, P. and {Mazzali}, P.~A.},
        title = "{The Carbon-rich Type Ic SN 2007gr: The Photospheric Phase}",
      journal = {\apjl},
         year = 2008,
        month = feb,
       volume = {673},
       number = {2},
        pages = {L155},
          doi = {10.1086/527672},
archivePrefix = {arXiv},
       eprint = {0712.1899},
 primaryClass = {astro-ph},
       adsurl = {https://ui.adsabs.harvard.edu/abs/2008ApJ...673L.155V}
}

@ARTICLE{Medler2021MNRAS,
       author = {{Medler}, K. and {Mazzali}, P.~A. and {Teffs}, J. and {Prentice}, S.~J. and {Ashall}, C. and {Amenouche}, M. and {Anderson}, J.~P. and {Burke}, J. and {Chen}, T.~W. and {Galbany}, L. and {Gromadzki}, M. and {Guti{\'e}rrez}, C.~P. and {Hiramatsu}, D. and {Howell}, D.~A. and {Inserra}, C. and {Kankare}, E. and {McCully}, C. and {M{\"u}ller-Bravo}, T.~E. and {Nicholl}, M. and {Pellegrino}, C. and {Sollerman}, J.},
        title = "{SN 2020cpg: an energetic link between Type IIb and Ib supernovae}",
      journal = {\mnras},
         year = 2021,
        month = sep,
       volume = {506},
       number = {2},
        pages = {1832-1849},
          doi = {10.1093/mnras/stab1761},
archivePrefix = {arXiv},
       eprint = {2106.09505},
 primaryClass = {astro-ph.HE},
       adsurl = {https://ui.adsabs.harvard.edu/abs/2021MNRAS.506.1832M}
}

@ARTICLE{Nagy2014AA,
       author = {{Nagy}, A.~P. and {Ordasi}, A. and {Vink{\'o}}, J. and {Wheeler}, J.~C.},
        title = "{A semianalytical light curve model and its application to type IIP supernovae}",
      journal = {\aap},
         year = 2014,
        month = nov,
       volume = {571},
          eid = {A77},
        pages = {A77},
          doi = {10.1051/0004-6361/201424237},
archivePrefix = {arXiv},
       eprint = {1409.6256},
 primaryClass = {astro-ph.HE},
       adsurl = {https://ui.adsabs.harvard.edu/abs/2014A&A...571A..77N}
}

@ARTICLE{Reguitti2025AA,
       author = {{Reguitti}, A. and {Pastorello}, A. and {Smartt}, S.~J. and {Valerin}, G. and {Pignata}, G. and {Campana}, S. and {Chen}, T. -W. and {Sankar}, A.~K. and {Moran}, S. and {Mazzali}, P.~A. and {Duarte}, J. and {Salmaso}, I. and {Anderson}, J.~P. and {Ashall}, C. and {Benetti}, S. and {Gromadzki}, M. and {Guti{\'e}rrez}, C.~P. and {Humina}, C. and {Inserra}, C. and {Kankare}, E. and {Kravtsov}, T. and {Muller-Bravo}, T.~E. and {Pessi}, P.~J. and {Sollerman}, J. and {Young}, D.~R. and {Chambers}, K. and {de Boer}, T. and {Gao}, H. and {Huber}, M. and {Lin}, C. -C. and {Lowe}, T. and {Magnier}, E. and {Minguez}, P. and {Smith}, I.~A. and {Smith}, K.~W. and {Srivastav}, S. and {Wainscoat}, R. and {Benedet}, M.},
        title = "{SN 2024abfo: A partially stripped type II supernova from a yellow supergiant}",
      journal = {\aap},
         year = 2025,
        month = jun,
       volume = {698},
          eid = {A129},
        pages = {A129},
          doi = {10.1051/0004-6361/202554388},
archivePrefix = {arXiv},
       eprint = {2503.03851},
 primaryClass = {astro-ph.HE},
       adsurl = {https://ui.adsabs.harvard.edu/abs/2025A&A...698A.129R}
}

@ARTICLE{Yoon2017ApJ,
       author = {{Yoon}, Sung-Chul and {Dessart}, Luc and {Clocchiatti}, Alejandro},
        title = "{Type Ib and IIb Supernova Progenitors in Interacting Binary Systems}",
      journal = {\apj},
         year = 2017,
        month = may,
       volume = {840},
       number = {1},
          eid = {10},
        pages = {10},
          doi = {10.3847/1538-4357/aa6afe},
archivePrefix = {arXiv},
       eprint = {1701.02089},
 primaryClass = {astro-ph.SR},
       adsurl = {https://ui.adsabs.harvard.edu/abs/2017ApJ...840...10Y}
}

@ARTICLE{Crockett2008MNRAS,
       author = {{Crockett}, R.~M. and {Eldridge}, J.~J. and {Smartt}, S.~J. and {Pastorello}, A. and {Gal-Yam}, A. and {Fox}, D.~B. and {Leonard}, D.~C. and {Kasliwal}, M.~M. and {Mattila}, S. and {Maund}, J.~R. and {Stephens}, A.~W. and {Danziger}, I.~J.},
        title = "{The type IIb SN 2008ax: the nature of the progenitor}",
      journal = {\mnras},
         year = 2008,
        month = nov,
       volume = {391},
       number = {1},
        pages = {L5-L9},
          doi = {10.1111/j.1745-3933.2008.00540.x},
archivePrefix = {arXiv},
       eprint = {0805.1913},
 primaryClass = {astro-ph},
       adsurl = {https://ui.adsabs.harvard.edu/abs/2008MNRAS.391L...5C}
}

@ARTICLE{Filippenko1997ARAA,
       author = {{Filippenko}, Alexei V.},
        title = "{Optical Spectra of Supernovae}",
      journal = {\araa},
         year = 1997,
        month = jan,
       volume = {35},
        pages = {309-355},
          doi = {10.1146/annurev.astro.35.1.309},
       adsurl = {https://ui.adsabs.harvard.edu/abs/1997ARA&A..35..309F}
}

@ARTICLE{Modjaz2019NatAs,
       author = {{Modjaz}, Maryam and {Guti{\'e}rrez}, Claudia P. and {Arcavi}, Iair},
        title = "{New regimes in the observation of core-collapse supernovae}",
      journal = {Nature Astronomy},
         year = 2019,
        month = aug,
       volume = {3},
        pages = {717-724},
          doi = {10.1038/s41550-019-0856-2},
archivePrefix = {arXiv},
       eprint = {1908.02476},
 primaryClass = {astro-ph.HE},
       adsurl = {https://ui.adsabs.harvard.edu/abs/2019NatAs...3..717M}
}

@ARTICLE{Smith2014ARAA,
       author = {{Smith}, Nathan},
        title = "{Mass Loss: Its Effect on the Evolution and Fate of High-Mass Stars}",
      journal = {\araa},
         year = 2014,
        month = aug,
       volume = {52},
        pages = {487-528},
          doi = {10.1146/annurev-astro-081913-040025},
archivePrefix = {arXiv},
       eprint = {1402.1237},
 primaryClass = {astro-ph.SR},
       adsurl = {https://ui.adsabs.harvard.edu/abs/2014ARA&A..52..487S}
}

@ARTICLE{Ritter1988AA,
       author = {{Ritter}, H.},
        title = "{Turning on and off mass transfer in cataclysmic binaries.}",
      journal = {\aap},
         year = 1988,
        month = aug,
       volume = {202},
        pages = {93-100},
       adsurl = {https://ui.adsabs.harvard.edu/abs/1988A&A...202...93R}
}

@ARTICLE{Filippenko1988AJ,
       author = {{Filippenko}, Alexei V.},
        title = "{Supernova 1987K: Type II in Youth, Type Ib in Old Age}",
      journal = {\aj},
         year = 1988,
        month = dec,
       volume = {96},
        pages = {1941},
          doi = {10.1086/114940},
       adsurl = {https://ui.adsabs.harvard.edu/abs/1988AJ.....96.1941F}
}

@ARTICLE{Chevalier2010ApJL,
       author = {{Chevalier}, Roger A. and {Soderberg}, Alicia M.},
        title = "{Type IIb Supernovae with Compact and Extended Progenitors}",
      journal = {\apjl},
         year = 2010,
        month = mar,
       volume = {711},
       number = {1},
        pages = {L40-L43},
          doi = {10.1088/2041-8205/711/1/L40},
archivePrefix = {arXiv},
       eprint = {0911.3408},
 primaryClass = {astro-ph.HE},
       adsurl = {https://ui.adsabs.harvard.edu/abs/2010ApJ...711L..40C}
}

@ARTICLE{Dessart2018AA,
       author = {{Dessart}, Luc and {Yoon}, Sung-Chul and {Livne}, Eli and {Waldman}, Roni},
        title = "{Supernovae from massive stars with extended tenuous envelopes}",
      journal = {\aap},
         year = 2018,
        month = apr,
       volume = {612},
          eid = {A61},
        pages = {A61},
          doi = {10.1051/0004-6361/201732363},
archivePrefix = {arXiv},
       eprint = {1801.02056},
 primaryClass = {astro-ph.HE},
       adsurl = {https://ui.adsabs.harvard.edu/abs/2018A&A...612A..61D}
}

@INPROCEEDINGS{Tody1986SPIE,
       author = {{Tody}, Doug},
        title = "{The IRAF Data Reduction and Analysis System}",
    booktitle = {Instrumentation in astronomy VI},
         year = 1986,
       editor = {{Crawford}, David L.},
       series = {Society of Photo-Optical Instrumentation Engineers (SPIE) Conference Series},
       volume = {627},
        month = jan,
        pages = {733},
          doi = {10.1117/12.968154},
       adsurl = {https://ui.adsabs.harvard.edu/abs/1986SPIE..627..733T}
}

@MISC{Becker2015ascl,
   author = {{Becker}, A.},
    title = "{HOTPANTS: High Order Transform of PSF ANd Template Subtraction}",
howpublished = {Astrophysics Source Code Library},
     year = 2015,
    month = apr,
archivePrefix = "ascl",
   eprint = {1504.004},
   adsurl = {http://adsabs.harvard.edu/abs/2015ascl.soft04004B}
}

@inbook{Turatto2003,
   title={Variety in Supernovae},
   year=2003,
   ISBN={3540004831},
   url={http://dx.doi.org/10.1007/10828549_26},
   DOI={10.1007/10828549_26},
   booktitle={From Twilight to Highlight: The Physics of Supernovae},
   publisher={Springer-Verlag},
   author={Turatto, Massimo and Benetti, Stefano and Cappellaro, Enrico},
   pages={200–209} }

@INPROCEEDINGS{Yuan2020SPIE11445E,
       author = {{Yuan}, Xiangyan and {Li}, Zhengyang and {Liu}, Xiaowei and {Niu}, Dongsheng and {Lu}, Qishui and {Jiang}, Fanghua and {Wang}, Yuefei and {Li}, Xiaoyan and {Liang}, YongJun and {Wang}, Hai and {Zhang}, Chao and {Wang}, Jinfeng and {Li}, Bo and {Tian}, Jie and {Lu}, Haiping and {Chen}, Bingqiu and {Huang}, Yang and {Liu}, Xiangkun and {Yao}, Zhengqiu and {Cui}, Xiangqun and {Li}, Guoping},
        title = "{Development of the Multi-channel Photometric Survey telescope}",
    booktitle = {Ground-based and Airborne Telescopes VIII},
         year = 2020,
       editor = {{Marshall}, Heather K. and {Spyromilio}, Jason and {Usuda}, Tomonori},
       series = {Society of Photo-Optical Instrumentation Engineers (SPIE) Conference Series},
       volume = {11445},
        month = dec,
          eid = {114457M},
        pages = {114457M},
          doi = {10.1117/12.2562334},
       adsurl = {https://ui.adsabs.harvard.edu/abs/2020SPIE11445E..7MY}
}

@ARTICLE{Shingles2021TNSAN,
       author = {{Shingles}, L. and {Smith}, K.~W. and {Young}, D.~R. and {Smartt}, S.~J. and {Tonry}, J. and {Denneau}, L. and {Heinze}, A. and {Weiland}, H. and {Flewelling}, H. and {Stalder}, B. and {Clocchiatti}, A. and {F{\"o}rster}, F. and {Pignata}, G. and {Rest}, A. and {Anderson}, J. and {Stubbs}, C. and {Erasmus}, N.},
        title = "{Release of the ATLAS Forced Photometry server for public use}",
      journal = {Transient Name Server AstroNote},
         year = 2021,
        month = jan,
       volume = {7},
        pages = {1-7},
       adsurl = {https://ui.adsabs.harvard.edu/abs/2021TNSAN...7....1S}
}

@ARTICLE{Smith2019RNAAS,
       author = {{Smith}, K.~W. and {Williams}, R.~D. and {Young}, D.~R. and {Ibsen}, A. and {Smartt}, S.~J. and {Lawrence}, A. and {Morris}, D. and {Voutsinas}, S. and {Nicholl}, M.},
        title = "{Lasair: The Transient Alert Broker for LSST:UK}",
      journal = {Research Notes of the American Astronomical Society},
         year = 2019,
        month = jan,
       volume = {3},
       number = {1},
          eid = {26},
        pages = {26},
          doi = {10.3847/2515-5172/ab020f},
       adsurl = {https://ui.adsabs.harvard.edu/abs/2019RNAAS...3...26S}
}

@ARTICLE{Forster2021AJ,
       author = {{F{\"o}rster}, F. and {Cabrera-Vives}, G. and {Castillo-Navarrete}, E. and {Est{\'e}vez}, P.~A. and {S{\'a}nchez-S{\'a}ez}, P. and {Arredondo}, J. and {Bauer}, F.~E. and {Carrasco-Davis}, R. and {Catelan}, M. and {Elorrieta}, F. and {Eyheramendy}, S. and {Huijse}, P. and {Pignata}, G. and {Reyes}, E. and {Reyes}, I. and {Rodr{\'\i}guez-Mancini}, D. and {Ruz-Mieres}, D. and {Valenzuela}, C. and {{\'A}lvarez-Maldonado}, I. and {Astorga}, N. and {Borissova}, J. and {Clocchiatti}, A. and {De Cicco}, D. and {Donoso-Oliva}, C. and {Hern{\'a}ndez-Garc{\'\i}a}, L. and {Graham}, M.~J. and {Jord{\'a}n}, A. and {Kurtev}, R. and {Mahabal}, A. and {Maureira}, J.~C. and {Mu{\~n}oz-Arancibia}, A. and {Molina-Ferreiro}, R. and {Moya}, A. and {Palma}, W. and {P{\'e}rez-Carrasco}, M. and {Protopapas}, P. and {Romero}, M. and {Sabatini-Gacitua}, L. and {S{\'a}nchez}, A. and {San Mart{\'\i}n}, J. and {Sep{\'u}lveda-Cobo}, C. and {Vera}, E. and {Vergara}, J.~R.},
        title = "{The Automatic Learning for the Rapid Classification of Events (ALeRCE) Alert Broker}",
      journal = {\aj},
         year = 2021,
        month = may,
       volume = {161},
       number = {5},
          eid = {242},
        pages = {242},
          doi = {10.3847/1538-3881/abe9bc},
archivePrefix = {arXiv},
       eprint = {2008.03303},
 primaryClass = {astro-ph.IM},
       adsurl = {https://ui.adsabs.harvard.edu/abs/2021AJ....161..242F}
}

@ARTICLE{Baal2023MNRAS,
       author = {{van Baal}, Bart F.~A. and {Jerkstrand}, Anders and {Wongwathanarat}, Annop and {Janka}, Hans-Thomas},
        title = "{Modelling supernova nebular lines in 3D with EXTRASS}",
      journal = {\mnras},
         year = 2023,
        month = jul,
       volume = {523},
       number = {1},
        pages = {954-973},
          doi = {10.1093/mnras/stad1488},
archivePrefix = {arXiv},
       eprint = {2305.08933},
 primaryClass = {astro-ph.HE},
       adsurl = {https://ui.adsabs.harvard.edu/abs/2023MNRAS.523..954V}
}

@ARTICLE{Clocchiatti1996ApJ,
       author = {{Clocchiatti}, A. and {Wheeler}, J.~C. and {Benetti}, S. and {Frueh}, M.},
        title = "{SN 1983N and the Nature of Stripped Envelope--Core Collapse Supernovae}",
      journal = {\apj},
         year = 1996,
        month = mar,
       volume = {459},
        pages = {547},
          doi = {10.1086/176919},
       adsurl = {https://ui.adsabs.harvard.edu/abs/1996ApJ...459..547C}
}

@ARTICLE{Roming2009ApJ,
       author = {{Roming}, P.~W.~A. and {Pritchard}, T.~A. and {Brown}, P.~J. and {Holland}, S.~T. and {Immler}, S. and {Stockdale}, C.~J. and {Weiler}, K.~W. and {Panagia}, N. and {Van Dyk}, S.~D. and {Hoversten}, E.~A. and {Milne}, P.~A. and {Oates}, S.~R. and {Russell}, B. and {Vandrevala}, C.},
        title = "{Multi-Wavelength Properties of the Type IIb SN 2008ax}",
      journal = {\apjl},
         year = 2009,
        month = oct,
       volume = {704},
       number = {2},
        pages = {L118-L123},
          doi = {10.1088/0004-637X/704/2/L118},
archivePrefix = {arXiv},
       eprint = {0909.0967},
 primaryClass = {astro-ph.HE},
       adsurl = {https://ui.adsabs.harvard.edu/abs/2009ApJ...704L.118R}
}

@ARTICLE{Taddia2018A&A,
       author = {{Taddia}, F. and {Stritzinger}, M.~D. and {Bersten}, M. and {Baron}, E. and {Burns}, C. and {Contreras}, C. and {Holmbo}, S. and {Hsiao}, E.~Y. and {Morrell}, N. and {Phillips}, M.~M. and {Sollerman}, J. and {Suntzeff}, N.~B.},
        title = "{The Carnegie Supernova Project I. Analysis of stripped-envelope supernova light curves}",
      journal = {\aap},
         year = 2018,
        month = feb,
       volume = {609},
          eid = {A136},
        pages = {A136},
          doi = {10.1051/0004-6361/201730844},
archivePrefix = {arXiv},
       eprint = {1707.07614},
 primaryClass = {astro-ph.HE},
       adsurl = {https://ui.adsabs.harvard.edu/abs/2018A&A...609A.136T}
}

@ARTICLE{Stritzinger2018A&A,
       author = {{Stritzinger}, M.~D. and {Anderson}, J.~P. and {Contreras}, C. and {Heinrich-Josties}, E. and {Morrell}, N. and {Phillips}, M.~M. and {Anais}, J. and {Boldt}, L. and {Busta}, L. and {Burns}, C.~R. and {Campillay}, A. and {Corco}, C. and {Castellon}, S. and {Folatelli}, G. and {Gonz{\'a}lez}, C. and {Holmbo}, S. and {Hsiao}, E.~Y. and {Krzeminski}, W. and {Salgado}, F. and {Ser{\'o}n}, J. and {Torres-Robledo}, S. and {Freedman}, W.~L. and {Hamuy}, M. and {Krisciunas}, K. and {Madore}, B.~F. and {Persson}, S.~E. and {Roth}, M. and {Suntzeff}, N.~B. and {Taddia}, F. and {Li}, W. and {Filippenko}, A.~V.},
        title = "{The Carnegie Supernova Project I. Photometry data release of low-redshift stripped-envelope supernovae}",
      journal = {\aap},
         year = 2018,
        month = feb,
       volume = {609},
          eid = {A134},
        pages = {A134},
          doi = {10.1051/0004-6361/201730842},
archivePrefix = {arXiv},
       eprint = {1707.07616},
 primaryClass = {astro-ph.HE},
       adsurl = {https://ui.adsabs.harvard.edu/abs/2018A&A...609A.134S}
}

@software{Thomas2013ascl,
       author = {{Thomas}, R.~C.},
        title = "{SYN++: Standalone SN spectrum synthesis}",
 howpublished = {Astrophysics Source Code Library, record ascl:1308.008},
         year = 2013,
        month = aug,
          eid = {ascl:1308.008},
archivePrefix = {ascl},
       eprint = {1308.008},
       adsurl = {https://ui.adsabs.harvard.edu/abs/2013ascl.soft08008T}
}

@article{Clocchiatti1997ApJ,
doi = {10.1086/304961},
url = {https://doi.org/10.1086/304961},
year = {1997},
month = {dec},
publisher = {},
volume = {491},
number = {1},
pages = {375},
author = {Clocchiatti, A. and Wheeler, J. C.},
title = {On the Light Curves of Stripped-Envelope Supernovae},
journal = {The Astrophysical Journal}
}

@ARTICLE{GOTO_dev_Steeghs2022MNRAS.511.2405S,
       author = {{Steeghs}, D. and {Galloway}, D.~K. and {Ackley}, K. and {Dyer}, M.~J. and {Lyman}, J. and {Ulaczyk}, K. and {Cutter}, R. and {Mong}, Y.-L. and {Dhillon}, V. and {O'Brien}, P. and {Ramsay}, G. and {Poshyachinda}, S. and {Kotak}, R. and {Nuttall}, L.~K. and {Pall{\'e}}, E. and {Breton}, R.~P. and {Pollacco}, D. and {Thrane}, E. and {Aukkaravittayapun}, S. and {Awiphan}, S. and {Burhanudin}, U. and {Chote}, P. and {Chrimes}, A. and {Daw}, E. and {Duffy}, C. and {Eyles-Ferris}, R. and {Gompertz}, B. and {Heikkil{\"a}}, T. and {Irawati}, P. and {Kennedy}, M.~R. and {Killestein}, T. and {Kuncarayakti}, H. and {Levan}, A.~J. and {Littlefair}, S. and {Makrygianni}, L. and {Marsh}, T. and {Mata-Sanchez}, D. and {Mattila}, S. and {Maund}, J. and {McCormac}, J. and {Mkrtichian}, D. and {Mullaney}, J. and {Noysena}, K. and {Patel}, M. and {Rol}, E. and {Sawangwit}, U. and {Stanway}, E.~R. and {Starling}, R. and {Str{\o}m}, P. and {Tooke}, S. and {West}, R. and {White}, D.~J. and {Wiersema}, K.},
        title = "{The Gravitational-wave Optical Transient Observer (GOTO): prototype performance and prospects for transient science}",
      journal = {\mnras},
         year = 2022,
        month = apr,
       volume = {511},
       number = {2},
        pages = {2405-2422},
          doi = {10.1093/mnras/stac013},
archivePrefix = {arXiv},
       eprint = {2110.05539},
 primaryClass = {astro-ph.IM},
       adsurl = {https://ui.adsabs.harvard.edu/abs/2022MNRAS.511.2405S}
}

@ARTICLE{Andrews2025TNSCR5179....1A,
       author = {{Andrews}, J. and {Hsu}, B. and {Shrestha}, M. and {Sand}, D. and {Bostroem}, A.},
        title = "{PASSTA Transient Classification Report for 2025-12-26}",
      journal = {Transient Name Server Classification Report},
         year = 2025,
        month = dec,
       volume = {2025-5179},
        pages = {1},
       adsurl = {https://ui.adsabs.harvard.edu/abs/2025TNSCR5179....1A}
}

@ARTICLE{Bitsakis2011A&A...533A.142B,
       author = {{Bitsakis}, T. and {Charmandaris}, V. and {da Cunha}, E. and {D{\'\i}az-Santos}, T. and {Le Floc'h}, E. and {Magdis}, G.},
        title = "{A mid-IR study of Hickson compact groups. II. Multiwavelength analysis of the complete GALEX-Spitzer sample}",
      journal = {\aap},
         year = 2011,
        month = sep,
       volume = {533},
          eid = {A142},
        pages = {A142},
          doi = {10.1051/0004-6361/201117355},
archivePrefix = {arXiv},
       eprint = {1107.3418},
 primaryClass = {astro-ph.CO},
       adsurl = {https://ui.adsabs.harvard.edu/abs/2011A&A...533A.142B}
}

@ARTICLE{SN2024aecx_Zou2026ApJ...997...77Z,
       author = {{Zou}, Xingzhu and {Kumar}, Brajesh and {Teja}, Rishabh Singh and {Sahu}, D.~K. and {Chen}, Xinlei and {Singh}, Avinash and {Lin}, Weikang and {Liu}, Xiangkun and {Liu}, Dezi and {Das}, Hrishav and {Singh}, Mridweeka and {Anupama}, G.~C. and {Pan}, Yu and {Du}, Guowang and {Guo}, Helong and {Wang}, Tao and {Zhu}, Xufeng and {Zhang}, Jujia and {Fang}, Yuan and {Liu}, Chenxu and {Chatterjee}, Kaushik and {Yang}, Yuan-Pei and {Li}, Liping and {Zhai}, Qian and {Lagioia}, Edoardo P. and {Du}, Xueling and {Er}, Xinzhong and {Lian}, Jianhui and {Li}, Ziwei and {Zhong}, Shiyan and {Liu}, Xiaowei},
        title = "{SN 2024aecx: A Double-peaked Rapidly Evolving Type IIb Supernova at 11 Mpc}",
      journal = {\apj},
         year = 2026,
        month = jan,
       volume = {997},
       number = {1},
          eid = {77},
        pages = {77},
          doi = {10.3847/1538-4357/ae21dd},
archivePrefix = {arXiv},
       eprint = {2505.19831},
 primaryClass = {astro-ph.HE},
       adsurl = {https://ui.adsabs.harvard.edu/abs/2026ApJ...997...77Z}
}

@ARTICLE{Yang2024ApJ...969..126Y,
       author = {{Yang}, Yuan-Pei and {Liu}, Xiangkun and {Pan}, Yu and {Er}, Xinzhong and {Liu}, Dezi and {Fang}, Yuan and {Du}, Guowang and {Cai}, Yongzhi and {Xu}, Xian and {Chen}, Xinlei and {Zou}, Xingzhu and {Guo}, Helong and {Liu}, Chenxu and {Cheng}, Yehao and {Kumar}, Brajesh and {Liu}, Xiaowei},
        title = "{Multiband Simultaneous Photometry of Type II SN 2023ixf with Mephisto and the Twin 50 cm Telescopes}",
      journal = {\apj},
         year = 2024,
        month = jul,
       volume = {969},
       number = {2},
          eid = {126},
        pages = {126},
          doi = {10.3847/1538-4357/ad4be3},
archivePrefix = {arXiv},
       eprint = {2405.08327},
 primaryClass = {astro-ph.HE},
       adsurl = {https://ui.adsabs.harvard.edu/abs/2024ApJ...969..126Y}
}

@ARTICLE{SN2024iss_Chen2026AA,
       author = {{Chen}, Liyang and {Wang}, Xiaofeng and {Wu}, Qinyu and {Andrews}, Moira and {Farah}, Joseph and {Ochner}, Paolo and {Reguitti}, Andrea and {Filippenko}, Alexei V. and {Brink}, Thomas G. and {Zhang}, Jujia and {Song}, Cuiying and {Liu}, Jialian and {Sand}, David J. and {Albanese}, Irene and {Alexander}, Kate D. and {Andrews}, Jennifer and {Bostroem}, K. Azalee and {Cai}, Y.-Z. and {Christy}, Collin and {Esamdin}, Ali and {Farina}, Andrea and {Franz}, Noah and {Andrew Howell}, D. and {Hsu}, Brian and {Hu}, Maokai and {Iskandar}, Abdusamatjan and {Li}, Liping and {Li}, Gaici and {Li}, Dongyue and {Li}, Wenxiong and {Liu}, Jinzhong and {Mao}, Yiming and {McCully}, Curtis and {Mo}, Jun and {Newsome}, Megan and {Ni}, Yuan Qi and {Pastorello}, Andrea and {Padilla Gonzalez}, Estefania and {Pearson}, Jeniveve and {Peng}, Haowei and {Ransome}, Conor and {Shrestha}, Manisha and {Smith}, Nathan and {Subrayan}, Bhagya and {Terreran}, Giacomo and {Valerin}, Giorgio and {Vink{\'o}}, J. and {Vasylyev}, Sergiy S. and {Wang}, Letian and {Wang}, Zhenyu and {Wang}, Hao and {Craig Wheeler}, J. and {Wynn}, Kathryn and {Xiang}, Danfeng and {Yan}, Shengyu and {Yuan}, Weimin and {Zhang}, Juan and {Zheng}, WeiKang and {Zhang}, Yu},
        title = "{SN 2024iss: A double-peaked Type IIb supernova with evidence of circumstellar interaction}",
      journal = {\aap},
         year = 2026,
        month = may,
       volume = {710},
          eid = {A33},
        pages = {A33},
          doi = {10.1051/0004-6361/202557843},
archivePrefix = {arXiv},
       eprint = {2510.22997},
 primaryClass = {astro-ph.HE},
       adsurl = {https://ui.adsabs.harvard.edu/abs/2026A&A...710A..33C}
}

@ARTICLE{Anderson2014MNRAS.441..671A,
       author = {{Anderson}, J.~P. and {Dessart}, L. and {Gutierrez}, C.~P. and {Hamuy}, M. and {Morrell}, N.~I. and {Phillips}, M. and {Folatelli}, G. and {Stritzinger}, M.~D. and {Freedman}, W.~L. and {Gonz{\'a}lez-Gait{\'a}n}, S. and et al.},
        title = "{Analysis of blueshifted emission peaks in Type II supernovae}",
      journal = {\mnras},
         year = 2014,
        month = jun,
       volume = {441},
       number = {1},
        pages = {671-680},
          doi = {10.1093/mnras/stu610},
archivePrefix = {arXiv},
       eprint = {1404.0581},
 primaryClass = {astro-ph.HE},
       adsurl = {https://ui.adsabs.harvard.edu/abs/2014MNRAS.441..671A}
}

@article{SN2022ngb_Zhao2026A&A,
	author = {{Zhao, J.-W.} and {Benetti, S.} and {Cai, Y.-Z.} and {Pastorello, A.} and {Elias-Rosa, N.} and {Reguitti, A.} and {Valerin, G.} and {Wang, Z.-Y.} and {Cappellaro, E.} and {Feng, G.-F.} and {Fiore, A.} and {Fitzpatrick, B.} and {Fraser, M.} and {Isern, J.} and {Kankare, E.} and {Kravtsov, T.} and {Kumar, B.} and {Lundqvist, P.} and {Matilainen, K.} and {Mattila, S.} and {Mazzali, P. A.} and {Moran, S.} and {Ochner, P.} and {Peng, Z.-H.} and {Reynolds, T. M.} and {Salmaso, I.} and {Srivastav, S.} and {Stritzinger, M. D.} and {Taubenberger, S.} and {Tomasella, L.} and {Vink\'o, J.} and {Wheeler, J. C.} and {Williams, S.} and {Pei, S.-P.} and {Yang, Y.-J.} and {Liu, X.-K.} and {Liu, X.-W.} and {Yang, Y.-P.}},
	title = {SN 2022ngb: A faint, slowly evolving Type IIb supernova with a low-mass envelope},
	DOI= "10.1051/0004-6361/202557619",
	url= "https://doi.org/10.1051/0004-6361/202557619",
	journal = {A\&A},
	year = 2026,
	volume = 706,
	pages = "A271",
}

@ARTICLE{SN2011dh_Arcavi2011ApJ...742L..18A,
       author = {{Arcavi}, Iair and {Gal-Yam}, Avishay and {Yaron}, Ofer and {Sternberg}, Assaf and {Rabinak}, Itay and {Waxman}, Eli and {Kasliwal}, Mansi M. and {Quimby}, Robert M. and {Ofek}, Eran O. and {Horesh}, Assaf and {Kulkarni}, Shrinivas R. and {Filippenko}, Alexei V. and {Silverman}, Jeffrey M. and {Cenko}, S. Bradley and {Li}, Weidong and {Bloom}, Joshua S. and {Sullivan}, Mark and {Nugent}, Peter E. and {Poznanski}, Dovi and {Gorbikov}, Evgeny and {Fulton}, Benjamin J. and {Howell}, D. Andrew and {Bersier}, David and {Riou}, Amedee and {Lamotte-Bailey}, Stephane and {Griga}, Thomas and {Cohen}, Judith G. and {Hachinger}, Stephan and {Polishook}, David and {Xu}, Dong and {Ben-Ami}, Sagi and {Manulis}, Ilan and {Walker}, Emma S. and {Maguire}, Kate and {Pan}, Yen-Chen and {Matheson}, Thomas and {Mazzali}, Paolo A. and {Pian}, Elena and {Fox}, Derek B. and {Gehrels}, Neil and {Law}, Nicholas and {James}, Philip and {Marchant}, Jonathan M. and {Smith}, Robert J. and {Mottram}, Chris J. and {Barnsley}, Robert M. and {Kandrashoff}, Michael T. and {Clubb}, Kelsey I.},
        title = "{SN 2011dh: Discovery of a Type IIb Supernova from a Compact Progenitor in the Nearby Galaxy M51}",
      journal = {\apjl},
         year = 2011,
        month = dec,
       volume = {742},
       number = {2},
          eid = {L18},
        pages = {L18},
          doi = {10.1088/2041-8205/742/2/L18},
archivePrefix = {arXiv},
       eprint = {1106.3551},
 primaryClass = {astro-ph.CO},
       adsurl = {https://ui.adsabs.harvard.edu/abs/2011ApJ...742L..18A}
}

@ARTICLE{SN2011hs_Bufano2014MNRAS.439.1807B,
       author = {{Bufano}, F. and {Pignata}, G. and {Bersten}, M. and {Mazzali}, P.~A. and {Ryder}, S.~D. and {Margutti}, R. and {Milisavljevic}, D. and {Morelli}, L. and {Benetti}, S. and {Cappellaro}, E. and {Gonzalez-Gaitan}, S. and {Romero-Ca{\~n}izales}, C. and {Stritzinger}, M. and {Walker}, E.~S. and {Anderson}, J.~P. and {Contreras}, C. and {de Jaeger}, T. and {F{\"o}rster}, F. and {Gutierrez}, C. and {Hamuy}, M. and {Hsiao}, E. and {Morrell}, N. and {Olivares E.}, F. and {Paillas}, E. and {Parker}, S. and {Pian}, E. and {Pickering}, T.~E. and {Sanders}, N. and {Stockdale}, C. and {Turatto}, M. and {Valenti}, S. and {Fesen}, R.~A. and {Maza}, J. and {Nomoto}, K. and {Phillips}, M.~M. and {Soderberg}, A.},
        title = "{SN 2011hs: a fast and faint Type IIb supernova from a supergiant progenitor}",
      journal = {\mnras},
         year = 2014,
        month = apr,
       volume = {439},
       number = {2},
        pages = {1807-1828},
          doi = {10.1093/mnras/stu065},
archivePrefix = {arXiv},
       eprint = {1401.2368},
 primaryClass = {astro-ph.SR},
       adsurl = {https://ui.adsabs.harvard.edu/abs/2014MNRAS.439.1807B}
}

@ARTICLE{SN2011dh_Ergon2014A&A...562A..17E,
       author = {{Ergon}, M. and {Sollerman}, J. and {Fraser}, M. and {Pastorello}, A. and {Taubenberger}, S. and {Elias-Rosa}, N. and {Bersten}, M. and {Jerkstrand}, A. and {Benetti}, S. and {Botticella}, M.~T. and {Fransson}, C. and {Harutyunyan}, A. and {Kotak}, R. and {Smartt}, S. and {Valenti}, S. and {Bufano}, F. and {Cappellaro}, E. and {Fiaschi}, M. and {Howell}, A. and {Kankare}, E. and {Magill}, L. and {Mattila}, S. and {Maund}, J. and {Naves}, R. and {Ochner}, P. and {Ruiz}, J. and {Smith}, K. and {Tomasella}, L. and {Turatto}, M.},
        title = "{Optical and near-infrared observations of SN 2011dh - The first 100 days}",
      journal = {\aap},
         year = 2014,
        month = feb,
       volume = {562},
          eid = {A17},
        pages = {A17},
          doi = {10.1051/0004-6361/201321850},
archivePrefix = {arXiv},
       eprint = {1305.1851},
 primaryClass = {astro-ph.SR},
       adsurl = {https://ui.adsabs.harvard.edu/abs/2014A&A...562A..17E}
}

@ARTICLE{SN2011ei_Milisavljevic2013ApJ...767...71M,
       author = {{Milisavljevic}, Dan and {Margutti}, Raffaella and {Soderberg}, Alicia M. and {Pignata}, Giuliano and {Chomiuk}, Laura and {Fesen}, Robert A. and {Bufano}, Filomena and {Sanders}, Nathan E. and {Parrent}, Jerod T. and {Parker}, Stuart and {Mazzali}, Paolo and {Pian}, Elena and {Pickering}, Timothy and {Buckley}, David A.~H. and {Crawford}, Steven M. and {Gulbis}, Amanda A.~S. and {Hettlage}, Christian and {Hooper}, Eric and {Nordsieck}, Kenneth H. and {O'Donoghue}, Darragh and {Husser}, Tim-Oliver and {Potter}, Stephen and {Kniazev}, Alexei and {Kotze}, Paul and {Romero-Colmenero}, Encarni and {Vaisanen}, Petri and {Wolf}, Marsha and {Bietenholz}, Michael F. and {Bartel}, Norbert and {Fransson}, Claes and {Walker}, Emma S. and {Brunthaler}, Andreas and {Chakraborti}, Sayan and {Levesque}, Emily M. and {MacFadyen}, Andrew and {Drescher}, Colin and {Bock}, Greg and {Marples}, Peter and {Anderson}, Joseph P. and {Benetti}, Stefano and {Reichart}, Daniel and {Ivarsen}, Kevin},
        title = "{Multi-wavelength Observations of Supernova 2011ei: Time-dependent Classification of Type IIb and Ib Supernovae and Implications for Their Progenitors}",
      journal = {\apj},
         year = 2013,
        month = apr,
       volume = {767},
       number = {1},
          eid = {71},
        pages = {71},
          doi = {10.1088/0004-637X/767/1/71},
archivePrefix = {arXiv},
       eprint = {1207.2152},
 primaryClass = {astro-ph.HE},
       adsurl = {https://ui.adsabs.harvard.edu/abs/2013ApJ...767...71M}
}

@ARTICLE{SW17_Sapir2017ApJ...838..130S,
       author = {{Sapir}, Nir and {Waxman}, Eli},
        title = "{UV/Optical Emission from the Expanding Envelopes of Type II Supernovae}",
      journal = {\apj},
         year = 2017,
        month = apr,
       volume = {838},
       number = {2},
          eid = {130},
        pages = {130},
          doi = {10.3847/1538-4357/aa64df},
archivePrefix = {arXiv},
       eprint = {1607.03700},
 primaryClass = {astro-ph.HE},
       adsurl = {https://ui.adsabs.harvard.edu/abs/2017ApJ...838..130S}
}

@ARTICLE{Hatano1999ApJS..121..233H,
       author = {{Hatano}, Kazuhito and {Branch}, David and {Fisher}, Adam and {Millard}, Jennifer and {Baron}, E.},
        title = "{Ion Signatures in Supernova Spectra}",
      journal = {\apjs},
         year = 1999,
        month = mar,
       volume = {121},
       number = {1},
        pages = {233-246},
          doi = {10.1086/313190},
archivePrefix = {arXiv},
       eprint = {astro-ph/9809236},
 primaryClass = {astro-ph},
       adsurl = {https://ui.adsabs.harvard.edu/abs/1999ApJS..121..233H}
}

@ARTICLE{Arcavi2017ApJ...837L...2A,
       author = {{Arcavi}, Iair and {Hosseinzadeh}, Griffin and {Brown}, Peter J. and {Smartt}, Stephen J. and {Valenti}, Stefano and {Tartaglia}, Leonardo and {Piro}, Anthony L. and {Sanchez}, Jos{\'e} L. and {Nicholls}, Brent and {Monard}, Berto L.~A.~G. and {Howell}, D. Andrew and {McCully}, Curtis and {Sand}, David J. and {Tonry}, John and {Denneau}, Larry and {Stalder}, Brian and {Heinze}, Ari and {Rest}, Armin and {Smith}, Ken W. and {Bishop}, David},
        title = "{Constraints on the Progenitor of SN 2016gkg from Its Shock-cooling Light Curve}",
      journal = {\apjl},
         year = 2017,
        month = mar,
       volume = {837},
       number = {1},
          eid = {L2},
        pages = {L2},
          doi = {10.3847/2041-8213/aa5be1},
archivePrefix = {arXiv},
       eprint = {1611.06451},
 primaryClass = {astro-ph.HE},
       adsurl = {https://ui.adsabs.harvard.edu/abs/2017ApJ...837L...2A}
}

@ARTICLE{P15_Piro2015ApJ...808L..51P,
       author = {{Piro}, Anthony L.},
        title = "{Using Double-peaked Supernova Light Curves to Study Extended Material}",
      journal = {\apjl},
         year = 2015,
        month = aug,
       volume = {808},
       number = {2},
          eid = {L51},
        pages = {L51},
          doi = {10.1088/2041-8205/808/2/L51},
archivePrefix = {arXiv},
       eprint = {1505.07103},
 primaryClass = {astro-ph.HE},
       adsurl = {https://ui.adsabs.harvard.edu/abs/2015ApJ...808L..51P}
}

@ARTICLE{dynesty_Speagle2020MNRAS.493.3132S,
       author = {{Speagle}, Joshua S.},
        title = "{DYNESTY: a dynamic nested sampling package for estimating Bayesian posteriors and evidences}",
      journal = {\mnras},
         year = 2020,
        month = apr,
       volume = {493},
       number = {3},
        pages = {3132-3158},
          doi = {10.1093/mnras/staa278},
archivePrefix = {arXiv},
       eprint = {1904.02180},
 primaryClass = {astro-ph.IM},
       adsurl = {https://ui.adsabs.harvard.edu/abs/2020MNRAS.493.3132S}
}

@ARTICLE{emcee_Foreman-Mackey2013PASP..125..306F,
       author = {{Foreman-Mackey}, Daniel and {Hogg}, David W. and {Lang}, Dustin and {Goodman}, Jonathan},
        title = "{emcee: The MCMC Hammer}",
      journal = {\pasp},
         year = 2013,
        month = mar,
       volume = {125},
       number = {925},
        pages = {306},
          doi = {10.1086/670067},
archivePrefix = {arXiv},
       eprint = {1202.3665},
 primaryClass = {astro-ph.IM},
       adsurl = {https://ui.adsabs.harvard.edu/abs/2013PASP..125..306F}
}

@ARTICLE{Nakar2010ApJ...725..904N,
       author = {{Nakar}, Ehud and {Sari}, Re'em},
        title = "{Early Supernovae Light Curves Following the Shock Breakout}",
      journal = {\apj},
         year = 2010,
        month = dec,
       volume = {725},
       number = {1},
        pages = {904-921},
          doi = {10.1088/0004-637X/725/1/904},
archivePrefix = {arXiv},
       eprint = {1004.2496},
 primaryClass = {astro-ph.HE},
       adsurl = {https://ui.adsabs.harvard.edu/abs/2010ApJ...725..904N}
}

@ARTICLE{Yoon2019ApJ...872..174Y,
       author = {{Yoon}, Sung-Chul and {Chun}, Wonseok and {Tolstov}, Alexey and {Blinnikov}, Sergey and {Dessart}, Luc},
        title = "{Type Ib/Ic Supernovae: Effect of Nickel Mixing on the Early-time Color Evolution and Implications for the Progenitors}",
      journal = {\apj},
         year = 2019,
        month = feb,
       volume = {872},
       number = {2},
          eid = {174},
        pages = {174},
          doi = {10.3847/1538-4357/ab0020},
archivePrefix = {arXiv},
       eprint = {1810.03108},
 primaryClass = {astro-ph.HE},
       adsurl = {https://ui.adsabs.harvard.edu/abs/2019ApJ...872..174Y}
}

@ARTICLE{Moriya2020MNRAS.497.1619M,
       author = {{Moriya}, Takashi J. and {Suzuki}, Akihiro and {Takiwaki}, Tomoya and {Pan}, Yen-Chen and {Blinnikov}, Sergei I.},
        title = "{Systematic investigation of the effect of $^{56}$Ni mixing in the early photospheric velocity evolution of stripped-envelope supernovae}",
      journal = {\mnras},
         year = 2020,
        month = sep,
       volume = {497},
       number = {2},
        pages = {1619-1626},
          doi = {10.1093/mnras/staa2060},
archivePrefix = {arXiv},
       eprint = {2007.04438},
 primaryClass = {astro-ph.HE},
       adsurl = {https://ui.adsabs.harvard.edu/abs/2020MNRAS.497.1619M}
}

@ARTICLE{SN2018gjx_Prentice2020MNRAS.499.1450P,
       author = {{Prentice}, S.~J. and {Maguire}, K. and {Boian}, I. and {Groh}, J. and {Anderson}, J. and {Barbarino}, C. and {Bostroem}, K.~A. and {Burke}, J. and {Clark}, P. and {Dong}, Y. and {Fraser}, M. and {Galbany}, L. and {Gromadzki}, M. and {Guti{\'e}rrez}, C.~P. and {Howell}, D.~A. and {Hiramatsu}, D. and {Inserra}, C. and {James}, P.~A. and {Kankare}, E. and {Kuncarayakti}, H. and {Mazzali}, P.~A. and {McCully}, C. and {M{\"u}ller-Bravo}, T.~E. and {Nichol}, M. and {Pellegrino}, C. and {Smartt}, S.~J. and {Sollerman}, J. and {Tartaglia}, L. and {Valenti}, S. and {Young}, D.~R.},
        title = "{SN 2018gjx reveals that some SNe Ibn are SNe IIb exploding in dense circumstellar material}",
      journal = {\mnras},
         year = 2020,
        month = nov,
       volume = {499},
       number = {1},
        pages = {1450-1467},
          doi = {10.1093/mnras/staa2947},
archivePrefix = {arXiv},
       eprint = {2009.10509},
 primaryClass = {astro-ph.HE},
       adsurl = {https://ui.adsabs.harvard.edu/abs/2020MNRAS.499.1450P}
}

@ARTICLE{SN2007Y_Stritzinger2009ApJ...696..713S,
       author = {{Stritzinger}, Maximilian and {Mazzali}, Paolo and {Phillips}, Mark M. and {Immler}, Stefan and {Soderberg}, Alicia and {Sollerman}, Jesper and {Boldt}, Luis and {Braithwaite}, Jonathan and {Brown}, Peter and {Burns}, Christopher R. and {Contreras}, Carlos and {Covarrubias}, Ricardo and {Folatelli}, Gast{\'o}n and {Freedman}, Wendy L. and {Gonz{\'a}lez}, Sergio and {Hamuy}, Mario and {Krzeminski}, Wojtek and {Madore}, Barry F. and {Milne}, Peter and {Morrell}, Nidia and {Persson}, S.~E. and {Roth}, Miguel and {Smith}, Mathew and {Suntzeff}, Nicholas B.},
        title = "{The He-Rich Core-Collapse Supernova 2007Y: Observations from X-Ray to Radio Wavelengths}",
      journal = {\apj},
         year = 2009,
        month = may,
       volume = {696},
       number = {1},
        pages = {713-728},
          doi = {10.1088/0004-637X/696/1/713},
archivePrefix = {arXiv},
       eprint = {0902.0609},
 primaryClass = {astro-ph.HE},
       adsurl = {https://ui.adsabs.harvard.edu/abs/2009ApJ...696..713S}
}

@ARTICLE{Dessart2015MNRAS.453.2189D,
       author = {{Dessart}, Luc and {Hillier}, D. John and {Woosley}, Stan and {Livne}, Eli and {Waldman}, Roni and {Yoon}, Sung-Chul and {Langer}, Norbert},
        title = "{Radiative-transfer models for supernovae IIb/Ib/Ic from binary-star progenitors}",
      journal = {\mnras},
         year = 2015,
        month = oct,
       volume = {453},
       number = {2},
        pages = {2189-2213},
          doi = {10.1093/mnras/stv1747},
archivePrefix = {arXiv},
       eprint = {1507.07783},
 primaryClass = {astro-ph.SR},
       adsurl = {https://ui.adsabs.harvard.edu/abs/2015MNRAS.453.2189D}
}

@ARTICLE{Maund2004Natur.427..129M,
       author = {{Maund}, Justyn R. and {Smartt}, Stephen J. and {Kudritzki}, Rolf P. and {Podsiadlowski}, Philipp and {Gilmore}, Gerard F.},
        title = "{The massive binary companion star to the progenitor of supernova 1993J}",
      journal = {\nat},
         year = 2004,
        month = jan,
       volume = {427},
       number = {6970},
        pages = {129-131},
          doi = {10.1038/nature02161},
archivePrefix = {arXiv},
       eprint = {astro-ph/0401090},
 primaryClass = {astro-ph},
       adsurl = {https://ui.adsabs.harvard.edu/abs/2004Natur.427..129M}
}

@ARTICLE{Brown2009AJ....137.4517B,
       author = {{Brown}, Peter J. and {Holland}, Stephen T. and {Immler}, Stefan and {Milne}, Peter and {Roming}, Peter W.~A. and {Gehrels}, Neil and {Nousek}, John and {Panagia}, Nino and {Still}, Martin and {Vanden Berk}, Daniel},
        title = "{Ultraviolet Light Curves of Supernovae with the Swift Ultraviolet/Optical Telescope}",
      journal = {\aj},
         year = 2009,
        month = may,
       volume = {137},
       number = {5},
        pages = {4517-4525},
          doi = {10.1088/0004-6256/137/5/4517},
archivePrefix = {arXiv},
       eprint = {0803.1265},
 primaryClass = {astro-ph},
       adsurl = {https://ui.adsabs.harvard.edu/abs/2009AJ....137.4517B}
}

@ARTICLE{Brown2014Ap&SS.354...89B,
       author = {{Brown}, Peter J. and {Breeveld}, Alice A. and {Holland}, Stephen and {Kuin}, Paul and {Pritchard}, Tyler},
        title = "{SOUSA: the Swift Optical/Ultraviolet Supernova Archive}",
      journal = {\apss},
         year = 2014,
        month = nov,
       volume = {354},
       number = {1},
        pages = {89-96},
          doi = {10.1007/s10509-014-2059-8},
archivePrefix = {arXiv},
       eprint = {1407.3808},
 primaryClass = {astro-ph.HE},
       adsurl = {https://ui.adsabs.harvard.edu/abs/2014Ap&SS.354...89B}
}

@ARTICLE{Rehemtulla2026TNSCR.196....1R,
       author = {{Rehemtulla}, N.},
        title = "{ZTF Transient Classification Report for 2026-01-16}",
      journal = {Transient Name Server Classification Report},
         year = 2026,
        month = jan,
       volume = {2026-196},
        pages = {1},
       adsurl = {https://ui.adsabs.harvard.edu/abs/2026TNSCR.196....1R}
}

@ARTICLE{Cai2022A&A...667A...4C,
       author = {{Cai}, Y.-Z. and {Pastorello}, A. and {Fraser}, M. and {Wang}, X.-F. and {Filippenko}, A.~V. and {Reguitti}, A. and {Patra}, K.~C. and {Goranskij}, V.~P. and {Barsukova}, E.~A. and {Brink}, T.~G. and {Elias-Rosa}, N. and {Stevance}, H.~F. and {Zheng}, W. and {Yang}, Y. and {Atapin}, K.~E. and {Benetti}, S. and {de Boer}, T.~J.~L. and {Bose}, S. and {Burke}, J. and {Byrne}, R. and {Cappellaro}, E. and {Chambers}, K.~C. and {Chen}, W.-L. and {Emami}, N. and {Gao}, H. and {Hiramatsu}, D. and {Howell}, D.~A. and {Huber}, M.~E. and {Kankare}, E. and {Kelly}, P.~L. and {Kotak}, R. and {Kravtsov}, T. and {Lander}, V. Yu. and {Li}, Z.-T. and {Lin}, C.-C. and {Lundqvist}, P. and {Magnier}, E.~A. and {Malygin}, E.~A. and {Maslennikova}, N.~A. and {Matilainen}, K. and {Mazzali}, P.~A. and {McCully}, C. and {Mo}, J. and {Moran}, S. and {Newsome}, M. and {Oparin}, D.~V. and {Padilla Gonzalez}, E. and {Reynolds}, T.~M. and {Shatsky}, N.~I. and {Smartt}, S.~J. and {Smith}, K.~W. and {Stritzinger}, M.~D. and {Tatarnikov}, A.~M. and {Terreran}, G. and {Uklein}, R.~I. and {Valerin}, G. and {Vallely}, P.~J. and {Vozyakova}, O.~V. and {Wainscoat}, R. and {Yan}, S.-Y. and {Zhang}, J.-J. and {Zhang}, T.-M. and {Zheltoukhov}, S.~G. and {Dastidar}, R. and {Fulton}, M. and {Galbany}, L. and {Gangopadhyay}, A. and {Ge}, H.-W. and {Guti{\'e}rrez}, C.~P. and {Lin}, H. and {Misra}, K. and {Ou}, Z.-W. and {Salmaso}, I. and {Tartaglia}, L. and {Xiao}, L. and {Zhang}, X.-H.},
        title = "{Forbidden hugs in pandemic times. III. Observations of the luminous red nova AT 2021biy in the nearby galaxy NGC 4631}",
      journal = {\aap},
         year = 2022,
        month = nov,
       volume = {667},
          eid = {A4},
        pages = {A4},
          doi = {10.1051/0004-6361/202244393},
archivePrefix = {arXiv},
       eprint = {2207.00734},
 primaryClass = {astro-ph.SR},
       adsurl = {https://ui.adsabs.harvard.edu/abs/2022A&A...667A...4C}
}

@ARTICLE{Dessart2024A&A...685A.169D,
       author = {{Dessart}, Luc and {Guti{\'e}rrez}, Claudia P. and {Ercolino}, Andrea and {Jin}, Harim and {Langer}, Norbert},
        title = "{A sequence of Type Ib, IIb, II-L, and II-P supernovae from binary-star progenitors with varying initial separations}",
      journal = {\aap},
         year = 2024,
        month = may,
       volume = {685},
          eid = {A169},
        pages = {A169},
          doi = {10.1051/0004-6361/202349066},
archivePrefix = {arXiv},
       eprint = {2402.12977},
 primaryClass = {astro-ph.SR},
       adsurl = {https://ui.adsabs.harvard.edu/abs/2024A&A...685A.169D}
}

@ARTICLE{SN2017iro_Kumar2022ApJ...927...61K,
       author = {{Kumar}, Brajesh and {Singh}, Avinash and {Sahu}, D.~K. and {Anupama}, G.~C.},
        title = "{Investigating the Observational Properties of Type Ib Supernova SN 2017iro}",
      journal = {\apj},
         year = 2022,
        month = mar,
       volume = {927},
       number = {1},
          eid = {61},
        pages = {61},
          doi = {10.3847/1538-4357/ac4bb9},
archivePrefix = {arXiv},
       eprint = {2201.03260},
 primaryClass = {astro-ph.HE},
       adsurl = {https://ui.adsabs.harvard.edu/abs/2022ApJ...927...61K}
}

@ARTICLE{Brennan2022A&A...667A..62B,
       author = {{Brennan}, S.~J. and {Fraser}, M.},
       title = "{The AUTOmated Photometry Of Transients pipeline (AutoPhOT)}",
      journal = {\aap},
         year = 2022,
        month = nov,
       volume = {667},
          eid = {A62},
        pages = {A62},
          doi = {10.1051/0004-6361/202243067},
archivePrefix = {arXiv},
       eprint = {2201.02635},
 primaryClass = {astro-ph.IM},
       adsurl = {https://ui.adsabs.harvard.edu/abs/2022A&A...667A..62B}
}

@software{Photoutils_Bradley202514889440,
  author       = {Larry Bradley and
                  Brigitta Sip{\H o}cz and
                  Thomas Robitaille and
                  Erik Tollerud and
                  Z\`e Vin{\'{\i}}cius and
                  Christoph Deil and
                  Kyle Barbary and
                  Tom J Wilson and
                  Ivo Busko and
                  Axel Donath and
                  Hans Moritz G{\"u}nther and
                  Mihai Cara and
                  P. L. Lim and
                  Sebastian Me{\ss}linger and
                  Zach Burnett and
                  Simon Conseil and
                  Michael Droettboom and
                  Azalee Bostroem and
                  E. M. Bray and
                  Lars Andersen Bratholm and
                  William Jamieson and
                  Adam Ginsburg and
                  Geert Barentsen and
                  Matt Craig and
                  Sergio Pascual and
                  Shivangee Rathi and
                  Marshall Perrin and
                  Brett M. Morris},
  title        = {astropy/photutils: 2.2.0},
  month        = feb,
  year         = 2025,
  publisher    = {Zenodo},
  version      = {2.2.0},
  doi          = {10.5281/zenodo.14889440},
  url          = {https://doi.org/10.5281/zenodo.14889440},
  swhid        = {swh:1:dir:11159107f27a28985192ed1118b1f2055709d093
                   ;origin=https://doi.org/10.5281/zenodo.596036;visi
                   t=swh:1:snp:ae8c4a55d349d43e53cfe9ce92e678fcfe840f
                   3b;anchor=swh:1:rel:0117f67e8888adcdfc85308287dd9c
                   854b466389;path=astropy-photutils-ffb96c5
                  },
}

@ARTICLE{Lang2010AJ....139.1782L,
       author = {{Lang}, Dustin and {Hogg}, David W. and {Mierle}, Keir and {Blanton}, Michael and {Roweis}, Sam},
        title = "{Astrometry.net: Blind Astrometric Calibration of Arbitrary Astronomical Images}",
      journal = {\aj},
         year = 2010,
        month = may,
       volume = {139},
       number = {5},
        pages = {1782-1800},
          doi = {10.1088/0004-6256/139/5/1782},
archivePrefix = {arXiv},
       eprint = {0910.2233},
 primaryClass = {astro-ph.IM},
       adsurl = {https://ui.adsabs.harvard.edu/abs/2010AJ....139.1782L}
}

@software{Bertin2010ascl.soft10063B,
       author = {{Bertin}, Emmanuel},
        title = "{SCAMP: Automatic Astrometric and Photometric Calibration}",
 howpublished = {Astrophysics Source Code Library, record ascl:1010.063},
         year = 2010,
        month = oct,
          eid = {ascl:1010.063},
archivePrefix = {ascl},
       eprint = {1010.063},
       adsurl = {https://ui.adsabs.harvard.edu/abs/2010ascl.soft10063B}
}

@software{GAIAXPY_Daniela202517943031,
  author       = {Daniela Ruz-Mieres and
                  zuzannakr and
                  Francesca De Angeli},
  title        = {gaia-dpci/GaiaXPy: GaiaXPy v2.1.4},
  month        = dec,
  year         = 2025,
  publisher    = {Zenodo},
  version      = {2.1.4},
  doi          = {10.5281/zenodo.17943031},
  url          = {https://doi.org/10.5281/zenodo.17943031},
}

@ARTICLE{Li2025A&A...704A.233L,
       author = {{Li}, L.-H. and {Benetti}, S. and {Cai}, Y.-Z. and {Wang}, B. and {Pastorello}, A. and {Elias-Rosa}, N. and {Reguitti}, A. and {Borsato}, L. and {Cappellaro}, E. and {Fiore}, A. and {Fraser}, M. and {Gromadzki}, M. and {Harmanen}, J. and {Isern}, J. and {Kangas}, T. and {Kankare}, E. and {Lundqvist}, P. and {Mattila}, S. and {Ochner}, P. and {Peng}, Z.-H. and {Reynolds}, T.~M. and {Salmaso}, I. and {Srivastav}, S. and {Stritzinger}, M.~D. and {Tomasella}, L. and {Valerin}, G. and {Wang}, Z.-Y. and {Zhang}, J.-J. and {Wu}, C.-Y.},
        title = "{SN 2017ckj: A linearly declining type IIb supernova with a relatively massive hydrogen envelope}",
      journal = {\aap},
         year = 2025,
        month = dec,
       volume = {704},
          eid = {A233},
        pages = {A233},
          doi = {10.1051/0004-6361/202556873},
archivePrefix = {arXiv},
       eprint = {2510.22989},
 primaryClass = {astro-ph.SR},
       adsurl = {https://ui.adsabs.harvard.edu/abs/2025A&A...704A.233L}
}

@ARTICLE{ONeill2025TNSTR5147....1O,
       author = {{O'Neill}, D. and {Ackley}, K. and {Dyer}, M. and {Lyman}, J. and {Ulaczyk}, K. and {Steeghs}, D. and {Galloway}, D. and {Dhillon}, V. and {O'Brien}, P. and {Ramsay}, G. and {Noysena}, K. and {Kotak}, R. and {Breton}, R. and {Casares}, J. and {Nuttall}, L. and {Starling}, R. and {Gompertz}, B. and {Godson}, B. and {Killestein}, T. and {Kumar}, A. and {Pursiainen}, M.},
        title = "{GOTO Transient Discovery Report for 2025-12-24}",
      journal = {Transient Name Server Discovery Report},
         year = 2025,
        month = dec,
       volume = {2025-5147},
        pages = {1},
       adsurl = {https://ui.adsabs.harvard.edu/abs/2025TNSTR5147....1O}
}

@ARTICLE{Ensman1988ApJ...333..754E,
       author = {{Ensman}, Lisa M. and {Woosley}, S.~E.},
        title = "{Explosions in Wolf-Rayet Stars and Type Ib Supernovae. I. Light Curves}",
      journal = {\apj},
         year = 1988,
        month = oct,
       volume = {333},
        pages = {754},
          doi = {10.1086/166785},
       adsurl = {https://ui.adsabs.harvard.edu/abs/1988ApJ...333..754E}
}

@INPROCEEDINGS{Woosley1997ASIC..486..821W,
       author = {{Woosley}, S.~E. and {Eastman}, R.~G.},
        title = "{Type Ib and Ic supernovae: models and spectra}",
    booktitle = {Thermonuclear Supernovae},
         year = 1997,
       editor = {{Ruiz-Lapuente}, P. and {Canal}, R. and {Isern}, J.},
       series = {NATO Advanced Study Institute (ASI) Series C},
       volume = {486},
        month = jan,
        pages = {821},
          doi = {10.1007/978-94-011-5710-0_51},
       adsurl = {https://ui.adsabs.harvard.edu/abs/1997ASIC..486..821W}
}

@ARTICLE{Maund2009ApJ...705.1139M,
       author = {{Maund}, Justyn R. and {Wheeler}, J. Craig and {Baade}, Dietrich and {Patat}, Ferdinando and {H{\"o}flich}, Peter and {Wang}, Lifan and {Clocchiatti}, Alejandro},
        title = "{The Early Asymmetries of Supernova 2008D/XRF 080109}",
      journal = {\apj},
         year = 2009,
        month = nov,
       volume = {705},
       number = {2},
        pages = {1139-1151},
          doi = {10.1088/0004-637X/705/2/1139},
archivePrefix = {arXiv},
       eprint = {0908.2841},
 primaryClass = {astro-ph.SR},
       adsurl = {https://ui.adsabs.harvard.edu/abs/2009ApJ...705.1139M}
}

@ARTICLE{Dessart2012MNRAS.424.2139D,
       author = {{Dessart}, Luc and {Hillier}, D. John and {Li}, Chengdong and {Woosley}, Stan},
        title = "{On the nature of supernovae Ib and Ic}",
      journal = {\mnras},
         year = 2012,
        month = aug,
       volume = {424},
       number = {3},
        pages = {2139-2159},
          doi = {10.1111/j.1365-2966.2012.21374.x},
archivePrefix = {arXiv},
       eprint = {1205.5349},
 primaryClass = {astro-ph.SR},
       adsurl = {https://ui.adsabs.harvard.edu/abs/2012MNRAS.424.2139D}
}

@ARTICLE{Baron2000ApJ...545..444B,
       author = {{Baron}, E. and {Branch}, David and {Hauschildt}, Peter H. and {Filippenko}, Alexei V. and {Kirshner}, R.~P. and {Challis}, P.~M. and {Jha}, S. and {Chevalier}, R. and {Fransson}, Claes and {Lundqvist}, Peter and {Garnavich}, Peter and {Leibundgut}, Bruno and {McCray}, R. and {Michael}, E. and {Panagia}, Nino and {Phillips}, M.~M. and {Pun}, C.~S.~J. and {Schmidt}, Brian and {Sonneborn}, George and {Suntzeff}, N.~B. and {Wang}, L. and {Wheeler}, J.~C.},
        title = "{Preliminary Spectral Analysis of the Type II Supernova 1999EM}",
      journal = {\apj},
         year = 2000,
        month = dec,
       volume = {545},
       number = {1},
        pages = {444-448},
          doi = {10.1086/317795},
archivePrefix = {arXiv},
       eprint = {astro-ph/0010614},
 primaryClass = {astro-ph},
       adsurl = {https://ui.adsabs.harvard.edu/abs/2000ApJ...545..444B}
}

@ARTICLE{Silva2026arXiv260723690S,
       author = {{Silva-Farf{\'a}n}, Javier and {Maeda}, Keiichi and {Pastorello}, Andrea and {Kuncarayakti}, Hanindyo and {Nagao}, Takashi and {Valerin}, Giorgio and {Reguitti}, Andrea and {Pierluigi Fugazza}, Dino and {Pignata}, Giuliano and {Sollerman}, Jesper and {Ashall}, Chris and {M{\"u}ller-Bravo}, Tom{\'a}s E. and {Anderson}, Joseph P. and {Guti{\'e}rrez}, Claudia P. and {Pessi}, Priscila J. and {Kankare}, Erkki and {Pyykkinen}, Niko and {Gromadzki}, Mariusz and {Chen}, Ting-Wan and {Reynolds}, Thomas M. and {Salmaso}, Irene and {Gal-Yam}, Avishay},
        title = "{Dust Without Na I D Trace: The Case of Highly Attenuated Type Ib SN 2024vjc}",
      journal = {arXiv e-prints},
         year = 2026,
        month = jul,
          eid = {arXiv:2607.23690},
        pages = {arXiv:2607.23690},
archivePrefix = {arXiv},
       eprint = {2607.23690},
 primaryClass = {astro-ph.HE},
       adsurl = {https://ui.adsabs.harvard.edu/abs/2026arXiv260723690S}
}

@article{Pilyugin2005ApJ,
doi = {10.1086/432408},
url = {https://doi.org/10.1086/432408},
year = {2005},
month = {sep},
publisher = {},
volume = {631},
number = {1},
pages = {231},
author = {Pilyugin, Leonid S. and Thuan, Trinh X.},
title = {Oxygen Abundance Determination in H II Regions: The Strong Line Intensities-Abundance Calibration Revisited},
journal = {The Astrophysical Journal}
}

@ARTICLE{Pilyugin2003A&A...399.1003P,
       author = {{Pilyugin}, L.~S.},
        title = "{Abundance determinations in H II regions. Model fitting versus T$_{e}$-method}",
      journal = {\aap},
         year = 2003,
        month = mar,
       volume = {399},
        pages = {1003-1007},
          doi = {10.1051/0004-6361:20021669},
archivePrefix = {arXiv},
       eprint = {astro-ph/0211319},
 primaryClass = {astro-ph},
       adsurl = {https://ui.adsabs.harvard.edu/abs/2003A&A...399.1003P}
}

@ARTICLE{Marino2013A&A...559A.114M,
       author = {{Marino}, R.~A. and {Rosales-Ortega}, F.~F. and {S{\'a}nchez}, S.~F. and {Gil de Paz}, A. and {V{\'\i}lchez}, J. and {Miralles-Caballero}, D. and {Kehrig}, C. and {P{\'e}rez-Montero}, E. and {Stanishev}, V. and {Iglesias-P{\'a}ramo}, J. and {D{\'\i}az}, A.~I. and {Castillo-Morales}, A. and {Kennicutt}, R. and {L{\'o}pez-S{\'a}nchez}, A.~R. and {Galbany}, L. and {Garc{\'\i}a-Benito}, R. and {Mast}, D. and {Mendez-Abreu}, J. and {Monreal-Ibero}, A. and {Husemann}, B. and {Walcher}, C.~J. and {Garc{\'\i}a-Lorenzo}, B. and {Masegosa}, J. and {Del Olmo Orozco}, A. and {Mour{\~a}o}, A.~M. and {Ziegler}, B. and {Moll{\'a}}, M. and {Papaderos}, P. and {S{\'a}nchez-Bl{\'a}zquez}, P. and {Gonz{\'a}lez Delgado}, R.~M. and {Falc{\'o}n-Barroso}, J. and {Roth}, M.~M. and {van de Ven}, G. and {CALIFA Team}},
        title = "{The O3N2 and N2 abundance indicators revisited: improved calibrations based on CALIFA and T$_{e}$-based literature data}",
      journal = {\aap},
         year = 2013,
        month = nov,
       volume = {559},
          eid = {A114},
        pages = {A114},
          doi = {10.1051/0004-6361/201321956},
archivePrefix = {arXiv},
       eprint = {1307.5316},
 primaryClass = {astro-ph.CO},
       adsurl = {https://ui.adsabs.harvard.edu/abs/2013A&A...559A.114M}
}

@ARTICLE{Pettini2004MNRAS.348L..59P,
       author = {{Pettini}, Max and {Pagel}, Bernard E.~J.},
        title = "{[OIII]/[NII] as an abundance indicator at high redshift}",
      journal = {\mnras},
         year = 2004,
        month = mar,
       volume = {348},
       number = {3},
        pages = {L59-L63},
          doi = {10.1111/j.1365-2966.2004.07591.x},
archivePrefix = {arXiv},
       eprint = {astro-ph/0401128},
 primaryClass = {astro-ph},
       adsurl = {https://ui.adsabs.harvard.edu/abs/2004MNRAS.348L..59P}
}

@ARTICLE{Yuan_EP2025SCPMA..6839501Y,
       author = {{Yuan}, Weimin and {Dai}, Lixin and {Feng}, Hua and {Jin}, Chichuan and {Jonker}, Peter and {Kuulkers}, Erik and {Liu}, Yuan and {Nandra}, Kirpal and {O'Brien}, Paul and {Piro}, Luigi and {Rau}, Arne and {Rea}, Nanda and {Sanders}, Jeremy and {Tao}, Lian and {Wang}, Junfeng and {Wu}, Xuefeng and {Zhang}, Bing and {Zhang}, Shuangnan and {Ai}, Shunke and {Buchner}, Johannes and {Bulbul}, Esra and {Chen}, Hechao and {Chen}, Minghua and {Chen}, Yong and {Chen}, Yu-Peng and {Coleiro}, Alexis and {Coti Zelati}, Francesco and {Dai}, Zigao and {Fan}, Xilong and {Fan}, Zhou and {Friedrich}, Susanne and {Gao}, He and {Ge}, Chong and {Ge}, Mingyu and {Geng}, Jinjun and {Ghirlanda}, Giancarlo and {Gianfagna}, Giulia and {Gou}, Lijun and {Guillot}, S{\'e}bastien and {Hou}, Xian and {Hu}, Jingwei and {Huang}, Yongfeng and {Ji}, Long and {Jia}, Shumei and {Komossa}, S. and {Kong}, Albert K.~H. and {Lan}, Lin and {Li}, An and {Li}, Ang and {Li}, Chengkui and {Li}, Dongyue and {Li}, Jian and {Li}, Zhaosheng and {Ling}, Zhixing and {Liu}, Ang and {Liu}, Jinzhong and {Liu}, Liangduan and {Liu}, Zhu and {Luo}, Jiawei and {Ma}, Ruican and {Maggi}, Pierre and {Maitra}, Chandreyee and {Marino}, Alessio and {Ng}, Stephen Chi-Yung and {Pan}, Haiwu and {Rukdee}, Surangkhana and {Soria}, Roberto and {Sun}, Hui and {Tam}, Pak-Hin Thomas and {Thakur}, Aishwarya Linesh and {Tian}, Hui and {Troja}, Eleonora and {Wang}, Wei and {Wang}, Xiangyu and {Wang}, Yanan and {Wei}, Junjie and {Wen}, Sixiang and {Wu}, Jianfeng and {Wu}, Ting and {Xiao}, Di and {Xu}, Dong and {Xu}, Renxin and {Xu}, Yanjun and {Xu}, Yu and {Yang}, Haonan and {You}, Bei and {Yu}, Heng and {Yu}, Yunwei and {Zhang}, Binbin and {Zhang}, Chen and {Zhang}, Guobao and {Zhang}, Liang and {Zhang}, Wenda and {Zhang}, Yu and {Zhou}, Ping and {Zou}, Zecheng},
        title = "{Science objectives of the Einstein Probe mission}",
      journal = {Science China Physics, Mechanics, and Astronomy},
         year = 2025,
        month = mar,
       volume = {68},
       number = {3},
          eid = {239501},
        pages = {239501},
          doi = {10.1007/s11433-024-2600-3},
archivePrefix = {arXiv},
       eprint = {2501.07362},
 primaryClass = {astro-ph.HE},
       adsurl = {https://ui.adsabs.harvard.edu/abs/2025SCPMA..6839501Y}
}

@ARTICLE{Asplund2021A&A...653A.141A,
       author = {{Asplund}, M. and {Amarsi}, A.~M. and {Grevesse}, N.},
        title = "{The chemical make-up of the Sun: A 2020 vision}",
      journal = {\aap},
         year = 2021,
        month = sep,
       volume = {653},
          eid = {A141},
        pages = {A141},
          doi = {10.1051/0004-6361/202140445},
archivePrefix = {arXiv},
       eprint = {2105.01661},
 primaryClass = {astro-ph.SR},
       adsurl = {https://ui.adsabs.harvard.edu/abs/2021A&A...653A.141A}
}

@ARTICLE{Andrievsky2001A&A...367..605A,
       author = {{Andrievsky}, S.~M. and {Kovtyukh}, V.~V. and {Korotin}, S.~A. and {Spite}, M. and {Spite}, F.},
        title = "{Magellanic Clouds elemental abundances from F supergiants: Revisited results for the Large Magellanic Cloud}",
      journal = {\aap},
         year = 2001,
        month = feb,
       volume = {367},
        pages = {605-612},
          doi = {10.1051/0004-6361:20000407},
       adsurl = {https://ui.adsabs.harvard.edu/abs/2001A&A...367..605A}
}

@ARTICLE{Ahumada2020ApJS..249....3A,
       author = {{Ahumada}, Romina and {Allende Prieto}, Carlos and {Almeida}, Andr{\'e}s and {Anders}, Friedrich and {Anderson}, Scott F. and {Andrews}, Brett H. and {Anguiano}, Borja and {Arcodia}, Riccardo and {Armengaud}, Eric and {Aubert}, Marie and {Avila}, Santiago and {Avila-Reese}, Vladimir and {Badenes}, Carles and {Balland}, Christophe and {Barger}, Kat and {Barrera-Ballesteros}, Jorge K. and {Basu}, Sarbani and {Bautista}, Julian and {Beaton}, Rachael L. and {Beers}, Timothy C. and {Benavides}, B. Izamar T. and {Bender}, Chad F. and {Bernardi}, Mariangela and {Bershady}, Matthew and {Beutler}, Florian and {Bidin}, Christian Moni and {Bird}, Jonathan and {Bizyaev}, Dmitry and {Blanc}, Guillermo A. and {Blanton}, Michael R. and {Boquien}, M{\'e}d{\'e}ric and {Borissova}, Jura and {Bovy}, Jo and {Brandt}, W.~N. and {Brinkmann}, Jonathan and {Brownstein}, Joel R. and {Bundy}, Kevin and {Bureau}, Martin and {Burgasser}, Adam and {Burtin}, Etienne and {Cano-D{\'\i}az}, Mariana and {Capasso}, Raffaella and {Cappellari}, Michele and {Carrera}, Ricardo and {Chabanier}, Sol{\`e}ne and {Chaplin}, William and {Chapman}, Michael and {Cherinka}, Brian and {Chiappini}, Cristina and {Doohyun Choi}, Peter and {Chojnowski}, S. Drew and {Chung}, Haeun and {Clerc}, Nicolas and {Coffey}, Damien and {Comerford}, Julia M. and {Comparat}, Johan and {da Costa}, Luiz and {Cousinou}, Marie-Claude and {Covey}, Kevin and {Crane}, Jeffrey D. and {Cunha}, Katia and {Ilha}, Gabriele da Silva and {Dai}, Yu Sophia and {Damsted}, Sanna B. and {Darling}, Jeremy and {Davidson}, Jr., James W. and {Davies}, Roger and {Dawson}, Kyle and {De}, Nikhil and {de la Macorra}, Axel and {De Lee}, Nathan and {Queiroz}, Anna B{\'a}rbara de Andrade and {Deconto Machado}, Alice and {de la Torre}, Sylvain and {Dell'Agli}, Flavia and {du Mas des Bourboux}, H{\'e}lion and {Diamond-Stanic}, Aleksandar M. and {Dillon}, Sean and {Donor}, John and {Drory}, Niv and {Duckworth}, Chris and {Dwelly}, Tom and {Ebelke}, Garrett and {Eftekharzadeh}, Sarah and {Davis Eigenbrot}, Arthur and {Elsworth}, Yvonne P. and {Eracleous}, Mike and {Erfanianfar}, Ghazaleh and {Escoffier}, Stephanie and {Fan}, Xiaohui and {Farr}, Emily and {Fern{\'a}ndez-Trincado}, Jos{\'e} G. and {Feuillet}, Diane and {Finoguenov}, Alexis and {Fofie}, Patricia and {Fraser-McKelvie}, Amelia and {Frinchaboy}, Peter M. and {Fromenteau}, Sebastien and {Fu}, Hai and {Galbany}, Llu{\'\i}s and {Garcia}, Rafael A. and {Garc{\'\i}a-Hern{\'a}ndez}, D.~A. and {Garma Oehmichen}, Luis Alberto and {Ge}, Junqiang and {Geimba Maia}, Marcio Antonio and {Geisler}, Doug and {Gelfand}, Joseph and {Goddy}, Julian and {Gonzalez-Perez}, Violeta and {Grabowski}, Kathleen and {Green}, Paul and {Grier}, Catherine J. and {Guo}, Hong and {Guy}, Julien and {Harding}, Paul and {Hasselquist}, Sten and {Hawken}, Adam James and {Hayes}, Christian R. and {Hearty}, Fred and {Hekker}, S. and {Hogg}, David W. and {Holtzman}, Jon A. and {Horta}, Danny and {Hou}, Jiamin and {Hsieh}, Bau-Ching and {Huber}, Daniel and {Hunt}, Jason A.~S. and {Ider Chitham}, J. and {Imig}, Julie and {Jaber}, Mariana and {Jimenez Angel}, Camilo Eduardo and {Johnson}, Jennifer A. and {Jones}, Amy M. and {J{\"o}nsson}, Henrik and {Jullo}, Eric and {Kim}, Yerim and {Kinemuchi}, Karen and {Kirkpatrick}, IV, Charles C. and {Kite}, George W. and {Klaene}, Mark and {Kneib}, Jean-Paul and {Kollmeier}, Juna A. and {Kong}, Hui and {Kounkel}, Marina and {Krishnarao}, Dhanesh and {Lacerna}, Ivan and {Lan}, Ting-Wen and {Lane}, Richard R. and {Law}, David R. and {Le Goff}, Jean-Marc and {Leung}, Henry W. and {Lewis}, Hannah and {Li}, Cheng and {Lian}, Jianhui and {Lin}, Lihwai and {Long}, Dan and {Longa-Pe{\~n}a}, Pen{\'e}lope and {Lundgren}, Britt and {Lyke}, Brad W. and {Mackereth}, J. Ted and {MacLeod}, Chelsea L. and {Majewski}, Steven R. and {Manchado}, Arturo and {Maraston}, Claudia and {Martini}, Paul and {Masseron}, Thomas and {Masters}, Karen L. and {Mathur}, Savita and {McDermid}, Richard M. and {Merloni}, Andrea and {Merrifield}, Michael and {M{\'e}sz{\'a}ros}, Szabolcs and {Miglio}, Andrea and {Minniti}, Dante and {Minsley}, Rebecca and {Miyaji}, Takamitsu and {Mohammad}, Faizan Gohar and {Mosser}, Benoit and {Mueller}, Eva-Maria and {Muna}, Demitri and {Mu{\~n}oz-Guti{\'e}rrez}, Andrea and {Myers}, Adam D. and {Nadathur}, Seshadri and {Nair}, Preethi and {Nandra}, Kirpal and {Correa do Nascimento}, Janaina and {Nevin}, Rebecca Jean and {Newman}, Jeffrey A. and {Nidever}, David L. and {Nitschelm}, Christian and {Noterdaeme}, Pasquier and {O'Connell}, Julia E. and {Olmstead}, Matthew D. and {Oravetz}, Daniel and {Oravetz}, Audrey and {Osorio}, Yeisson and {Pace}, Zachary J. and {Padilla}, Nelson and {Palanque-Delabrouille}, Nathalie and {Palicio}, Pedro A.},
        title = "{The 16th Data Release of the Sloan Digital Sky Surveys: First Release from the APOGEE-2 Southern Survey and Full Release of eBOSS Spectra}",
      journal = {\apjs},
         year = 2020,
        month = jul,
       volume = {249},
       number = {1},
          eid = {3},
        pages = {3},
          doi = {10.3847/1538-4365/ab929e},
archivePrefix = {arXiv},
       eprint = {1912.02905},
 primaryClass = {astro-ph.GA},
       adsurl = {https://ui.adsabs.harvard.edu/abs/2020ApJS..249....3A}
}

@ARTICLE{Poole2008MNRAS.383..627P,
       author = {{Poole}, T.~S. and {Breeveld}, A.~A. and {Page}, M.~J. and {Landsman}, W. and {Holland}, S.~T. and {Roming}, P. and {Kuin}, N.~P.~M. and {Brown}, P.~J. and {Gronwall}, C. and {Hunsberger}, S. and {Koch}, S. and {Mason}, K.~O. and {Schady}, P. and {vanden Berk}, D. and {Blustin}, A.~J. and {Boyd}, P. and {Broos}, P. and {Carter}, M. and {Chester}, M.~M. and {Cucchiara}, A. and {Hancock}, B. and {Huckle}, H. and {Immler}, S. and {Ivanushkina}, M. and {Kennedy}, T. and {Marshall}, F. and {Morgan}, A. and {Pandey}, S.~B. and {de Pasquale}, M. and {Smith}, P.~J. and {Still}, M.},
        title = "{Photometric calibration of the Swift ultraviolet/optical telescope}",
      journal = {\mnras},
         year = 2008,
        month = jan,
       volume = {383},
       number = {2},
        pages = {627-645},
          doi = {10.1111/j.1365-2966.2007.12563.x},
archivePrefix = {arXiv},
       eprint = {0708.2259},
 primaryClass = {astro-ph},
       adsurl = {https://ui.adsabs.harvard.edu/abs/2008MNRAS.383..627P}
}

@software{Hosseinzadeh_2024_11405219,
  author       = {Hosseinzadeh, Griffin and
                  Bostroem, K. Azalee and
                  Ben-Ami, Tom and
                  Gomez, Sebastian},
  title        = {Light Curve Fitting v0.10.0},
  month        = may,
  year         = 2024,
  publisher    = {Zenodo},
  version      = {v0.10.0},
  doi          = {10.5281/zenodo.11405219},
  url          = {https://doi.org/10.5281/zenodo.11405219},
}

@ARTICLE{SN2011dh_Bersten2012ApJ...757...31B,
       author = {{Bersten}, Melina C. and {Benvenuto}, Omar G. and {Nomoto}, Ken'ichi and {Ergon}, Mattias and {Folatelli}, Gast{\'o}n and {Sollerman}, Jesper and {Benetti}, Stefano and {Botticella}, Maria Teresa and {Fraser}, Morgan and {Kotak}, Rubina and {Maeda}, Keiichi and {Ochner}, Paolo and {Tomasella}, Lina},
        title = "{The Type IIb Supernova 2011dh from a Supergiant Progenitor}",
      journal = {\apj},
         year = 2012,
        month = sep,
       volume = {757},
       number = {1},
          eid = {31},
        pages = {31},
          doi = {10.1088/0004-637X/757/1/31},
archivePrefix = {arXiv},
       eprint = {1207.5975},
 primaryClass = {astro-ph.HE},
       adsurl = {https://ui.adsabs.harvard.edu/abs/2012ApJ...757...31B}
}

\begin{appendix}
\nolinenumbers

\section{Data observation and reduction}
\label{sec:app:dataobs}
We carried out multi-band optical follow-up observations of SN~2025aico in the Johnson-Cousins \textit{BV}, Sloan \textit{ugriz}, and Mephisto \textit{uvgriz} bands after its discovery. Optical data are collected from:
The 1.6m Mephisto telescope with mosaic camera at Lijiang Observatory of Yunnan Astronomical Observatories, Chinese Academy of Sciences, China; The 2.0m Himalayan Chandra Telescope (HCT) with Himalayan Faint Object Spectrograph Camera (HFOSC) of the Indian Astronomical Observatory, Hanle, India; The 0.7m GROWTH-India telescope (GIT) of Indian Astronomical Observatory, Hanle, India; The 0.67\,m/0.92\,m Schmidt telescope with a Moravian camera at Padova Astronomical Observatory, Asiago, Italy; The 1.82\,m Copernico Telescope with the Asiago Faint Object Spectrograph and Camera (AFOSC), Padova Astronomical Observatory, Asiago, Italy; The 2.56\,m Nordic Optical Telescope (NOT), at Observatorio Roque de Los Muchachos, La Palma, Spain, with the Alhambra Faint Object Spectrograph and Camera (ALFOSC).

The initial reduction of all raw photometric data was performed using standard procedures within the \textsc{iraf} \citep{Tody1986SPIE} environment. This process included bias subtraction, overscan correction, image trimming, and flat-fielding. When multiple images in a given band were obtained on the same night, we median-combined them into a single frame to increase the S/N. We updated the world coordinate system solutions by applying \textsc{astrometry.net} \citep{Lang2010AJ....139.1782L} and \textsc{scamp} \citep{Bertin2010ascl.soft10063B}. Subsequently, we performed photometric measurements on these science images using the dedicated pipeline \textsc{autophot}\footnote{\textsc{autophot} is a package for transient photometry using PSF fitting and template subtraction. See \url{https://github.com/Astro-Sean/autophot}} \citep{Brennan2022A&A...667A..62B}, which integrates several photometric packages, such as \textsc{photutils} \citep{Photoutils_Bradley202514889440} for source extraction and flux measurements, and \textsc{hotpants} \citep{Becker2015ascl} for image subtraction. Because SN~2025aico is located in a region contaminated by host-galaxy emission, we employed template subtraction to remove this background flux. Specifically, the instrumental magnitudes of the SN were measured using point spread function (PSF) fitting. The PSF model was constructed by fitting the profiles of isolated and unsaturated stars within the SN field. The resulting PSF model was subtracted from the science frames, and the residuals at the position of the SN were evaluated to ensure fitting quality. In cases where the SN was not detected, an upper magnitude limit was derived.

The instrumental magnitudes of the SN were calibrated by determining the instrumental zero points (ZPs) and colour terms (CTs). Specifically, the calibration for most bands was performed via synthetic photometry derived from Gaia XP spectra utilizing the \textsc{gaiaxpy}\footnote{\textsc{gaiaxpy} is a package to facilitate handling Gaia BP/RP spectra as distributed from the Gaia archive. See \url{https://gaia-dpci.github.io/GaiaXPy-website/}} \citep{GAIAXPY_Daniela202517943031} package. For the Mephisto $ugriz$ bands, we adopted a calibration methodology similar to those described in previous literature \citep{Chen2024apjl, SN2024aecx_Zou2026ApJ...997...77Z, Yang2024ApJ...969..126Y}. Furthermore, to correct the ZPs on non-photometric nights and enhance the overall photometric accuracy, we applied corrections derived from a local sequence of standard stars in the vicinity of the SN.

Additionally, we obtained archival data from several public surveys, including ATLAS, ZTF, and \textit{Swift}/UVOT. The ATLAS \textit{c}- and \textit{o}-band light curves were retrieved from the ATLAS forced photometry service \citep{Shingles2021TNSAN}. ZTF $g$- and $r$-band data were obtained through the ALeRCE \citep{Forster2021AJ} and Lasair \citep{Smith2019RNAAS} brokers. Data from \textit{Swift}/UVOT were reduced using \textsc{heasoft} version 6.35.2, by subtracting the baseline counts from historical data \citep{Brown2009AJ....137.4517B, Brown2014Ap&SS.354...89B}. We used an extraction radius of 5'' and a background radius of 20'' to estimate the fluxes. For the $UBV$ bands, the fluxes were converted to the standard Johnson $UBV$ system by applying the colour corrections by \cite{Poole2008MNRAS.383..627P}. Note that only $UVW2$-, $UVW1$- and $U$- bands are template-subtracted; the other bands lack a template image.

The spectroscopic data were obtained using the following telescopes and instruments: 
The 2.0m Himalayan Chandra Telescope (HCT) with Himalayan Faint Object Spectrograph Camera (HFOSC) of the Indian Astronomical Observatory, Hanle, India; the 1.82\,m Copernico Telescope with AFOSC; the 2.56\,m NOT with ALFOSC; the 3.6\,m Telescopio Nazionale Galileo (TNG) at Observatorio Roque de Los Muchachos, La Palma, Spain, with its Low Resolution Spectrograph (LRS); and the 6.5\,m Magellan Baade Telescope at Las Campanas Observatory, Chile, with the Inamori-Magellan Areal Camera \& Spectrograph (IMACS). Also, we retrieved several public spectroscopic data in TNS, which were reported by Gemini North Telescope in Maunakea Observatory, Hawaii, USA and Multiple Mirror Telescope (MMT) in Fred Lawrence Whipple Observatory, Arizona, USA \citep{Andrews2025TNSCR5179....1A, Rehemtulla2026TNSCR.196....1R}.

The raw spectroscopic data of SN~2025aico were processed using standard routines within the \textsc{iraf} package, following the procedures described in \cite{Cai2022A&A...667A...4C}. Initial calibration steps, including bias subtraction, overscan correction, image trimming, and flat-fielding, were performed following an approach similar to that used for the photometric data reduction. One-dimensional spectra were then optimally extracted from the processed 2D frames. Wavelength calibration was achieved using arc-lamp spectra obtained on the same night, while flux calibration was performed using spectrophotometric standard stars. To verify the accuracy of the flux calibration, the calibrated spectra were compared and scaled against contemporaneous photometry of SN~2025aico. Finally, observations of standard stars were utilized to remove telluric absorption features from the SN spectra. A detailed log of the spectroscopic observations is presented in Table~\ref{apptab:spec:obslog}.

\onecolumn
\section{Complementary tables and figures}
\label{sec:app:ctables}
\begin{figure*}[htbp]
\centering
\includegraphics[width=\linewidth]{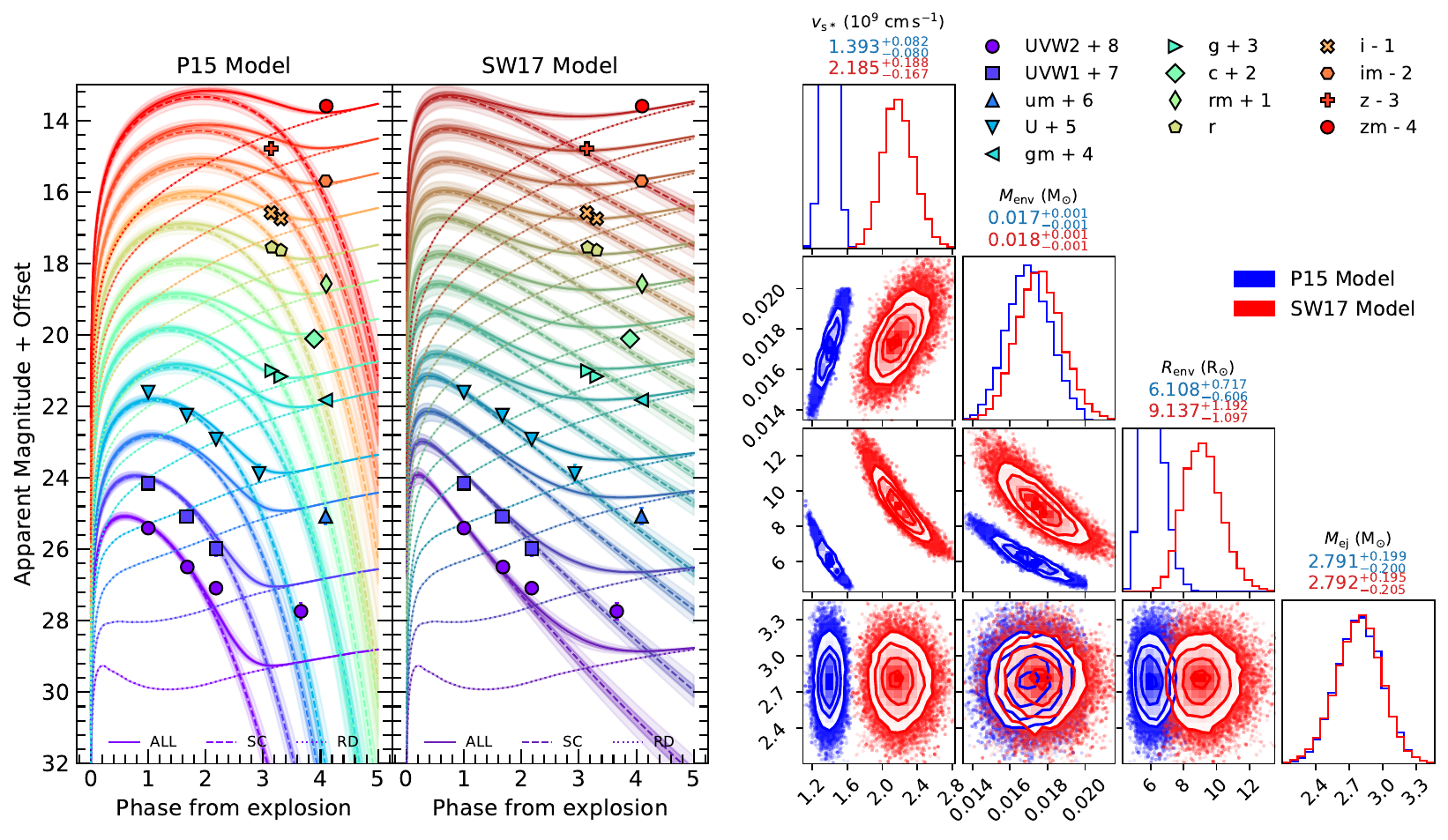}
\caption{Multi-band light curve fits and posterior distributions for SN~2025aico. Left panel: Multi-band modelling of the early-phase shock cooling emission. The dotted lines denote the radioactive-powered diffusion component, while the dashed lines represent the shock-cooling models. Right panel: Corner plots illustrating the parameter constraints. The blue and red regions correspond to the posterior distributions derived from the models of \cite{P15_Piro2015ApJ...808L..51P} and \cite{SW17_Sapir2017ApJ...838..130S}, respectively.}
\label{fig:phot:earlylcmdlcorner}
\end{figure*}
\begin{table}[htbp]
\caption{Spectroscopic observation log of SN~2025aico.}
\label{apptab:spec:obslog}
\small
\setlength{\tabcolsep}{5pt}
\begin{tabular}{cccccccc}
\hline\hline
Date & MJD & Phase$^{a}$ & Instrumental setup & Grism/Grating & Spectral range & Exposure time & Resolution \\
     &     & (d) & & & (\AA) & (s) & (\AA) \\
\hline
2025-12-25 & 61034.62 & $-$20.37 & GEMINI+GMOS-N & B480+G5309 & 3600--7200 & 600 & R$\sim$750\\
2025-12-26 & 61035.20 & $-$19.79 & NOT+ALFOSC & Grism\#4 & 3450--8200 & 1200 & 15\\
2025-12-28 & 61037.93 & $-$17.06 & HCT+HFOSC & Grism\#7+Grism\#8 & 3800--7800 & 2400 & R$\sim$1000\\
2025-12-29 & 61038.98 & $-$16.01 & HCT+HFOSC & Grism\#7+Grism\#8 & 5080--9300 & 2400 & R$\sim$1200\\
2025-12-31 & 61040.10 & $-$14.89 & NOT+ALFOSC & Grism\#4 & 3400--9680 & 1800 & 15\\
2026-01-01 & 61041.86 & $-$13.13 & HCT+HFOSC & Grism\#7+Grism\#8 & 3800--9300 & 2400 & R$\sim$1000\\
2026-01-02 & 61042.89 & $-$12.10 & HCT+HFOSC & Grism\#7+Grism\#8 & 3800--7800 & 2400 & R$\sim$1000\\
2026-01-04 & 61044.88 & $-$10.11 & HCT+HFOSC & Grism\#7+Grism\#8 & 3800--9300 & 2400 & R$\sim$1000\\
2026-01-05 & 61045.80 & $-$9.19  & HCT+HFOSC & Grism\#7+Grism\#8 & 3800--9300 & 2700 & R$\sim$1000\\
2026-01-12 & 61052.86 & $-$2.13  & HCT+HFOSC & Grism\#7+Grism\#8 & 3800--9300 & 1800 & R$\sim$1000\\
2026-01-14 & 61054.11 & $-$0.88  & TNG+LRS & LRB+LRR & 3400--10580 & 1200 & 12\\
2026-01-15 & 61055.78 & $+$0.79  & HCT+HFOSC & Grism\#7+Grism\#8 & 3800--9300 & 1800 & R$\sim$1000\\
2026-01-16 & 61056.54 & $+$1.55  & MMT+BINOSPEC & G270 & 3810--9170 & 180 & R$\sim$1340\\
2026-01-20 & 61060.04 & $+$5.05  & Copernico+AFOSC & VPH6+VPH7 & 3600--8950 & 1200 & 14\\
2026-01-20 & 61060.87 & $+$5.88  & HCT+HFOSC & Grism\#7+Grism\#8 & 3800--9300 & 1800 & R$\sim$1000\\
2026-01-29 & 61069.04 & $+$14.05 & NOT+ALFOSC & Grism\#4 & 3400--9670 & 1800 & 15\\
2026-02-14 & 61085.05 & $+$30.06 & TNG+LRS & LRB & 3240--8010 & 1200 & 12\\
2026-02-14 & 61085.14 & $+$30.15 & NOT+ALFOSC & Grism\#4 & 3400--9680 & 1800 & 15\\
2026-02-22 & 61093.92 & $+$38.93 & Copernico+AFOSC & VPH6 & 6200--10140 & 1800 & 14\\
2026-02-27 & 61098.15 & $+$43.16 & Magellan+IMACS & Gri 300 & 4250--9430 & 900 & 7\\
\hline\hline
\end{tabular}
\vspace{0.3em}
\begin{minipage}{\linewidth}
\footnotesize
$^{a}$Phases are relative to $r_\mathrm{M}$-band maximum luminosity (MJD=61054.99) in the observer frame.
\end{minipage}
\end{table}
\newpage
\begin{figure*}[htbp]
\centering
\includegraphics[width=\linewidth]{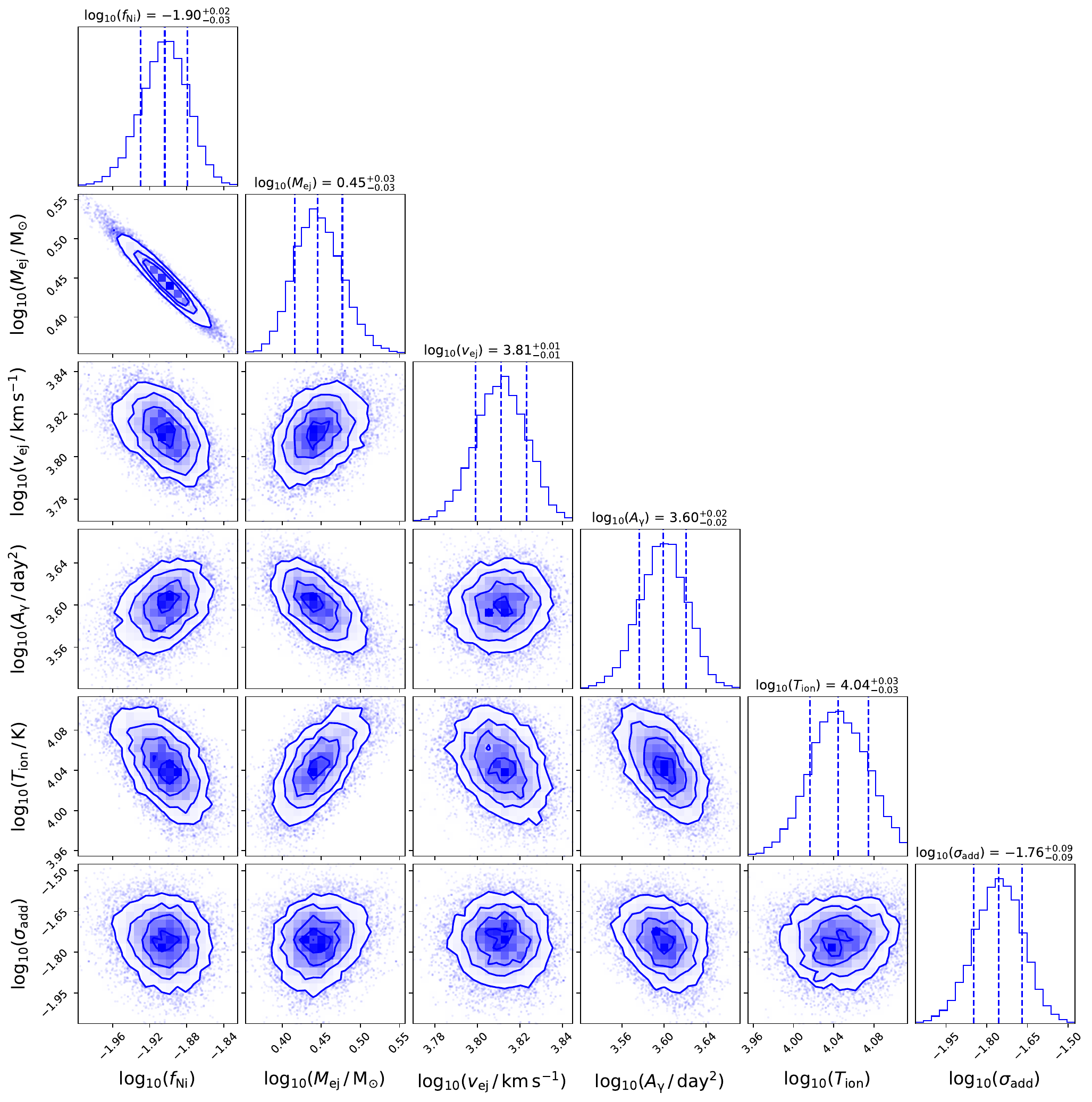}
\caption{Corner plot and posterior distribution of our radioactive-powered light curve fitting result. A confidence interval of $\sim1\sigma$ is marked for each posterior, and the result is marked on each subplot.}
\label{fig:phot:rdmodelcorner}
\end{figure*}
\newpage
\begin{table*}[htbp]
    \centering
    \caption{Apparent light curve parameters of SN~2025aico}
    \small
    \begin{tabular}{cccccccc}
    \hline\hline
        Filter & System & $t_{peak}$ & $m_{peak}$ & $\gamma_{t\lesssim15}$ & $\gamma_{15<t<100}$ \\
        && (MJD) & (mag) & ($\mathrm{mag\times(100~days)^{-1}}$) & ($\mathrm{mag\times(100~days)^{-1}}$)\\
        \hline
        $u_\mathrm{M}$ & AB & $>61045.62$ & $<16.62$ & -- & -- \\
        $g_\mathrm{M}$ & AB & $61052.90 \pm 0.28$ & $15.72 \pm 0.02$ & $8.53 \pm 0.49$ & $1.88 \pm 0.23$ \\
        $r_\mathrm{M}$ & AB & $61054.99 \pm 0.48$ & $15.60 \pm 0.02$ & $6.33 \pm 0.39$ & $2.76 \pm 0.20$ \\
        $i_\mathrm{M}$ & AB & $61055.17 \pm 0.62$ & $15.78 \pm 0.02$ & $3.97 \pm 0.61$ & $2.11 \pm 0.19$ \\
        $z_\mathrm{M}$ & AB & $61056.78 \pm 2.78$ & $15.81 \pm 0.23$ & $3.13 \pm 0.90$ & $3.19 \pm 0.55$ \\
        $u$ & AB & $61050.85 \pm 0.50$ & $16.47 \pm 0.03$ & $32.94 \pm 2.62$ & $1.08 \pm 0.57$ \\
        $g$ & AB & $61052.51 \pm 0.17$ & $15.82 \pm 0.01$ & $12.10 \pm 0.84$ & $2.42 \pm 0.28$ \\
        $r$ & AB & $61055.22 \pm 0.30$ & $15.62 \pm 0.01$ & $6.56 \pm 0.38$ & $2.81 \pm 0.31$ \\
        $i$ & AB & $61055.48 \pm 0.42$ & $15.67 \pm 0.02$ & $3.51 \pm 1.06$ & $2.67 \pm 0.22$ \\
        $z$ & AB & $61055.50 \pm 0.47$ & $15.82 \pm 0.03$ & $3.23 \pm 0.92$ & $1.91 \pm 0.44$ \\
        $B$ & Vega & $61052.46 \pm 0.27$ & $15.98 \pm 0.03$ & $14.55 \pm 0.49$ & $2.18 \pm 0.30$ \\
        $V$ & Vega & $61053.91 \pm 0.28$ & $15.61 \pm 0.03$ & $8.79 \pm 0.80$ & $2.70 \pm 0.24$ \\
        \hline
        \hline
    \end{tabular}
    \label{apptab:phot:apparentlc}
    \vspace{1em}
    \scriptsize 
    \raggedright
\end{table*}
\begin{table*}[htbp]
    \centering
    \caption{Bolometric light curve fitting result of SN~2025aico, using different density profiles and shock-cooling models}
    \begin{tabular}{ccllc}
        \hline
        \hline
        Parameters & Unit & Prior type & Prior & Posterior \\
        \hline
        \multicolumn{4}{c}{\textbf{Radioactive powered \citep{Arnett1989apj}}} \\
        \hline
        $f_{\mathrm{Ni}}$&--&
        $\mathrm{LogUniformPrior}$&
        (0.001,\,1)&
        $0.012^{+0.001}_{-0.001}$\\
        
        $M_{\mathrm{ej}}$&$M_\odot$&
        $\mathrm{LogUniformPrior}$&
        (0.1,\,10)&
        $2.79^{+0.21}_{-0.18}$\\
        
        $v_{\mathrm{ej}}$&$\mathrm{km\,s^{-1}}$&
        $\mathrm{Gaussian}$&
        (6450,\,180)&
        $6470^{+180}_{-170}$\\
        
        $A_{\mathrm{\gamma}}$&$\mathrm{day^2}$&
        $\mathrm{LogUniformPrior}$&
        $(10^{3},\,10^{5})$&
        $3975^{+206}_{-203}$\\
        
        $T_{\mathrm{ion}}$&$\mathrm{kK}$&
        $\mathrm{LogUniformPrior}$&
        $(8,\, 13)$&
        $11.08^{+0.79}_{-0.69}$\\
        
        $\sigma_\mathrm{add}$&$\mathrm{erg\,s^{-1}}$&
        $\mathrm{LogUniformPrior}$&
        $(10^{-6},\,10)$&
        $0.018^{+0.004}_{-0.003}$\\
        
        \hline
        \multicolumn{4}{c}{\textbf{P15 Shock-cooling model \citep{P15_Piro2015ApJ...808L..51P}}} \\
        \hline
        $v_\mathrm{s*}$&$10^{9}\,\mathrm{cm\,s^{-1}}$&
        $\mathrm{LogUniformPrior}$&(0.5, 1.8)&
        $1.393^{+0.082}_{-0.080}$\\
        
        $M_\mathrm{env}$&$M_\odot$&
        $\mathrm{LogUniformPrior}$&(0.001, 0.02)&
        $0.017^{+0.001}_{-0.001}$\\
        
        $R_\mathrm{env}$&$R_\odot$&
        $\mathrm{LogUniformPrior}$&(1, 30)&
        $6.1^{+0.7}_{-0.6}$\\
        
        $M_\mathrm{ej}$&$M_\odot$&
        $\mathrm{Gaussian}$&(2.79, 0.20)&
        $2.795^{+0.204}_{-0.197}$\\
        
        \hline
        \multicolumn{4}{c}{\textbf{SW17 shock-cooling model \citep{SW17_Sapir2017ApJ...838..130S}}} \\
        \hline
        $v_\mathrm{s*}$&$10^{8.5}\,\mathrm{cm\,s^{-1}}$&
        $\mathrm{LogUniformPrior}$&
        $(2, 10)$&
        $6.909^{+0.596}_{-0.528}$\\
        
        $M_\mathrm{env}$&$M_\odot$&
        $\mathrm{LogUniformPrior}$&(0.002, 0.03)&
        $0.018^{+0.001}_{-0.001}$\\
        
        $R_\mathrm{env}$&$R_\odot$&
        $\mathrm{LogUniformPrior}$&(1, 30)&
        $9.1^{+1.1}_{-1.1}$\\
        
        $M_\mathrm{ej}$&$M_\odot$&
        $\mathrm{Gaussian}$&(2.79, 0.20)&
        $2.790^{+0.198}_{-0.201}$\\
        \hline \hline
    \end{tabular}
    \label{apptab:lcfitting}
    \vspace{1em}
    \scriptsize 
    \raggedright
\end{table*}

\newpage
\section{Acknowledgments}
{\small
{We thank the anonymous referee for his/her insightful comments and suggestions that improved the paper.}

This work is supported by the National Key Research and Development Program of China (Grant No. 2024YFA1611603), the National Natural Science Foundation of China (NSFC, Grant Nos. 12303054, 12473047), the Yunnan Fundamental Research Projects (Grant Nos. 202401AU070063, 202501AS070078), the Yunnan Key Laboratory of Survey Science (No. 202449CE340002), and {the International Centre of Supernovae (ICESUN), Yunnan Key Laboratory of Supernova Research (No. 202505AV340004). This publication has emanated from research conducted with the financial support of Taighde \'Eireann -- Research Ireland under Grant number 87581.}
B.K. is supported by the ``Special Project for High-End Foreign Experts,'' Xingdian Funding from Yunnan Province.
Y.-Z. Cai, A.R., and G.V. acknowledge financial support from the SOXS project (PI S. Campana).
A.P., A.R., N.E.R., and G.V. acknowledge support from the PRIN-INAF 2022, ``Shedding light on the nature of gap transients: from the observations to the models.''
A.R. was also supported by the GRAWITA Large Program Grant (PI P. D'Avanzo).
G.C.A. acknowledges support from the Indian National Science Academy (INSA) under their Senior Scientist Programme.
N.E.R. acknowledges support from the Spanish Ministerio de Ciencia e Innovaci\'on (MCIN) and the Agencia Estatal de Investigaci\'on (AEI) 10.13039/501100011033 under the program Unidad de Excelencia Mar\'ia de Maeztu CEX2020-001058-M.
M.D. Stritzinger is funded by the Independent Research Fund Denmark (IRFD, grant number 10.46540/2032-00022B).

We acknowledge the support of the staffs of the various observatories at which data were obtained.
{J.-W~Zhao thanks D.-Z.~Liu for kindly providing part of the data reduction pipeline for Mephisto.}
Mephisto is developed at and operated by the South-Western Institute for Astronomy Research of Yunnan University (SWIFAR-YNU), funded by the ``Yunnan University Development Plan for World-Class University'' and ``Yunnan University Development Plan for World-Class Astronomy Discipline.''
We thank the staff of IAO, Hanle, and CREST, Hosakote, that made these observations possible. The facilities at IAO and CREST are operated by the Indian Institute of Astrophysics, Bangalore.
The GROWTH India Telescope (GIT) is a 70-cm telescope with a 0.7-degree field of view, set up by the Indian Institute of Astrophysics and the Indian Institute of Technology Bombay with support from the Indo-US Science and Technology Forum (IUSSTF) and the Science and Engineering Research Board (SERB) of the Department of Science and Technology (DST), Government of India. It is located at the Indian Astronomical Observatory (Hanle), operated by the Indian Institute of Astrophysics (IIA). We acknowledge funding by the IITB alumni batch of 1994, which partially supports the operations of the telescope.
Based in part on observations made with the Nordic Optical Telescope, owned in collaboration by the University of Turku and Aarhus University, and operated jointly by Aarhus University, the University of Turku, and the University of Oslo, representing Denmark, Finland, and Norway, the University of Iceland, and Stockholm University at the Observatorio del Roque de los Muchachos, La Palma, Spain, of the Instituto de Astrofisica de Canarias.
Observations from the NOT were obtained through the NUTS2 collaboration, which is supported in part by the Instrument Centre for Danish Astrophysics (IDA), and the Finnish Centre for Astronomy with ESO (FINCA) via Academy of Finland grant nr 306531. The data presented here were obtained in part with ALFOSC, which is provided by the Instituto de Astrofisica de Andalucia (IAA) under a joint agreement with the University of Copenhagen and NOTSA.
This article is also based on observations made with the Italian Telescopio Nazionale Galileo (TNG), operated on the island of La Palma by the Fundaci\'on Galileo Galilei of the INAF (Istituto Nazionale di Astrofisica) at the Spanish Observatorio del Roque de los Muchachos of the Instituto de Astrof\'{\i}sica de Canarias, under the program A50TAC\_41 (PI: G. Valerin).
Based in part on observations collected at Copernico and Schmidt telescopes (Asiago Mount Ekar, Italy) of the INAF -- Osservatorio Astronomico di Padova.
This paper includes data gathered with the 6.5 meter Magellan Telescopes located at Las Campanas Observatory, Chile.

This work has made use of data from the Asteroid Terrestrial-impact Last Alert System (ATLAS) project. The Asteroid Terrestrial-impact Last Alert System (ATLAS) project is primarily funded to search for near-Earth objects (NEOs) through National Aeronautics and Space Administration (NASA) grants NN12AR55G, 80NSSC18K0284, and 80NSSC18K1575; byproducts of the NEO search include images and catalogs from the survey area. This work was partially funded by Kepler/K2 grant J1944/80NSSC19K0112 and HST GO-15889, and STFC grants ST/T000198/1 and ST/S006109/1. The ATLAS science products have been made possible through the contributions of the University of Hawaii Institute for Astronomy, the Queen's University Belfast, the Space Telescope Science Institute, the South African Astronomical Observatory, and The Millennium Institute of Astrophysics (MAS), Chile.

ZTF is supported by the National Science Foundation under Grants No. AST-1440341 and AST-2034437 and a collaboration including current partners Caltech, IPAC, the Oskar Klein Center at Stockholm University, the University of Maryland, University of California, Berkeley, the University of Wisconsin at Milwaukee, University of Warwick, Ruhr University, Cornell University, Northwestern University and Drexel University. Operations are conducted by COO, IPAC, and UW.

We acknowledge the use of public data from the \textit{Swift} data archive. SDSS is managed by the Astrophysical Research Consortium for the Participating Institutions of the SDSS Collaboration including the Brazilian Participation Group, the Carnegie Institution for Science, Carnegie Mellon University, Center for Astrophysics -- Harvard \& Smithsonian (CfA), the Chilean Participation Group, the French Participation Group, Instituto de Astrof\'isica de Canarias, The Johns Hopkins University, Kavli Institute for the Physics and Mathematics of the Universe (IPMU) / University of Tokyo, the Korean Participation Group, Lawrence Berkeley National Laboratory, Leibniz Institut f\"{u}r Astrophysik Potsdam (AIP), Max-Planck-Institut f\"{u}r Astronomie (MPIA Heidelberg), Max-Planck-Institut f\"{u}r Astrophysik (MPA Garching), Max-Planck-Institut f\"{u}r Extraterrestrische Physik (MPE), National Astronomical Observatories of China, New Mexico State University, New York University, University of Notre Dame, Observat\'orio Nacional / MCTI, The Ohio State University, Pennsylvania State University, Shanghai Astronomical Observatory, United Kingdom Participation Group, Universidad Nacional Aut\'onoma de M\'exico, University of Arizona, University of Colorado Boulder, University of Oxford, University of Portsmouth, University of Utah, University of Virginia, University of Washington, University of Wisconsin, Vanderbilt University, and Yale University. This research has made use of the NASA/IPAC Extragalactic Database (NED), which is operated by the Jet Propulsion Laboratory, California Institute of Technology, under contract with NASA.
}
\end{appendix}

\end{document}